\documentclass[superscriptaddress, aps, prd, reprint, nofootinbib]{revtex4-2}

\usepackage{graphicx}
\usepackage{tikz}
\usepackage{amsmath,amssymb,mathtools,mathrsfs,booktabs,multirow}
\usepackage{dcolumn}
\usepackage{bm}
\usepackage{xcolor}
\usepackage{soul}
\usepackage{braket}
\usepackage[separate-uncertainty=true,multi-part-units=single]{siunitx}
\usepackage{scalerel}
\usepackage{enumitem}
\usepackage{dsfont}
\usepackage{comment}

\definecolor{Mathematica1}{rgb}{0.368417, 0.506779, 0.709798}
\definecolor{Mathematica2}{rgb}{0.880722, 0.611041, 0.142051}
\definecolor{Mathematica3}{rgb}{0.560181, 0.691569, 0.194885}
\definecolor{darkred}{rgb}{0.545,0,0}
\definecolor{dullblue}{rgb}{0,0.298,0.49}
\definecolor{blue3}{RGB}{31, 119, 180}
\definecolor{andreapurple}{RGB}{130,0,180}
\usepackage[colorlinks=true
,urlcolor=dullblue
,anchorcolor=blue3
,citecolor=dullblue
,filecolor=blue3
,linkcolor=darkred
,menucolor=blue3
,pagecolor=blue3
,linktocpage=true
,pdfproducer=medialab
]{hyperref}

\makeatletter
\def\doauthor#1#2#3{%
  \global\let\@extramark\@empty
  \ignorespaces#1\unskip\@listcomma
  \begingroup
   #3%
  \@if@empty{#2}{\endgroup{}{}}{\endgroup{\comma@space}{}\frontmatter@footnote{#2}}%
  \@extramark
  \space \@listand
}%
\newcommand{\extranote}[1]{%
  \texorpdfstring{\gdef\@extramark{\comma@space\frontmatter@footnote{#1}}}{}%
}
\makeatother

\newcommand{\dd}{\mathop{\mathrm{d}\!}{}}

\renewcommand{\vec}[1]{\boldsymbol{\mathbf{#1}}}

\DeclareMathOperator\artanh{artanh}

\DeclareMathOperator\tr{tr}
\DeclareMathOperator\sint{si}
\DeclareMathOperator\Si{Si}
\DeclareMathOperator\Ci{Ci}

\renewcommand{\Im}{\operatorname{Im}}
\renewcommand{\Re}{\operatorname{Re}}

\DeclareMathOperator\diag{diag}

\newcommand{\ped}[1]{_{\mathrm{#1}}}
\renewcommand{\ap}[1]{^{\mathrm{#1}}}
\newcommand{\sped}[1]{_{\textsc{#1}}}
\newcommand{\sap}[1]{^{\textsc{#1}}}
\DeclarePairedDelimiter{\abs}{\lvert}{\rvert}

\def\beq{\begin{equation}}
\def\eeq{\end{equation}}

\begin{document}

\title{Evolution of Binaries Under Stochastic Perturbations}

\author{Andrea Caputo}
\thanks{These authors contributed equally.}
\affiliation{Department of Theoretical Physics, CERN, 1217 Geneva, Switzerland}
\affiliation{Dipartimento di Fisica, ``Sapienza'' Università di Roma \& Sezione INFN Roma 1, 00185 Roma, Italy}
\affiliation{Department of Particle Physics and Astrophysics, Weizmann Institute of Science, Rehovot 7610001, Israel}

\author{Giovanni Maria Tomaselli\extranote{\href{mailto:tomaselli@ias.edu}{tomaselli@ias.edu}}}
\thanks{These authors contributed equally.}
\affiliation{School of Natural Sciences, Institute for Advanced Study, Princeton, NJ 08540, USA}

\author{Chris Hamilton}
\affiliation{School of Natural Sciences, Institute for Advanced Study, Princeton, NJ 08540, USA}
\affiliation{Department of Astrophysical Sciences, Princeton University, Princeton, NJ 08544, USA}

\begin{abstract}

We develop a general Fokker--Planck framework describing the dynamical evolution of Keplerian binaries subjected to stochastic perturbations. The formalism provides an algorithmic way to obtain the Fokker--Planck drift and diffusion coefficients of any set of orbital variables given the statistics of the perturbations. We apply the method to three physically distinct regimes: adiabatic tidal perturbations, white-noise tidal perturbations, and impulsive encounters with a third body of arbitrary density profile. In each regime we provide explicit drift and diffusion coefficients for all six orbital elements, derive the associated evolution timescale, and obtain analytic steady-state distribution functions. Our results extend previous treatments by including the evolution of the binary's orientation, retaining the complete tensor structure of tidal correlators, treating non-pointlike perturbers, and resolving the exact geometry of impulsive encounters. The latter correction leads to a steady-state eccentricity distribution that is slightly sub-thermal. We also show how these equations can be applied directly in several astrophysical scenarios, including binaries perturbed by dark matter subhaloes, ultralight dark matter, and the interstellar medium. This work delivers both a complete mathematical framework and a practical toolkit for stochastic binary evolution, providing ready-to-evaluate equations to be applied directly to binary population data.

\end{abstract}

\maketitle

\begingroup
\setcounter{tocdepth}{2} 
\makeatletter
\let\oldcontentsline\contentsline
\renewcommand{\contentsline}[4]{%
  \def\tempa{#1}%
  \def\tempb{subsubsection}%
  \def\tempc{section}%
  \ifx\tempa\tempb
  \else
    \ifx\tempa\tempc
      \def\temptitle{#2}%
      \ifx\temptitle\tempd
      \else
        \oldcontentsline{#1}{#2}{#3}{#4}%
      \fi
    \else
      \oldcontentsline{#1}{#2}{#3}{#4}%
    \fi
  \fi
}
\makeatother
\tableofcontents
\endgroup

\section{Introduction}

Binary systems are central to much active theoretical and observational research in astrophysics. Astronomers study binaries for their role in shaping stellar evolution \cite{El-Badry:2024vjt}, their capacity to affect the long-term evolution of star clusters \cite{Sigurdsson:1993tui,2003gmbp.book.....H}, the tendency of (compact object) binaries to merge and produce gravitational wave sources \cite{LIGOScientific:2016aoc}, and many other reasons. The dynamical properties of these binaries---the evolution of their semi-major axes, eccentricities, orientations, etc.---are critical ingredients when attacking these problems. At the same time, binaries often serve as sensitive probes of astrophysical \textit{environments}, and so are of interest in order to constrain, e.g., the nature and distribution of dark matter and other unseen objects in the Milky Way \cite{1985ApJ...290...15B,Ramirez:2022mys} and dwarf galaxies \cite{Penarrubia:2016ltr,Shariat:2025dxs}, the properties of interstellar medium turbulence and star formation regions \cite{Xu_2023,Mathew_2024}, and even the nature of gravity \cite{Banik:2018ydl}. In many of these cases, it is necessary that we have a solid understanding of the dynamical evolution of binaries when they are subjected to gravitational perturbations.

Gravitational perturbations to binaries can be divided into two main categories. Perturbations that maintain coherence over a timescale comparable to or longer than the evolution timescale they induce on the binary can be considered \emph{deterministic}. Examples include von Zeipel--Lidov--Kozai oscillations driven by a bound tertiary companion \citep{1910AN....183..345V,Kozai:1962zz,Lidov:1962wjn} and torques due to the Galactic tide \citep{1986Icar...65...13H,Hamilton_2022,Modak_2023,Stegmann:2024rnk} and cluster tides \cite{Hamilton_2019,Rasskazov:2023tal}. On the other hand, perturbations that lose coherence faster than the binary evolves are best considered \emph{stochastic}. 

Here, we focus on stochastic perturbations and the resulting binary evolution. This problem has previously been studied in many specific cases. For instance, Refs.~\cite{Heggie:1975tg,1987ApJ...312..367W,Jiang:2009ax,2024MNRAS.532.2425H,Bhalla:2025xce} considered the effect of passing stars (or any other pointlike object) upon the binary semi-major axis and eccentricity evolution, Ref.~\cite{Penarrubia:2019wei} studied the effect of forcing by random tidal fields, and Refs.~\cite{Blas:2021mpc, Blas:2021mqw} focused on the evolution driven by a stochastic gravitational-wave background. In all these cases, under the assumption of weak perturbations, a diffusive dynamics emerges, where the evolution of an ensemble of binaries is governed by a Fokker--Planck equation on the space of orbital parameters.

\begin{figure*}[p]
\centering
\begin{tikzpicture}[seclink/.style={fill=white, fill opacity=0.85,
    text opacity=1, rounded corners=1.2pt, inner sep=1.8pt, font=\small}]
\node[anchor=south west, inner sep=0pt, outer sep=0pt] (regimesimg)
  {\includegraphics[width=0.73\textwidth]{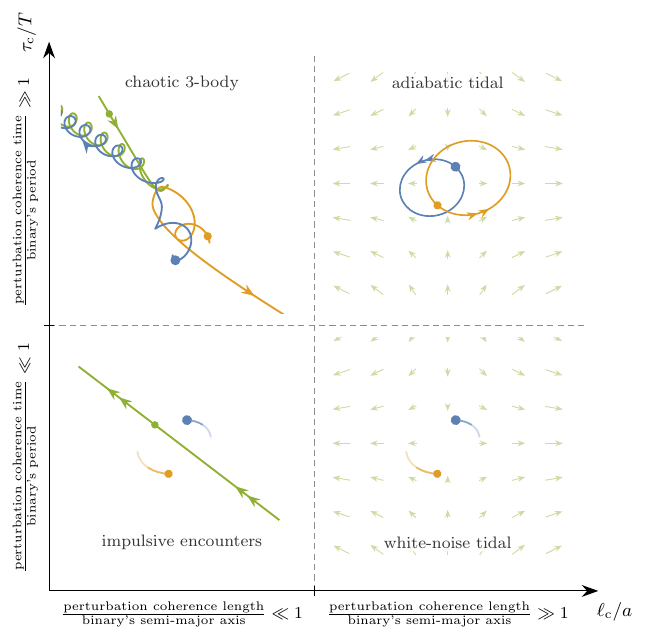}};
\begin{scope}[x={(regimesimg.south east)}, y={(regimesimg.north west)}]
\node[seclink] at (0.6939,0.8351) {Sec.~\ref{sec:adiabatic}};
\node[seclink] at (0.6939,0.1173) {Sec.~\ref{sec:white-noise}};
\node[seclink] at (0.2820,0.1173) {Sec.~\ref{sec:impulsive}};
\end{scope}
\end{tikzpicture}
\caption{Classification of the types of stochastic perturbations considered in this paper, based on how their coherence length $\ell\ped{c}$ and time $\tau\ped{c}$ compare to the binary's semi-major axis $a$ and period $T$. In this paper, we study the adiabatic tidal, white-noise tidal, and impulsive encounter regimes. We do not study chaotic three-body interactions.}
\label{fig:regimes}
\end{figure*}
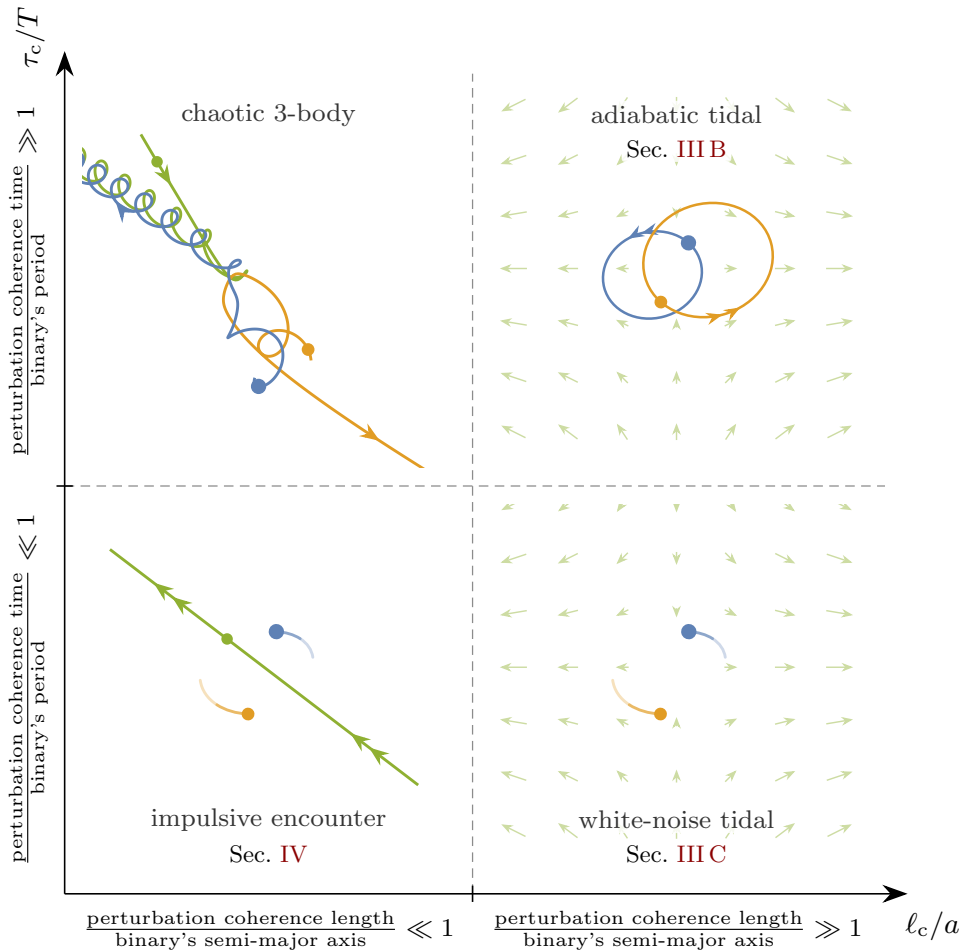

In this work, we unify all these cases into a single theoretical framework. We present a general derivation of the Fokker--Planck equation, giving particular emphasis to all the assumptions that underlie various physical applications. We show how the Fokker--Planck coefficients can be obtained in a simple and algorithmic way, once the statistics of the perturbing forces are specified and a set of orbital variables is chosen. We then study three particular limits of the general problem---adiabatic tidal perturbations, white-noise tidal perturbations, and impulsive encounters, as sketched in Fig.~\ref{fig:regimes}---and write down the Fokker--Planck coefficients and the typical evolution timescale for each of them. 
We also demonstrate explicitly how our results can be readily applied to a number of interesting astrophysical contexts, including binaries perturbed by fluctuating density fields like ultralight dark matter or the interstellar medium, and by flybys from dark matter substructures.
The reader who is less interested in the formal side of the problem may then use our paper as a \emph{dictionary} of closed-form, ready-to-evaluate expressions of the Fokker--Planck coefficients in various physical cases, with these examples as a guide. Table~\ref{tab:users-summary} provides a user's summary of these results.

\begin{table*}[p]
\centering
\small
\setlength{\tabcolsep}{3pt}
\renewcommand{\arraystretch}{1.5}
\begin{tabular*}{\textwidth}{@{\extracolsep{\fill}}ccccc@{}}
\toprule
Regime & Coeffs.\ $(a,e)$ & Coeffs.\ (orientation) & $T\sped{d}$ & Steady-state \\
\midrule
Adiabatic tidal
& \eqref{eqn:Ba-adiabatic}--\eqref{eqn:Dee-adiabatic}
& \eqref{eqn:Dhatehate-adiabatic}--\eqref{eqn:DM0M0-adiabatic}
& \eqref{eqn:Td-tidal}, \eqref{eqn:Td-Prho}
& \eqref{eqn:f-thermal} \\
White-noise tidal
& \eqref{eqn:Ba-white-noise}--\eqref{eqn:Dee-white-noise}
& \eqref{eqn:Dhatehate-white-noise}--\eqref{eqn:DM0M0-white-noise}
& \eqref{eqn:Td-tidal}, \eqref{eqn:Td-Prho}
& \eqref{eqn:f-zero-flux}, \eqref{eqn:f-steady-state} \\
Impulsive encounters (arbitrary profile)
& \eqref{eqn:Ba-impulsive}--\eqref{eqn:Dee-impulsive}
& \eqref{eqn:Dhatehate-impulsive}--\eqref{eqn:DM0M0-impulsive}
& \eqref{eqn:Td-impulsive}
& --- \\
Impulsive encounters (pointlike perturbers)
& \eqref{eqn:Ba-point-mass}--\eqref{eqn:Dee-point-mass}
& \eqref{eqn:Dhatehate-point-mass}--\eqref{eqn:DM0M0-point-mass}
& \eqref{eqn:Td*}
& \eqref{eqn:f-ss-point-mass} \\
\bottomrule
\end{tabular*}%
\caption{User's summary of the main results. We provide hyperlinks to the Fokker--Planck coefficients for the $(a,e)$ (semi-major axis and eccentricity) and for the orbital orientation sectors, as well as to formulae for the characteristic diffusion time $T\sped{d}$ and for the steady-state distributions.}
\label{tab:users-summary}
\end{table*}

Introducing a general and mechanical methodology is functional beyond its purely academic interest. Indeed, we use it to generalize in different directions the results previously available in the literature. For instance, we study the evolution of all orbital elements, including the ones that determine the binary's orientation in space: to the best of our knowledge, this was previously done only for gravitational-wave backgrounds \cite{Blas:2021mpc, Blas:2021mqw}, or partially in other cases \cite{Merritt:2001dc}. We also retain the full tensor structure of tidal perturbations (which previous treatments failed to do), generalize the case of impulsive encounters to perturbers of arbitrary density profile, and carry out the impulsive calculation for pointlike perturbers without any kinematical or geometrical approximations. This last correction reveals that the steady-state eccentricity distribution of binaries subjected to impulsive kicks from pointlike perturbers slightly deviates from the thermal distribution, contrary to what was previously thought.

This study is particularly timely because astronomers increasingly have access to large and accurate datasets of binary orbits, mostly thanks to Gaia~\cite{Gaia:2016zol}. The number of binaries with known orbital parameters is expected to increase dramatically following the fourth Gaia data release (DR4) in December 2026, making the present time particularly exciting for related theoretical predictions.
Given suitable initial conditions, solutions of the Fokker--Planck equations provided in this paper can be directly compared with such observational data. 
Alternatively, if the initial conditions are uncertain (as is the case, e.g., for wide stellar binaries \cite{Hwang_2022,2024MNRAS.532.2425H,Mathew_2024}), then our Fokker--Planck approach can be used for rapid forward modeling, allowing the space of initial conditions to be searched efficiently.

The paper is organized as follows. In Sec.~\ref{sec:formalism} we present the formalism and the assumptions underlying the Fokker--Planck description of binary evolution under stochastic perturbations. In Secs.~\ref{sec:tidal} and~\ref{sec:impulsive} we specialize to tidal perturbations and impulsive encounters, respectively. We then apply the results to concrete physical examples in Sec.~\ref{sec:applications}, compare them with previous works and other competing physical effects in Sec.~\ref{sec:discussion}, and conclude in Sec.~\ref{sec:conclusions}. Appendices \ref{sec:kramers-moyal}, \ref{sec:coeffs-euler-angles}, \ref{sec:first-principles}, and \ref{sec:numerical-check} contain additional technical details and verifications of our results.

We publicly release on GitHub symbolic calculation scripts and numerical integrators, which can be used to reproduce all the analytical and numerical results of this paper, including the figures \cite{stochastic-binaries:github}.

\clearpage

\section{General formalism}
\label{sec:formalism}

In this section, we set up a general formalism for the stochastic evolution of binary orbital elements, and identify the three key asymptotic regimes that will be the focus of Secs.~\ref{sec:tidal}--\ref{sec:impulsive}.

\subsection{Equations of motion and timescales}

Consider a Keplerian binary with total mass $m$, period $T=2\pi\sqrt{a^3/Gm}$, and specific energy $\mathcal E=-Gm/(2a)$, immersed within a given environment. We are interested in how the binary parameters evolve under the environmental perturbations. Such perturbations will generally include both a ``smooth'' deterministic part (e.g., galactic tides), as well as a ``noisy'' stochastic part. In this work we focus on the stochastic terms and ignore deterministic forces. If those are known and relevant, they can simply be added alongside stochastic forces. We will return to a brief discussion of deterministic perturbations in Sec.~\ref{sec:deterministic}.

We denote the internal state of the binary with the vector $\vec w^{*}$. This vector is generically 6-dimensional, and multiple parametrizations are possible. For example, we can use the Keplerian orbital elements $(M,a,e,\Omega,i,\omega)$ relative to some fixed cartesian axes $(x,y,z)$: mean anomaly $M$, semi-major axis $a$, eccentricity $e$, longitude of ascending node $\Omega$, inclination $i$, and argument of pericenter $\omega$. Alternatively, we could use Delaunay variables, where $(a,e,i)$ are replaced by the actions $(L,J,J_z)$ conjugate to the angles $(M,\omega,\Omega)$.\footnote{The Delaunay variables can be expressed in terms of the Keplerian orbital elements as $L=\sqrt{Gma}$, $J=L\sqrt{1-e^2}$, $J_z=J\cos i$.} The chosen parametrization is not relevant for the purpose of the general discussion of this section; we will make a choice when discussing specific cases later.

The equation of motion for $\vec w^{*}$ is
\beq
\dot{\vec w}^{*}=\vec F^{*}(\vec w^{*}, t)=\vec F^{*}\sped{k}(\vec w^{*})+\vec F^{*}\sped{s}(\vec w^{*}, t)\,.
\label{eqn:eom}
\eeq
The term $\vec F^{*}\sped{k}$ encodes the unperturbed Keplerian evolution, and reads $\vec F^{*}\sped{k}=(2\pi/T,0,0,0,0,0)$ in either Keplerian or Delaunay variables. The other term, $\vec F^{*}\sped{s}(\vec w^{*}, t)$, corresponds to the stochastic perturbations. This force depends on the specific realization of the noise, hence the explicit dependence on time. Furthermore, $\vec F^{*}\sped{s}(\vec w^{*}, t)$ exhibits a characteristic \emph{coherence time} $\tau\ped{c}$ and \emph{coherence length} $\ell\ped{c}$. We can define $\tau\ped{c}$ as the typical time over which the correlator $\braket{\vec F^{*}\sped{s}(t)\vec F^{*}\sped{s}(t+\tau)}$ decays as a function of $\tau$. Similarly, $\ell\ped{c}$ is the typical physical length over which the \emph{force field} giving rise to $\vec F^{*}\sped{s}$ auto-decorrelates at fixed time. Here we focus on the timescales only---we will return to a discussion of the length scales in Sec.~\ref{sec:regimes}.

There are three distinct timescales in the problem described by Eq.~\eqref{eqn:eom}. First, the binary's orbital period $T$ governs the evolution of the orbital phase. Second, we have the coherence time $\tau\ped{c}$ introduced above, which depends on the properties of the environment. Third, all orbital variables except for the fast orbital phase evolve under the stochastic perturbation over a timescale that we denote by $T\sped{d}$.\footnote{At this stage, $T\sped{d}$ simply denotes the evolution timescale of the slow orbital elements. We will later identify it with the diffusion timescale of a Fokker--Planck equation and write down explicit expressions for it.} Throughout the paper, we assume that
\beq
\max\{T,\tau\ped{c}\}\ll T\sped{d}\,.
\label{eqn:Ttauc<<Td}
\eeq
The assumption $T\ll T\sped{d}$ means that the perturbation to the binary is \emph{weak}, which allows us to apply the standard techniques of perturbation theory. The assumption $\tau\ped{c}\ll T\sped{d}$ means that the stochastic force can be treated as a \emph{random noise} over the longer diffusion timescale.  We can then assume that the time average of $\vec F^{*}\sped{s}$ vanishes, as any nonzero average can be incorporated in the deterministic part of the perturbation, which we are ignoring here. 

Given the separation of timescales, it is useful to isolate the orbital phase, which evolves over the timescale $T$, from the remaining parameters, which change only over the much longer diffusion timescale $T\sped{d}$. In particular, it is convenient to decompose the mean anomaly $M$ into a rapidly advancing orbital phase and a slowly varying phase offset. The latter coincides with the mean anomaly at a reference epoch in the unperturbed problem. We therefore write $M=\mathcal M+M_0\,$, where $\mathcal M$ tracks the fast orbital phase, with $\dot{\mathcal M}=2\pi/T$, while $M_0$ evolves slowly under the action of the stochastic perturbation. We then collect the six slow variables into the vector $\vec w\equiv(a,e,\Omega,i,\omega,M_0)\,$. In this parametrization, the unperturbed Keplerian motion leaves $\vec w$ unchanged. The perturbing force can nevertheless depend on both fast and slow variables so that the evolution of the slow variables can be generically written as $\dot{\vec w}=\vec F\sped{s}(\vec w^*,t)$. We will interpret $\vec w^*$ as an extended set of variables, $\vec w^*=(\mathcal M,\vec w)$. Note that this parametrization is intentionally redundant, since the physical state depends on $\mathcal M$ and $M_0$ only through their sum $M=\mathcal M+M_0$.

\makeatletter\@secpenalty=\@M\makeatother 
\subsection{Ensemble average}
\label{sec:ensemble-average}
\makeatletter\@secpenalty=-300\makeatother

We now imagine a large ensemble of binaries, each having the same total mass and being subject to a different realization of the stochastic force $\vec F^{*}\sped{s}(\vec w^{*}, t)$. Consider an interval of time $\Delta t$ after a reference time $t=0$, such that $\max(T,\tau\ped{c})\ll\Delta t\ll T\sped{d}$, long enough to average over many orbital periods and/or many coherence times, but short enough that the slow variables barely evolve. To first order in $\Delta t$, the variation of the $\mu$-th component of $\vec w$ over such interval is
\begin{widetext}
\beq
\begin{split}
\Delta w^\mu&=\int_0^{\Delta t}F^\mu\sped{s}(\mathcal M(t),\vec w(t), t)\dd t\\
&=\int_0^{\Delta t}\big(F^\mu\sped{s}(\mathcal M(t),\vec w(0), t)+\partial_{w^\nu}F^\mu\sped{s}(\mathcal M(t),\vec w(0), t)(w^\nu(t)-w^\nu(0))\big)\dd t+\mathcal O((\Delta t)^2)\\
&=\int_0^{\Delta t}F^\mu\sped{s}(\mathcal M(t),\vec w(0), t)\dd t+\int_0^{\Delta t}\int_0^{t}\partial_{w^\nu}F^\mu\sped{s}(\mathcal M(t),\vec w(0), t)F^\nu\sped{s}(\mathcal M(t'),\vec w(0), t')\dd t'\dd t+\mathcal O((\Delta t)^2)\,.
\end{split}
\label{eqn:deltaw*}
\eeq
\end{widetext}
We now take the ensemble average, which we denote with $\braket{\cdots}$. Under the assumption that the ensemble average of $\vec F\sped{s}$ is zero, the first term vanishes. This must be true, e.g., for any ergodic $\vec F\sped{s}$, in which case time and ensemble averages are equal. As for the second term, it is useful to distinguish the cases $T\ll\tau\ped{c}$, which we will refer to as ``adiabatic'', and $\tau\ped{c}\ll T$, which we call either ``white-noise'' or ``impulsive''.

In the adiabatic case ($T\ll\tau\ped{c}$), the stochastic force retains coherence over many orbital periods. Over one coherence time, the dependence on the rapidly evolving orbital phase $\mathcal M$ is therefore averaged, so that we can replace $F^\mu\sped{s}(\mathcal M,\vec w,t)$ with its orbital average,
\beq
\bar F^\mu\sped{s}(\vec w,t)\equiv\frac1{2\pi}\int_0^{2\pi} F^\mu\sped{s}(\mathcal M,\vec w,t)\dd\mathcal M\,,
\eeq
within each of the two integrals. We can then change variables from $t'$ to $\tau=t'-t$ and use $\tau\ped{c}\ll\Delta t$ to extend the limits of the inner integral to infinity,
\beq
\braket{\Delta w^\mu}=\frac{\Delta t}2\int_{-\infty}^\infty\braket{\partial_{w^\nu}\bar F^\mu\sped{s}(\vec w,t)\bar F^\nu\sped{s}(\vec w,t+\tau)}\dd\tau\,,
\label{eqn:Deltaw1}
\eeq
where $\vec w$ is intended to be evaluated at $t=0$, and we assumed that the correlator is an even function of $\tau$.\footnote{More generally, if the mixed correlator is not an even function of $\tau$, the drift should be written as a one-sided integral over $\tau>0$ without the factor of $1/2$.}

Conversely, in the white-noise/impulsive case ($\tau\ped{c}\ll T$), the force loses coherence in a time much shorter than one orbit. Over the time interval where the correlator has support, the orbital phase changes only by a small amount. We can thus first transform the inner integral as described above, obtaining an $\mathcal M$-dependent correlator. Then, we can use $\Delta t \gg T$ to average the outer integral over many orbits, and replace it by its orbital average, finally arriving at
\beq
\braket{\Delta w^\mu}=\frac{\Delta t}2\int_{-\infty}^\infty\overline{\braket{\partial_{w^\nu}F^\mu\sped{s}(\vec w^*,t)F^\nu\sped{s}(\vec w^*,t+\tau)}}\dd\tau\,.
\label{eqn:Deltaw2}
\eeq

The variation $\braket{\Delta w^\mu}$ is needed to turn the stochastic differential equation $\dot w^\mu=F^\mu\sped{s}(\vec w^*,t)$ into a partial differential equation for the distribution of $w^\mu$. Alongside $\braket{\Delta w^\mu}$, we also need to compute the quadratic terms
$\braket{\Delta w^\mu\Delta w^\nu}$. Working again to first order in $\Delta t$, we have
\beq
\begin{split}
\Delta w^\mu\Delta w^\nu={}&\int_0^{\Delta t}F^\mu\sped{s}(\mathcal M(t),\vec w(t), t)\dd t\\
&\times\int_0^{\Delta t}F^\nu\sped{s}(\mathcal M(t'),\vec w(t'), t')\dd t'\,.
\end{split}
\eeq
As before, after taking an ensemble average and manipulating the integrals, we arrive at
\beq
\braket{\Delta w^\mu\Delta w^\nu}=\Delta t\int_{-\infty}^\infty\braket{\bar F^\mu\sped{s}(\vec w,t)\bar F^\nu\sped{s}(\vec w,t+\tau)}\dd\tau
\label{eqn:Deltaww1}
\eeq
for $T\ll\tau\ped{c}$, and
\beq
\braket{\Delta w^\mu\Delta w^\nu}=\Delta t\int_{-\infty}^\infty\overline{\braket{F^\mu\sped{s}(\vec w^*,t)F^\nu\sped{s}(\vec w^*,t+\tau)}}\dd\tau
\label{eqn:Deltaww2}
\eeq
for $\tau\ped{c}\ll T$.

\subsection{Fokker--Planck equation}

We define the probability density function of the slow orbital variables of the ensemble of binaries as $f(\vec w,t)=\braket{\delta(\vec w-\vec w(t))}$, normalized so that $\int f(\vec w,t)\dd\vec w=1$. On timescales much longer than both the orbital period and the coherence time of the perturbation, but still short compared to the slow variables' evolution timescale, the stochastic evolution can be described in terms of the first- and second-order ensemble-averaged increments, $\braket{\Delta w^\mu}$ and $\braket{\Delta w^\mu\Delta w^\nu}$. As reviewed in Appendix~\ref{sec:kramers-moyal}, this leads to a Fokker--Planck equation for $f(\vec w,t)$,
\beq
\frac{\partial f}{\partial t}=-\frac{\partial}{\partial w^\mu}\big(B^\mu f\big)+\frac12\frac{\partial^2}{\partial w^\mu\partial w^\nu}\big(D^{\mu\nu}f\big)\,,
\label{eqn:fp}
\eeq
where the drift and diffusion coefficients are defined by
\beq
B^\mu=\frac{\braket{\Delta w^\mu}}{\Delta t}\,,\qquad D^{\mu\nu}=\frac{\braket{\Delta w^\mu\Delta w^\nu}}{\Delta t}\,.
\label{eqn:drift-diffusion-def}
\eeq
Once the stochastic force $F^\mu\sped{s}$ and its two-point correlators are specified, the coefficients $B^\mu$ and $D^{\mu\nu}$ follow directly from Eqs.~\eqref{eqn:Deltaw1} and \eqref{eqn:Deltaww1} in the adiabatic regime, or from Eqs.~\eqref{eqn:Deltaw2} and \eqref{eqn:Deltaww2} in the white-noise/impulsive regime.

It is useful to see how the Fokker--Planck equation transforms under a change of variables $w^\mu\to\tilde w^{\tilde\mu}$. Define the transformed probability density $\tilde f=\abs{\partial w/\partial\tilde w}f$, where $\abs{\partial w/\partial\tilde w}$ is the Jacobian of the transformation. Then $\tilde f$ satisfies a Fokker--Planck equation with drift coefficients
\beq
\tilde B^{\tilde \mu}=\frac{\partial\tilde w^{\tilde\mu}}{\partial w^\mu}B^\mu+\frac12\frac{\partial^2\tilde w^{\tilde\mu}}{\partial w^\mu\partial w^\nu}D^{\mu\nu}\,,
\label{eqn:B-transform}
\eeq
and diffusion coefficients
\beq
\tilde D^{\tilde\mu\tilde\nu}=\frac{\partial\tilde w^{\tilde\mu}}{\partial w^\mu}\frac{\partial\tilde w^{\tilde\nu}}{\partial w^\nu}D^{\mu\nu}.
\label{eqn:D-transform}
\eeq
Notice that the diffusion tensor transforms like an ordinary contravariant rank-2 tensor. This is because it is already built from a quadratic moment, so under a change of variables its leading contribution comes from the linear part of the coordinate transformation. By contrast, the drift is a first moment, and since the stochastic increments satisfy $(\Delta w)^2\sim \Delta t$, the quadratic part of the coordinate transformation contributes at the same order as the original drift. This gives rise to the second term in \eqref{eqn:B-transform}, known as the It\^o correction \cite{gardiner2004handbook}.

We will later use \eqref{eqn:B-transform} and \eqref{eqn:D-transform} repeatedly, so we find it useful to illustrate the origin of the It\^o correction with a simple one-dimensional example. Consider a variable $x$ undergoing Brownian motion, with drift $B^x=0$ and constant diffusion coefficient $D^x=D$. Now make the nonlinear change of variable $x\to y=x^2$. Using the transformation laws \eqref{eqn:B-transform} and \eqref{eqn:D-transform}, we obtain $B^y=D$ and $D^y=4Dy$. We thus see that $y$ acquires a nonzero drift even though the original variable had $B^x=0$. This result has a simple physical interpretation: while $x$ wanders symmetrically around zero, the amplitude of its fluctuations grows, so $\braket{y}=\braket{x^2}$ must increase with time.

\subsection{Different regimes}
\label{sec:regimes}

In the remainder of the paper, we study stochastic perturbations of binaries in different regimes, which we summarize visually in Fig.~\ref{fig:regimes}. We distinguish different cases based both on how the perturbation coherence time $\tau\ped{c}$ compares to the binary orbital period $T$, and on how the typical coherence length $\ell\ped{c}$ compares to the semi-major axis $a$. We consider three (partially overlapping) regimes.
\begin{itemize}
\item \textbf{Adiabatic tidal perturbations (Sec.~\ref{sec:adiabatic}).} 
\beq
T\ll\tau\ped{c}\qquad\text{and}\qquad a \ll \ell\ped{c}.
\eeq
Both the coherence length $\ell\ped{c}$ and coherence time $\tau\ped{c}$ of the perturbations are much larger than the binary's internal scales $a$ and $T$. The external force can thus be approximated via a tidal field that varies slowly in time.
\item \textbf{White-noise tidal perturbations
(Sec.~\ref{sec:white-noise}).} 
\beq
\tau\ped{c} \ll T\qquad\text{and}\qquad a \ll \ell\ped{c}.
\eeq
The external perturbation has a large coherence length so it can be approximated as a tidal field, but it varies on short timescales compared to the binary's orbital period.
\item \textbf{Impulsive encounters (Sec.~\ref{sec:impulsive}).} 
\beq
\text{fast perturbers }(\tau\ped{c} \ll T).
\eeq
Here we consider gravitational perturbations due to (spherical) bodies moving on straight-line orbits, passing within the vicinity of the binary. They are ``fast'' in the sense that they interact with the binary for a time much shorter than its period (hence $\tau\ped{c} \ll T$). The size of the perturber can be either large ($a\ll \ell\ped{c}$) or small ($\ell\ped{c} \ll a$) compared to the binary's semi-major axis.
\end{itemize}

Note that while the words ``white-noise'' and ``impulsive'' both indicate the regime $\tau\ped{c}\ll T$, in an attempt to better distinguish the three cases we reserve ``white-noise'' for strictly tidal perturbations and ``impulsive'' for encounters with a third spherical body. In the limit where the spherical body's characteristic radius is much smaller than the binary's semi-major axis, this usage of ``impulsive encounter'' is the traditional one illustrated in the bottom-left quadrant of Fig.~\ref{fig:regimes}. In the opposite limit, where the body is so large that it engulfs the binary, we actually recover the results of the white-noise tidal regime, as we will see in Sec.~\ref{sec:large-perturbers}.

We have left out of this categorization encounters with a third body that last for a time comparable to or larger than the binary's period (top-left quadrant of Fig.~\ref{fig:regimes}). These typically result in a chaotic interaction, which can be studied through both numerical simulations and statistical analytical techniques \cite{Quinlan:1996vp,Sesana:2006xw,Stone:2019qvl,Ginat:2021ypu,2023ApJ...955..134R,2024ApJ...970..112A,Tomaselli:2026uqg}. At least for three bodies of comparable mass, these processes cannot be studied with our Fokker--Planck approach because the perturbations to the binary's orbital elements are not small, so we exclude them from the scope of this paper.

We note that the formalism could be generalized to any finite value of $\tau\ped{c}/T$ and $a/\ell\ped{c}$ when the complete temporal and spatial power spectrum of the perturbation is available. Those regimes might include resonant phenomena when, e.g., such power spectrum features peaks around $\tau\ped{c}\sim T$. However, in the limits discussed above, the Fokker--Planck coefficients take on a \emph{universal} form, which depends only on $T\sped{d}$ rather than on the precise shape of the power spectrum of the perturbation. Given our aim to provide general and ready-to-apply formulae, we focus on those limiting cases.

Finally, we stress that throughout this work we consider perturbations mediated purely by gravity. Possible additional non-gravitational interactions of the binary components with the surrounding medium or dark sector are beyond the scope of our analysis.

\section{Tidal perturbations}
\label{sec:tidal}

In this section, we apply the formalism of Sec.~\ref{sec:formalism} to the case of a perturbation induced by an external stochastic tidal field. We set up the framework of the calculation in Sec.~\ref{sec:tidal-force}. We then specialize to the adiabatic and white-noise regimes in Secs.~\ref{sec:adiabatic} and~\ref{sec:white-noise} respectively.

\subsection{Stochastic tidal force}
\label{sec:tidal-force}

We denote the cartesian components of the tidal field by $T_{ij}(t)=\partial_i\partial_j\Phi$, where $\Phi$ is the stochastic part of the gravitational potential of the environment. Note that this definition explicitly \emph{excludes} any deterministic part of the potential, but it \emph{includes} the trace of the tidal tensor.\footnote{The trace of the tidal tensor encodes the monopole part of the acceleration caused by density fluctuations. There are however applications where the trace ought to be subtracted, such as binaries evolving under perturbations from stochastic gravitational waves. We will show in Sec.~\ref{sec:lit-gw-background} how the results change in that case.} The tidal acceleration acting on the binary is then $a_i=-T_{ij}r_j$, where $r_i$ is the separation vector. For simplicity, we assume that the tidal field is isotropic, in the sense that its probability distribution is invariant under spatial rotations. We define the \emph{diffusion time} $T\sped{d}$ as follows. The Fokker--Planck coefficients $B^\mu$ and $D^{\mu\nu}$ have dimensions of $(\text{time})^{-1}$, and they are proportional to the time-integrated correlator of the tidal field $\int\braket{T_{ij}T_{kl}}\dd\tau$ through \eqref{eqn:Deltaw1}--\eqref{eqn:Deltaw2} and \eqref{eqn:Deltaww1}--\eqref{eqn:Deltaww2}. The only dimensionally correct equation where this correlator is rescaled by a combination of the binary's parameters to give an inverse time is
\beq
\int_{-\infty}^\infty\braket{T_{ij}(t)T_{kl}(t+\tau)}\dd\tau\equiv\frac{Gm}{a^3T\sped{d}}\,\frac{\mathds{1}_{ijkl}}{15}\,,
\label{eqn:Td-tidal}
\eeq
where $\mathds{1}_{ijkl}\equiv\delta_{ij}\delta_{kl}+\delta_{ik}\delta_{jl}+\delta_{il}\delta_{jk}$. This tensorial structure is fixed by the isotropy assumption, while the normalizing factor of $15=\mathds{1}_{ijij}$ is chosen in order to make later expressions simpler. We will see in Sec.~\ref{sec:density-fluctuations} how both of them arise naturally in physical cases.

We choose to work with the Keplerian orbital elements $(a,e,\Omega,i,\omega,M_0)$. The stochastic force $F^\mu\sped{s}(\vec w^*,t)$ can then be read off, e.g., from Gauss' planetary equations for the Keplerian orbital elements,
\begin{align}
\label{eqn:dot-a}
\dot a={}&\frac{2a^{3/2}}{\sqrt{Gm(1-e^2)}}(e\sin\phi\,a^r+(1+e\cos\phi)a^\phi)\,,\\
\label{eqn:dot-e}
\dot e={}&\sqrt{\frac{a(1-e^2)}{Gm}}(\sin\phi\,a^r+(\cos\phi+\cos E)a^\phi)\,,\\
\label{eqn:dot-Omega}
\dot\Omega={}&\frac{\sin(\omega+\phi)\,r\,a^z}{\sqrt{Gma(1-e^2)}\sin i}\,,\\
\label{eqn:dot-i}
\dot i={}&\frac{\cos(\omega+\phi)\,r\,a^z}{\sqrt{Gma(1-e^2)}}\,,\\
\label{eqn:dot-omega}
\begin{split}
\dot\omega={}&{-\sqrt{\frac{a(1-e^2)}{Gme^2}}}\bigg[\cos\phi\,a^r\!-\frac{2\!+\!e\cos\phi}{1\!+\!e\cos\phi}\sin\phi\,a^\phi\bigg]\\
&-\frac{\cos i\sin(\omega+\phi)\,r\,a^z}{\sqrt{Gma(1-e^2)}\sin i}\,,
\end{split}
\end{align}
\vskip -12pt
\beq
\label{eqn:dot-M0}
\begin{split}
\dot M_0={}&\sqrt{\frac{a}{Gm}}\frac{1-e^2}e\bigg(\bigg(\!\cos\phi-\frac{2e(1-e\cos E)}{1-e^2}\bigg)a^r\\
&-\bigg(1+\frac{1-e\cos E}{1-e^2}\bigg)\sin\phi\,a^\phi\bigg)\,,
\end{split}
\eeq
where $\phi$ and $E$ are the true and eccentric anomalies respectively. In these equations, $a^r=-T_{ij}r_i\hat r_j$, $a^\phi=-T_{ij}r_i\hat\phi_j$, and $a^z=-T_{ij}r_i\hat z_j$ are the radial, tangential, and axial components of the tidal acceleration, where
\begin{align}
\hat r&=\vec R(\Omega,i,\omega)(\cos\phi,\sin\phi,0)\,,\\
\hat\phi&=\vec R(\Omega,i,\omega)(-\sin\phi,\cos\phi,0)\,,\\
\hat z&=\vec R(\Omega,i,\omega)(0,0,1)\,,
\end{align}
and $\vec R(\Omega,i,\omega)=\vec R_z(\Omega)\vec R_x(i)\vec R_z(\omega)$ is a matrix that rotates the cartesian axes $(\hat x,\hat y,\hat z)$ onto the axes $(\hat e,\hat q,\hat J)$ of the binary's orbit, where $\hat e$ is the direction of the eccentricity vector, $\hat J$ is the direction of the angular momentum, and $\hat q=\hat J\times\hat e$.

By factoring out the tidal field components, we can put the force in the form
\beq
F^\mu\sped{s}(\vec w^*,t)=g^\mu_{ij}(\vec w^*)T_{ij}(t)\,,
\label{eqn:Fs-tidal}
\eeq
where $g^\mu_{ij}$ can be read off from \eqref{eqn:dot-a}--\eqref{eqn:dot-M0}. From here, we can compute the drift and diffusion coefficients $B^\mu$ and $D^{\mu\nu}$ by simply plugging \eqref{eqn:Fs-tidal} into \eqref{eqn:Deltaw1} and \eqref{eqn:Deltaww1} for adiabatic perturbations, or in \eqref{eqn:Deltaw2} and \eqref{eqn:Deltaww2} for white-noise perturbations. Because \eqref{eqn:Fs-tidal} separates the functional dependence on $\vec w$ and $t$, Eqs.~\eqref{eqn:Deltaw1} and \eqref{eqn:Deltaww1} can be written as
\begin{align}
\label{eqn:Bmu}
B^\mu&=\frac12\partial_{w^\nu}\bar g^\mu_{ij}(\vec w)\bar g^\nu_{kl}(\vec w)\frac{\mathds{1}_{ijkl}}{15}\frac{Gm}{a^3T\sped{d}}\,,\\
\label{eqn:Dmunu}
D^{\mu\nu}&=\bar g^\mu_{ij}(\vec w)\bar g^\nu_{kl}(\vec w)\frac{\mathds{1}_{ijkl}}{15}\frac{Gm}{a^3T\sped{d}}
\end{align}
where we replaced the time integral with the definition of the diffusion time $T\sped{d}$, given in \eqref{eqn:Td-tidal}. Analogous transformations hold for \eqref{eqn:Deltaw2} and \eqref{eqn:Deltaww2}, in which case the orbital average is not performed on the individual $g^\mu_{ij}$, but over the entire expression at the end.

In practice, to compute the orbit average of any function $X(\phi,E)$, we first express the true anomaly in terms of the eccentric anomaly via $\cos\phi=(\cos E-e)/(1-e\cos E)$, and then compute the average as
\beq
\bar X\equiv\frac1{2\pi}\int_0^{2\pi}X(\phi(E),E)(1-e\cos E)\dd E\,.
\label{eqn:orbit-average}
\eeq

An important subtlety arises in the computation of the drift coefficients $B^\mu$. As derived in Sec.~\ref{sec:ensemble-average}, the partial derivatives $\partial_{w^\nu}$ in \eqref{eqn:Deltaw1} and \eqref{eqn:Deltaw2} are taken at fixed orbital phase $\mathcal M$. However, the force components $g^\mu_{ij}$, as read off from Gauss' planetary equations \eqref{eqn:dot-a}--\eqref{eqn:dot-M0}, are naturally expressed in terms of the true and eccentric anomalies $\phi$ and $E$. These anomalies are functions of $e$ and $M_0$ at fixed $\mathcal M$, through Kepler's equation $\mathcal M+M_0=E-e\sin E$. At fixed $\mathcal M$, we find
\beq
\frac{\partial E}{\partial e}=\frac{\sin E}{1-e\cos E}\,,\qquad
\frac{\partial E}{\partial M_0}=\frac{1}{1-e\cos E}\,,
\label{eqn:dEdedM0}
\eeq
and the derivatives of $\phi$ follow by the chain rule. These contributions must be included whenever $\partial_{w^\nu}g^\mu_{ij}$ is evaluated at fixed $\mathcal M$: the full derivative is $\partial_{w^\nu}g^\mu_{ij}|_{\mathcal M}=\partial_{w^\nu}g^\mu_{ij}|_E+(\partial_E g^\mu_{ij})(\partial_{w^\nu}E|_{\mathcal M})$.

Finally, we note that the tensor $\mathds{1}_{ijkl}$ is invariant under rotations,
\beq
R_{ia}R_{jb}R_{kc}R_{ld}\mathds{1}_{ijkl}=\mathds{1}_{abcd}\,.
\label{eqn:identity-rotation}
\eeq
This identity can be used to simplify the computation of the diffusion coefficients \eqref{eqn:Dmunu}. On the other hand, the rotation matrices appearing in the drift coefficients \eqref{eqn:Bmu} are hit by derivatives with respect to the Euler angles $\{\Omega,i,\omega\}$. We can then still use \eqref{eqn:identity-rotation} in the form $R_{jb}R_{kc}R_{ld}\mathds{1}_{ijkl}=R_{ia}\mathds{1}_{abcd}$, and expand the derivative as
\beq
\begin{split}
B^\mu\propto{}&\partial_{w^\nu}\bar g^\mu_{ij}\bar g^\nu_{kl}\mathds{1}_{ijkl}\\
&+(\partial_{w^\nu}R_{ai})R_{ab}\bar g^\mu_{ij}\bar g^\nu_{kl}\mathds{1}_{bjkl}\\
&+(\partial_{w^\nu}R_{aj})R_{ab}\bar g^\mu_{ij}\bar g^\nu_{kl}\mathds{1}_{ibkl}\,.
\end{split}
\eeq

\subsection{Adiabatic tidal perturbations}
\label{sec:adiabatic}

We now specialize to the adiabatic case, $T\ll\tau\ped{c}$. As per equations \eqref{eqn:Deltaw1} and \eqref{eqn:Deltaww1}, we first average the force over the orbit following \eqref{eqn:orbit-average}. For example, the only nonzero cartesian components of $\bar g^a_{ij}$ and $\bar g^e_{ij}$ are
\begin{align}
\bar g^a_{xy}&=-\bar g^a_{yx}=\sqrt{\frac{a^5(1-e^2)}{Gm}}\,,\\
\label{eqn:gbar-components}
\bar g^e_{yx}&=\frac52a^{3/2}e\sqrt{\frac{1-e^2}{Gm}}\,.
\end{align}
Similar expressions hold for the other components of $\bar g^\mu_{ij}(\vec w)$, which we do not report here. Using these, we calculate the drift and diffusion coefficients from \eqref{eqn:Bmu}--\eqref{eqn:Dmunu}. The diffusion matrix $D^{\mu\nu}$ has a block structure: the semi-major axis and eccentricity $(a,e)$ decouple from all other variables, while the mean anomaly at epoch $M_0$ couples to the orbital orientation through a single off-diagonal element $D^{\omega M_0}$. It is thus useful to discuss the $(a,e)$ sector and the $(\Omega,i,\omega,M_0)$ sector separately.

\subsubsection{Semi-major axis and eccentricity}

For the $(a,e)$ subspace, we find
\begin{align}
\label{eqn:Ba-adiabatic}
B^a&=0\,,\\
B^e&=\frac{5e(3-5e^2)}{24T\sped{d}}\,,\\
D^{aa}&=0\,,\\
D^{ae}&=0\,,\\
D^{ee}&=\frac{5e^2(1-e^2)}{12T\sped{d}}\,,
\label{eqn:Dee-adiabatic}
\end{align}
where $T\sped{d}$ is the diffusion time defined in \eqref{eqn:Td-tidal}. We can write these in terms of energy $\mathcal E=-Gm/(2a)$ and angular momentum $J=\sqrt{Gma(1-e^2)}$ using \eqref{eqn:B-transform} and \eqref{eqn:D-transform}, obtaining
\begin{align}
B^J&=-\frac{5Je^2(4-5e^2)}{24(1-e^2)T\sped{d}}\,,\\
D^{JJ}&=\frac{5Gmae^4}{12T\sped{d}}\,,
\end{align}
with $B^{\mathcal E}=D^{\mathcal E\mathcal E}=D^{\mathcal EJ}=0$.

The semi-major axis $a$ does not evolve. This is expected, as in the adiabatic limit the tidal field is approximately constant over one binary period, implying by energy conservation that $a$ remains constant. By contrast, the eccentricity shows interesting and nontrivial dynamics. To keep expressions easily readable, we abuse notation (both here and in the remainder of the paper) and denote with $f$ the probability density function marginalized over the variables we are not studying, which will be clear from the context. In this case, we marginalize over $\Omega$, $i$, $\omega$, $M_0$, as well as the non-evolving $a$. We can then write the one-dimensional Fokker--Planck equation as
\beq
\frac{\partial f}{\partial t}=-\partial_eJ^e\,,
\label{eqn:fp-e-adiabatic}
\eeq
where
\beq
\begin{split}
J^e(e)&=B^e f-\frac12\partial_e(D^{ee}f)\\
&=-\frac{5}{24T\sped{d}}e^3(1-e^2)\frac{\partial}{\partial e}\bigg(\frac{f}{e}\bigg)\,,
\end{split}
\label{eqn:Je-adiabatic}
\eeq
is the probability current. The equation $\partial f/\partial t=0$ admits the unique normalizable solution
\beq
f\ped{th}(e)=2e\,,
\label{eqn:f-thermal}
\eeq
the \emph{thermal eccentricity distribution}, corresponding to $J^e=0$.

\begin{figure}
\centering
\includegraphics[width=0.48\textwidth]{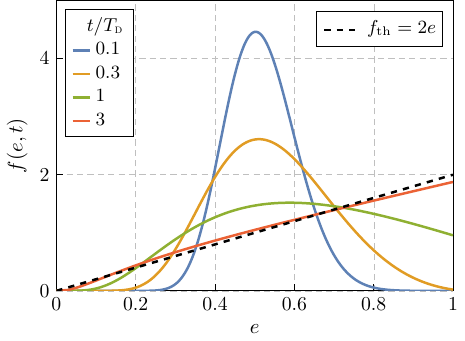}
\caption{Numerical solution of the Fokker--Planck equation for the eccentricity \eqref{eqn:fp-e-adiabatic} in the case of adiabatic tidal perturbations ($T\ll\tau\ped{c}$), starting from the initial condition $f(e,0)=\delta(e-0.5)$. The eccentricity distribution relaxes to the thermal $f\ped{th}(e)=2e$ in a characteristic time $T\sped{d}$.}
\label{fig:f-adiabatic}
\end{figure}

Equation~\eqref{eqn:fp-e-adiabatic} does not admit a simple, closed-form, analytical solution. We therefore solve it numerically, with initial condition $f(e,0)=\delta(e-e_0)$. The solution for arbitrary initial conditions can be obtained with a Green's function approach, i.e., linear superposition of solutions with different values of $e_0$. We show in Fig.~\ref{fig:f-adiabatic} the result for $e_0=0.5$. The initially peaked distribution diffuses towards both higher and lower eccentricities, and relaxes to the thermal distribution $f\ped{th}=2e$ on a timescale of order $T\sped{d}$. Both boundaries $e=0$ and $e=1$ are inaccessible, as the probability current $J^e$ vanishes at both endpoints for any regular $f(e)$.

\subsubsection{Orbital orientation and mean anomaly at epoch}
\label{sec:orientation-adiabatic}

We now move on to discuss the evolution of the orbital orientation and its coupling to the mean anomaly at epoch, $M_0$. In the Euler angle parametrization $(\Omega,i,\omega)$, the drift and diffusion coefficients are generically all nonzero, except for $D^{\Omega M_0}=D^{i M_0}=0$ (i.e., $M_0$ couples only to $\omega$). We report the explicit expressions in Appendix~\ref{sec:coeffs-euler-angles}. These complicated formulae are, however, purely the result of geometric factors. The physics becomes much clearer if we switch variables from Euler angles to the natural axes of the binary: the eccentricity direction $\hat e$, the angular momentum direction $\hat J$, and $\hat q\equiv\hat J\times\hat e$. We then consider $f$ to be a function of $M_0$ and $\vec R\in\text{SO}(3)$, where $\vec R$ rotates $(\hat x,\hat y,\hat z)$ into $(\hat e,\hat q,\hat J)$. An infinitesimal change in $\vec R$ can be parametrized with three angles $\Delta\theta^{\hat\mu}$ as $\vec R(t+\Delta t)=e^{\Delta\theta^{\hat\mu}\!\vec A_{\hat\mu}}\vec R(t)$, where $\vec A_{\hat\mu}\in\mathfrak{so}(3)$ generates the rotations around the instantaneous binary axis $\hat\mu$, with $\hat\mu\in\{\hat e,\hat q,\hat J\}$. It is then natural to define the \emph{left-invariant} (body-frame) vector fields,
\beq
J_{\hat \mu}(f)\equiv\frac{\dd}{\dd x}f\big(e^{x\vec A_{\hat \mu}}\vec R\big)\Big|_{x=0}\,.
\label{eqn:Jhatmu}
\eeq
When the parameter increments $\Delta w^\mu$ are replaced by the angles $\Delta\theta^{\hat\mu}$, the Kramers--Moyal expansion in Appendix~\ref{sec:kramers-moyal} that yields \eqref{eqn:fp}--\eqref{eqn:drift-diffusion-def} works exactly the same as before, except that we need to replace $\partial_{w^\mu}$ with the vector fields $J_{\hat\mu}$. The Fokker--Planck equation for the joint distribution $f(\vec R,M_0,t)$ in the body-frame basis then reads
\beq
\begin{split}
\frac{\partial f}{\partial t}={}&{-J_{\hat\mu}(B^{\hat\mu}f)}+\frac12J_{\hat \mu}J_{\hat \nu}\big(D^{\hat\mu\hat\nu}f\big)\\
&+J_{\hat\mu}\frac{\partial}{\partial M_0}\big(D^{\hat\mu M_0}f\big)\\
&-\frac{\partial}{\partial M_0}(B^{M_0}f)+\frac12\frac{\partial^2}{\partial M_0^2}\big(D^{M_0M_0}f\big)\,.
\end{split}
\label{eqn:fp-M0-orientation}
\eeq
To obtain the drift and diffusion coefficients, we need to convert the Euler angle rates to the body-frame angular velocities. The transformation reads $\partial_{w^\mu}=M^{\hat\mu}{}_\mu J_{\hat\mu}$, with $w^\mu\in\{\Omega,i,\omega\}$ and
\beq
M^{(\hat e,\hat q,\hat J)}{}_{(\Omega,i,\omega)}=\begin{pmatrix}
\sin i\sin\omega & \cos\omega & 0\\
\sin i\cos\omega & -\sin\omega & 0\\
\cos i & 0 & 1
\end{pmatrix}\,.
\label{eqn:change-of-basis-matrix}
\eeq
The transformations \eqref{eqn:B-transform} and \eqref{eqn:D-transform} are then
\begin{align}
B^{\hat\mu}&=M^{\hat\mu}{}_\mu B^\mu+\frac12\partial_{w^\mu}M^{\hat\mu}{}_\nu D^{\mu\nu}\,,\\
D^{\hat\mu\hat\nu}&=M^{\hat\mu}{}_\mu M^{\hat\nu}{}_\nu D^{\mu\nu}\,.
\label{eqn:Dhatmuhatmu}
\end{align}

The result is that $B^{\hat\mu}$ vanishes identically, while $D^{\hat\mu\hat\nu}$ is diagonal, with entries
\begin{align}
\label{eqn:Dhatehate-adiabatic}
D^{\hat e\hat e}&=\frac{(1+4e^2)^2}{60(1-e^2)T\sped{d}}\,,\\
\label{eqn:Dhatqhatq-adiabatic}
D^{\hat q\hat q}&=\frac{1-e^2}{60T\sped{d}}\,,\\
\label{eqn:DhatJhatJ-adiabatic}
D^{\hat J\hat J}&=\frac{43(1-e^2)}{60T\sped{d}}\,.
\end{align}
The cross-term coupling $M_0$ to the angular momentum direction is
\beq
D^{\hat J M_0}=-\frac{\sqrt{1-e^2}(67+43e^2)}{60T\sped{d}}\,,
\label{eqn:DM0hatJ-adiabatic}
\eeq
while $D^{\hat e M_0}=D^{\hat q M_0}=0$. The drift of the mean anomaly at epoch vanishes, $B^{M_0}=0$, and its diffusion coefficient is
\beq
D^{M_0M_0}=\frac{123+134e^2+43e^4}{60T\sped{d}}\,.
\label{eqn:DM0M0-adiabatic}
\eeq

These expressions are physically transparent. First of all, the absence of any drift confirms that there is no preferred direction for the evolution of the binary, consistent with the statistical isotropy of the tidal field. An isotropic probability density function $f$ is thus a steady-state solution. Then, in the $e\to1$ limit, we have $D^{\hat q\hat q}=D^{\hat J\hat J}=0$, while $D^{\hat e\hat e}$ diverges. This is indeed expected from a purely radial orbit: rotations around $\hat e$ become ``free'', while rotations around the other axes are impossible without first decreasing $e$. On the other hand, in the $e\to0$ limit, we have $D^{\hat e\hat e}=D^{\hat q\hat q}$, consistent with the non-existence of a preferred direction within the orbital plane of a circular orbit. 
Finally, if $f$ is independent of $M_0$, the last three terms in \eqref{eqn:fp-M0-orientation} vanish and one recovers a pure diffusion equation on SO(3) with coefficients \eqref{eqn:Dhatehate-adiabatic}, \eqref{eqn:Dhatqhatq-adiabatic}, and \eqref{eqn:DhatJhatJ-adiabatic}.

While the $(\hat e,\hat q,\hat J)$ basis allows for compact and physically transparent expressions of the diffusion coefficients, we warn the reader that the basis itself is a function of the orientation $\vec R\in\text{SO}(3)$, and it changes as the orbit rotates in space. As such, it does not allow for a straightforward numerical integration of the Fokker--Planck equation for $f(\vec R,M_0,t)$.

\subsection{White-noise tidal perturbations}
\label{sec:white-noise}

Next we consider the white-noise case, $\tau\ped{c}\ll T$. Opposite to the adiabatic regime, we first calculate $B^\mu$ and $D^{\mu\nu}$ at fixed $\phi$ and $E$, and only then perform an orbital average. Again, the $(a,e)$ sector decouples from the other variables, while $M_0$ couples to the orientation only via $D^{\omega M_0}$.

\subsubsection{Semi-major axis and eccentricity}

The drift and diffusion coefficients for $(a,e)$ are
\begin{align}
\label{eqn:Ba-white-noise}
B^a&=\frac{a(18+19e^2)}{30T\sped{d}}\,,\\
B^e&=\frac{28-51e^2-103e^4}{240eT\sped{d}}\,,\\
D^{aa}&=\frac{2a^2(2+e^2)}{15T\sped{d}}\,,\\
D^{ae}&=-\frac{2ae(1-e^2)}{15T\sped{d}}\,,\\
D^{ee}&=\frac{7(1-e^2)(4+5e^2)}{120T\sped{d}}\,,
\label{eqn:Dee-white-noise}
\end{align}
and the two-dimensional Fokker--Planck equation for $(a,e)$ is
\beq
\begin{split}
\frac{\partial f}{\partial t}={}&{-\frac{\partial}{\partial a}\big(B^a f\big)}-\frac{\partial}{\partial e}\big(B^e f\big)+\frac12\frac{\partial^2}{\partial a^2}\big(D^{aa}f\big)\\
&+\frac{\partial^2}{\partial a\partial e}\big(D^{ae}f\big)+\frac12\frac{\partial^2}{\partial e^2}\big(D^{ee}f\big)\,.
\end{split}
\label{eqn:fp-ae}
\eeq
In terms of energy and angular momentum, the coefficients read
\begin{align}
B^{\mathcal E}&=-\frac{\mathcal E(2+3e^2)}{6T\sped{d}}\,,
\label{eqn:BE-white-noise}\\
B^J&=\frac{J(8+40e^2+15e^4)}{240(1-e^2)T\sped{d}}\,,
\label{eqn:BJ-white-noise}\\
D^{\mathcal E\mathcal E}&=\frac{2\mathcal E^2(2+e^2)}{15T\sped{d}}\,,
\label{eqn:DEE-white-noise}\\
D^{\mathcal EJ}&=-\frac{\mathcal EJ(2+3e^2)}{15T\sped{d}}\,,
\label{eqn:DEJ-white-noise}\\
D^{JJ}&=\frac{Gma(8+40e^2+15e^4)}{120T\sped{d}}\,.
\label{eqn:DJJ-white-noise}
\end{align}
Equations \eqref{eqn:BE-white-noise}--\eqref{eqn:DJJ-white-noise} can be derived from \eqref{eqn:Ba-white-noise}--\eqref{eqn:Dee-white-noise} using \eqref{eqn:B-transform} and \eqref{eqn:D-transform}. However, in Appendix~\ref{sec:first-principles} we also present an alternative, first-principles derivation of these coefficients that does not require the use of Gauss' equations. 

Unlike the adiabatic case, the semi-major axis now evolves, as $B^a\ne0$ and $D^{aa}\ne0$. In particular, because $B^{\mathcal E}>0$, the tidal field always heats the binaries on average, regardless of the shape of the distribution function $f$. Given that $a$ is no longer a constant, it is important to remember that $T\sped{d}$ is itself a function of $a$, specifically $T\sped{d}^{-1}\propto a^3$, according to its definition in \eqref{eqn:Td-tidal}. Therefore, when solving the Fokker--Planck equation one must fix the diffusion time $T\sped{d,0}$ at a reference semi-major axis $a_0$, and set $T\sped{d}=(a_0/a)^3T\sped{d,0}$.

Similar to the case of adiabatic tidal perturbations, it is useful to write the Fokker--Planck equation as (cf.~Eq.~\eqref{eqn:fp-e-adiabatic}):
\beq
\frac{\partial f}{\partial t}=-\frac{\partial J^a}{\partial a}-\frac{\partial J^e}{\partial e}\,,
\eeq
where we defined the probability currents
\beq
\label{eqn:Ja-white-noise}
\begin{split}
J^a&=B^af-\frac12\partial_e(D^{ae}f)-\frac12\partial_a(D^{aa}f)\\
&=\frac{a}{15T\sped{d}}\bigg[\frac32e^2f-a(2+e^2)\partial_af+e(1-e^2)\partial_ef\bigg]\,,\!
\end{split}
\eeq
and
\beq
\label{eqn:Je-white-noise}
\begin{split}
J^e&=B^ef-\frac12\partial_a(D^{ae}f)-\frac12\partial_e(D^{ee}f)\\
&=\frac{e(1-e^2)}{15T\sped{d}}\bigg[a\partial_af-\frac{f}2-\frac{7(4+5e^2)}{16}\partial_e\bigg(\frac{f}e\bigg)\bigg]\,.\!
\end{split}
\eeq
Once again, the eccentricity current $J^e$ manifestly vanishes at both $e=0$ and $e=1$, making those boundaries inaccessible.\footnote{From \eqref{eqn:Je-white-noise}, we see that the vanishing of $J^e$ at $e=0$ requires that $f/e$ is regular for $e\to0$. This is always satisfied because the canonical phase space volume element is proportional to $e\dd e$ (e.g., \cite{2023dyps.book.....T}).} Meanwhile, the semi-major axis current $J^a$ vanishes at $a=0$, but not for $a\to\infty$. When solving the Fokker--Planck equation, we therefore need to make a choice of boundary conditions at $a\to\infty$. We enforce an absorbing boundary, which physically corresponds to binaries being unbound by tidal perturbations.

\begin{figure*}
\centering
\includegraphics[width=\textwidth]{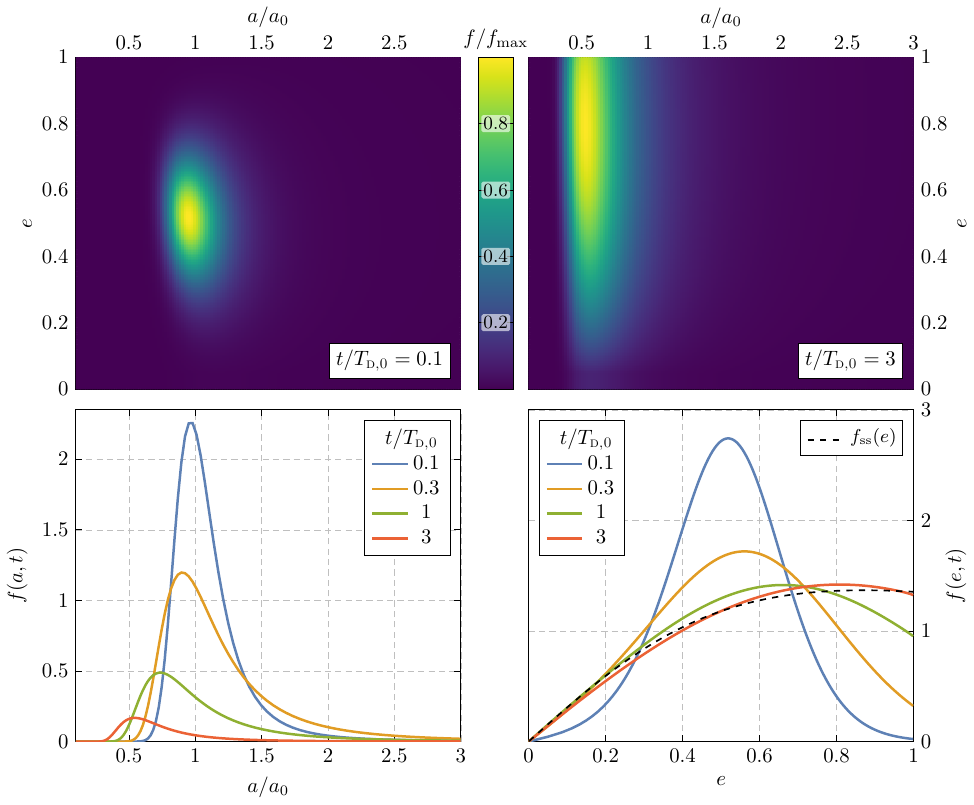}
\caption{Numerical solution of the $(a,e)$ sector of the Fokker--Planck equation \eqref{eqn:fp-ae} in the white-noise tidal regime ($\tau\ped{c}\ll T$), with the drift and diffusion coefficients \eqref{eqn:Ba-white-noise}--\eqref{eqn:Dee-white-noise}. The initial condition is a narrow two-dimensional Gaussian with $a/a_0$ centered at 1 with standard deviation $0.05$, and $e$ centered at $0.5$ with standard deviation $0.03$. The boundaries $a=0$, $e=0$ and $e=1$ are inaccessible, while at large $a$ we use absorbing boundary conditions. Time is in units of $T\sped{d,0}\equiv T\sped{d}(a=a_0)$. Top: heatmaps of $f(a,e)$ at selected times, with color scale normalized to each panel's range. Bottom-left: semi-major axis marginal $f(a,t)$, which progressively loses area as the binaries are unbound. Bottom-right: eccentricity marginal $f(e,t)$, \emph{renormalized to unit area} to display its shape. The surviving eccentricity distribution relaxes approximately to $f\ped{ss}(e)$ given in \eqref{eqn:f-steady-state}.}
\label{fig:f-white-noise}
\end{figure*}

We show in Fig.~\ref{fig:f-white-noise} a numerical solution to the Fokker--Planck equation \eqref{eqn:fp-ae}, using a square grid for the variables $(\log(a/a_0),e^2)$ over the intervals $[-3,6]\times[0,1]$, with absorbing boundary conditions at the large-$a$ boundary and reflective elsewhere (which does not matter in practice as those boundaries are inaccessible). The distribution function spreads over a characteristic time $T\sped{d}$. The marginalized distribution $f(a,t)$ shows a slightly counterintuitive behavior: even though the average semi-major axis increases, the peak of the distribution shifts to smaller $a$ over time. This is because the tail at larger $a$ is greatly stretched due to the faster diffusion, $T\sped{d}\propto a^{-3}$. The eccentricity distribution quickly loses memory of the initial peak at $e_0$ and relaxes to a quasi-stationary shape.

To understand the properties of the steady-state solutions of the two-dimensional Fokker--Planck equation \eqref{eqn:fp-ae}, we look for solutions with a separable ansatz $f(a,e)=a^\gamma h(e)$. The scaling $T\sped{d}^{-1}\propto a^3$ ensures that all terms in the stationary equation $\partial_a J^a+\partial_e J^e=0$ carry the common factor $a^{\gamma+3}$, which divides out. We remain with a second-order ODE for $h(e)$, parametrized by $\gamma$. Two values of $\gamma$ lead to closed-form solutions.

\begin{itemize}
\item \textbf{Detailed-balance solution ($\gamma=1/2$).} Setting both currents to zero, $J^a=J^e=0$, we find the non-normalizable stationary solution
\beq
f_0(a,e)=C_0\sqrt{a}\,e\,,
\label{eqn:f-zero-flux}
\eeq
where $C_0$ is an arbitrary constant. This is a detailed balance solution, with no net flux in either $a$ or $e$. The eccentricity distribution is thermal, analogous to the stationary solution previously found in the adiabatic tidal case, Eq.~\eqref{eqn:f-thermal}.
\item \textbf{Constant-current solution ($\gamma=-4$).} For $\gamma=-4$, $\partial_aJ^a$ vanishes identically, so the stationary equation reduces to $\partial_eJ^e=0$ alone. Similar to the stationary solution discussed in the adiabatic tidal regime (see Eq.~\eqref{eqn:Je-adiabatic}), the current $J^e(e)$ vanishes at the boundaries for any regular $h(e)$. The condition $\partial_eJ^e=0$ then forces $J^e=0$ everywhere, which can be solved analytically, yielding
\beq
f\ped{ss}(a,e)=C\ped{ss}\,\frac{e}{a^4(4+5e^2)^{36/35}}\,,
\label{eqn:f-steady-state}
\eeq
with $C\ped{ss}$ a normalization constant. While $J^e=0$, this solution carries a nonvanishing, $a$-independent outward flux $J^a>0$, describing a steady state in which probability flows uniformly from small to large~$a$. In other words, binaries get progressively disrupted.
\end{itemize}

Any linear combination $C_0 f_0+C\ped{ss} f\ped{ss}$ is also stationary and not normalizable in $a$. The numerical solution in Fig.~\ref{fig:f-white-noise} shows that the eccentricity marginal approaches $f\ped{ss}$ given in \eqref{eqn:f-steady-state}, but does not converge to it exactly. If we restrict instead to $a>a_0$, then convergence improves significantly, implying that wider binaries do converge to $f\ped{ss}$, but are also quickly unbound (and thus removed from the ensemble), leaving behind the slowly-evolving population that has not yet approached the steady state. Sustaining the steady state solution \eqref{eqn:f-steady-state} exactly would require a constant injection of binaries at the lower $a$-boundary. 

\subsubsection{Orbital orientation and mean anomaly at epoch}
\label{sec:orientation-white-noise}

As in Sec.~\ref{sec:orientation-adiabatic}, the drift and diffusion coefficients in the variables $(\Omega,i,\omega)$, which we report explicitly in Appendix~\ref{sec:coeffs-euler-angles}, contain complicated geometric factors. Switching variables to rotations around the axes $(\hat e,\hat q,\hat J)$, the drift vanishes and the orientation diffusion matrix is diagonal, with elements
\begin{align}
\label{eqn:Dhatehate-white-noise}
D^{\hat e\hat e}&=\frac{4+41e^2+18e^4}{120(1-e^2)T\sped{d}}\,,\\
D^{\hat q\hat q}&=\frac{4+3e^2}{120T\sped{d}}\,,\\
D^{\hat J\hat J}&=\frac{28+25e^2-46e^4}{120e^2T\sped{d}}\,.
\end{align}
The cross-term coupling $M_0$ to the orientation is
\beq
D^{\hat J M_0}=-\frac{\sqrt{1-e^2}(28+145e^2+44e^4)}{120e^2T\sped{d}}\,,
\label{eqn:DM0hatJ-white-noise}
\eeq
while $D^{\hat e M_0}=D^{\hat q M_0}=0$. The $M_0$ self-diffusion coefficient reads
\beq
D^{M_0M_0}=\frac{28+333e^2+349e^4+46e^6}{120e^2T\sped{d}}\,.
\label{eqn:DM0M0-white-noise}
\eeq
The Fokker--Planck equation for the joint distribution $f(\vec R,M_0,t)$ takes the same form as \eqref{eqn:fp-M0-orientation}, except with the coefficients above. In this case too we recover the intuitive equalities $D^{\hat e\hat e}=D^{\hat q\hat q}$ for $e=0$ and $D^{\hat q\hat q}=D^{\hat J\hat J}$ for $e=1$, as well as a diverging $D^{\hat e\hat e}$ for $e\to1$.

Note that $D^{\hat J\hat J}$, $D^{\hat J M_0}$, and $D^{M_0M_0}$ all diverge as $e\to0$. This is again a coordinate artifact: both $M_0$ and $\omega$ are individually ill-defined for a circular orbit. The well-defined quantity is the \emph{mean longitude at epoch}, $\lambda_0\equiv M_0+\omega$. Indeed, in the $e\to0$ limit the $(\hat J,M_0)$ diffusion block approaches
\beq
\frac{7}{30e^2T\sped{d}}\begin{pmatrix}1&-1\\-1&1\end{pmatrix}\,,
\eeq
which is degenerate, with eigenvalues $0$ and $7/(15e^2T\sped{d})$. The null eigenvalue corresponds to the direction $\lambda_0=M_0+\omega$, which has a finite diffusion coefficient in the circular limit. The divergent eigenvalue corresponds to the phase difference $M_0-\omega$.

Changing variables from $M_0$ to $\lambda_0$ in the Fokker--Planck equation does remove the $1/e^2$ divergences, but also introduces new cross-terms ($D^{\lambda_0\Omega}$, $D^{\lambda_0 i}$) and expressions involving $\sqrt{1-e^2}$ that are less compact than the present ones. We therefore refrain from writing down coefficients involving $\lambda_0$ here.

\section{Impulsive encounters}
\label{sec:impulsive}

In this section, we apply the formalism of Sec.~\ref{sec:formalism} to perturbations induced by impulsive encounters with a third body passing in the vicinity of the binary. We treat the case of generic, extended, spherically symmetric perturbers and only later restrict to point-mass ones. We set up the problem in Sec.~\ref{sec:impulsive-setup} and calculate the statistics of ``encounter kicks'' in Sec.~\ref{sec:kick-covariance-tensor}. We use this to write the general formulae for the drift and diffusion coefficients in Sec.~\ref{sec:impulsive-coefficients}. In Sec.~\ref{sec:large-perturbers}, we take the limit of large perturbers, recovering the ``tidal white-noise'' results of Sec.~\ref{sec:white-noise}. In Sec.~\ref{sec:point-mass}, we specialize instead to small or pointlike perturbers, giving the full and explicit expression of all coefficients, and discussing the ensemble evolution. We bridge the two limits in Sec.~\ref{sec:generic-size}.

\subsection{Setup}
\label{sec:impulsive-setup}

We consider a uniform and isotropic bath of ``perturbers'', each having mass $m_*$ and a spherically symmetric density profile $\rho_*(x)$, with $x$ the distance from their center, and denote with $n$ their number density. We consider ``impulsive'' encounters, where the interaction between the binary and perturber occurs on a timescale much shorter than the orbital period $T$. This corresponds to the $\tau\ped{c}\ll T$ regime described in Sec.~\ref{sec:formalism}.

By definition of impulsive encounter, the perturber flies by the binary at high velocity, and its trajectory is only weakly deflected. We then approximate the binary as \emph{frozen} during the encounter, and the perturber's motion as straight and uniform, $\vec r_*(t)=\vec b+\vec Vt$ with $\vec b\cdot\vec V=0$, where $\vec b$ is the impact parameter of the encounter with respect to the binary's most massive body and $\vec V$ is the velocity of the perturber. We then define the coherence time explicitly as $\tau\ped{c}\equiv b/V$ and require $\tau\ped{c}\ll T$. The trajectory parameters are sampled randomly from the distribution of perturbers. The total differential rate of encounters is
\beq
\dd\Gamma\equiv nVg(\vec V)\dd^3V\dd^2b\,,
\label{eqn:dGamma}
\eeq
where $g(\vec V)$ is the velocity distribution of perturbers; e.g., for a Maxwellian with one-dimensional velocity dispersion $\sigma$, it reads $g(\vec V)=(2\pi\sigma^2)^{-3/2}e^{-V^2/(2\sigma^2)}$.

The phase space ``force'' $\mathbf{F}$ created by a series of uncorrelated impulsive encounters can be approximated as a train of delta functions at Poisson-distributed times $t_k$. The variation of orbital elements depends linearly on the relative velocity kick $\Delta\vec v_k$ induced by the $k$-th encounter via Gauss' equations \eqref{eqn:dot-a}--\eqref{eqn:dot-M0}. We can thus write
\beq
F^\mu\sped{s}(\vec w^*,t)=\sum_kG^\mu_\alpha\,\Delta v^\alpha_k\,\delta(t-t_k)\,,\quad\;\; G^\mu_\alpha\equiv\frac{\partial \dot w^\mu}{\partial a^\alpha}\,,
\label{eqn:F-impulsive}
\eeq
where $G^\mu_\alpha$ can be read off \eqref{eqn:dot-a}--\eqref{eqn:dot-M0}, and $\alpha\in\{r,\phi,z\}$ denotes the vector components in the binary's frame. Because the number of encounters of a given kind in a time interval $\Delta t$ is Poisson-distributed with mean $\Delta t\dd\Gamma$, ensemble averages can be turned into averages over encounters as follows,
\begin{align}
\braket{\Delta w^\mu}&\to\Delta t\int\Delta w^\mu\dd\Gamma\,,\\
\braket{\Delta w^\mu\Delta w^\nu}&\to\Delta t\int\Delta w^\mu\Delta w^\nu\dd\Gamma\,.
\end{align}

The velocity kick $\Delta\vec v$ can be found by integrating the gravitational acceleration of the perturber along its trajectory. The result is~\cite{1985ApJ...295..374A, 2008gady.book.....B, Gonzalez-Morales:2012rfo}
\beq
\Delta\vec v=\frac{2Gm_*}{V}\vec y\,,\qquad\vec y\equiv\frac{\mathcal I(b_2)}{b_2^2}\vec b_2-\frac{\mathcal I(b_1)}{b_1^2}\vec b_1\,,
\label{eqn:Delta-v}
\eeq
where $\vec b_1=\vec b$ and $\vec b_2=\vec b-\vec r_\perp$ are the impact parameters with respect to each object, and we defined the projected binary separation $\vec r_\perp\equiv\vec r-(\vec r\cdot\hat V)\hat V$, where $\vec r$ is the binary separation vector. The \emph{form factor}
\beq
\mathcal I(b)\equiv\int_{-\infty}^{\infty}\frac{\int_0^{b\sqrt{1+u^2}}4\pi x^2\rho_*(x)\dd x}{2m_*(1+u^2)^{3/2}}\dd u
\label{eqn:imp-formfactor-def}
\eeq
encodes the dependence on the perturber's mass profile. We report in Table~\ref{tab:I(b)} the function $\mathcal I(b)$ for a few selected mass profiles.

\begin{table*}
\centering
\begin{tabular*}{1.0\textwidth}{@{\extracolsep{\fill}}cccc}
\toprule
\textbf{Density profile} & $\rho_*(x)$ & $\mathcal I(b)$ & $\tilde\rho_*(k)$\\
\midrule
Point mass & $m_*\delta^{(3)}(\vec x)$ & 1 & $1$\\
\midrule
Plummer & $\dfrac{3m_*}{4\pi R^3}\bigg(1+\dfrac{x^2}{R^2}\bigg)^{-5/2}$ & $\dfrac{b^2}{b^2+R^2}$ & $kR\,K_1(kR)$\\
\midrule
Gaussian & $\dfrac{m_*}{(2\pi)^{3/2}R^3}e^{-x^2/(2R^2)}$ & $1-e^{-b^2/(2R^2)}$ & $e^{-k^2R^2/2}$\\
\midrule
Homog.\ sphere & $\dfrac{3m_*}{4\pi R^3}\Theta(R-x)$ & $1-\bigg(1-\dfrac{b^2}{R^2}\bigg)^{3/2}\Theta(R-b)$ & $\dfrac{3(\sin kR-kR\cos kR)}{k^3R^3}$\\
\midrule
Hernquist & $\dfrac{m_* R}{2\pi x(x+R)^3}$ & $\dfrac{b^2}{R^2-b^2}\bigg(\dfrac{\artanh\sqrt{1-b^2/R^2}}{\sqrt{1-b^2/R^2}}-1\bigg)$ & $1+kR\big(\cos(kR)\sint(kR)-\sin(kR)\Ci(kR)\big)$\\
\bottomrule
\end{tabular*}
\caption{Perturber density profile $\rho_*(x)$, form factor $\mathcal I(b)$ (defined in \eqref{eqn:imp-formfactor-def}), and mass-normalized Fourier transform $\tilde\rho_*(k)$ (defined in \eqref{eqn:rho(k)}) for a few representative cases. In each case $R$ is the scale radius of the profile and $m_*$ its total mass. In all cases, $\mathcal I(b)\to1$ as $b\to\infty$, recovering the point-mass result. The Fourier transforms $\tilde\rho_*(k)$ of the Plummer and Hernquist profiles involve, respectively, the modified Bessel function of the second kind $K_1$, and the sine and cosine integrals $\sint(x)=\Si(x)-\pi/2$ and $\Ci(x)$.}
\label{tab:I(b)}
\end{table*}

We can find the drift and diffusion coefficients by plugging \eqref{eqn:F-impulsive} into \eqref{eqn:Deltaw2} and \eqref{eqn:Deltaww2}. Working in a fixed Cartesian frame, $i,j\in\{x,y,z\}$, the kick components $\Delta v^i_k$ of a given encounter are independent of the binary's state, so the derivative $\partial_{w^\nu}$, taken at fixed binary phase as in Sec.~\ref{sec:tidal-force}, acts only on $G^\mu_i$. (Had we used the orbit-aligned components $\Delta v^\alpha_k$, the derivative would also hit the terms depending on the frame orientation.) We thus get
\begin{align}
\label{eqn:Bmu-impulsive}
B^\mu&=\frac12\overline{\partial_{w^\nu}G^\mu_i G^\nu_j\mathcal Q^{ij}}\,,\\
\label{eqn:Dmunu-impulsive}
D^{\mu\nu}&=\overline{G^\mu_i G^\nu_j\mathcal Q^{ij}}\,,
\end{align}
where we defined the kick covariance tensor
\beq
\mathcal Q^{ij}\equiv\int\Delta v^i\Delta v^j\dd\Gamma\,.
\label{eqn:Q-tensor}
\eeq
Because the perturber bath is isotropic and the only vector $\Delta\vec v$ depends on is the binary separation $\vec r$, in the orbit-aligned frame $\alpha,\beta\in\{r,\phi,z\}$ the tensor $\mathcal Q^{\alpha\beta}$ must be diagonal with two identical entries,
\beq
\begin{split}
\mathcal Q^{\alpha\beta}&=Q_t\delta^{\alpha\beta}+(Q_r-Q_t)\hat r^\alpha\hat r^\beta\\
&=\diag(Q_r,Q_t,Q_t)\,,
\end{split}
\eeq
with trace $Q\equiv Q_r+2Q_t$.

The kick covariance tensor also provides a natural way to define the diffusion time $T\sped{d}$ for perturbations from impulsive three-body encounters. We write
\beq
\overline{Q_t(r)/r^2}=\frac{Gm}{15a^3T\sped{d}}\,,
\label{eqn:Td-impulsive}
\eeq
which is analogous to the definition given in \eqref{eqn:Td-tidal} for tidal perturbations. (In the case of impulsive encounters, however, the diffusion time $T\sped{d}$ may depend on the eccentricity, and it may not be easily factorized out of all Fokker--Planck coefficients.)

\subsection{Kick covariance tensor}
\label{sec:kick-covariance-tensor}

Given the properties of the binary and of the perturbers, to obtain the diffusion coefficients one must evaluate $\mathcal Q^{\alpha\beta}$. While conceptually straightforward, the calculation requires us to resolve the geometry of the encounter. The velocity kick $\Delta\vec v=2Gm_*\vec y/V$ (Eq.~\eqref{eqn:Delta-v}) is orthogonal to $\vec V$, so we can decompose it in an orthonormal basis $(\hat V,\hat e_\perp,\hat e_\parallel)$ as $\vec y=y_\perp\hat e_\perp+y_\parallel\hat e_\parallel$. Let us use spherical coordinates, $\dd^3V=V^2\dd V\,\sin\theta\dd\theta\dd\chi$ with $\hat V\cdot\hat r=\cos\theta$, and align $\hat e_\perp$ with $\vec r_\perp$, which gives $y^r=y_\perp\sin\theta$. Because the kick $\vec y$ depends only on $\vec b$ and $\vec r_\perp=r\sin\theta\,\hat e_\perp$, the integral over $\dd\chi$ is trivial and we can write
\begin{align}
\label{eqn:Qr}
Q_r(r)&=2G^2m_*^2n\braket{V^{-1}}\int_0^\pi\mathcal Y_\perp\sin^3\theta\dd\theta\,,\\
\label{eqn:Q}
Q(r)&=2G^2m_*^2n\braket{V^{-1}}\int_0^\pi(\mathcal Y_\perp+\mathcal Y_\parallel)\sin\theta\dd\theta\,,
\end{align}
where $\braket{V^{-1}}=\int_0^\infty4\pi V^2g(V)V^{-1}\dd V=\sqrt{2/\pi}\,\sigma^{-1}$ for a Maxwellian, and
\beq
\mathcal Y_\perp(r_\perp)\equiv\int y_\perp^2\dd^2b\,,\qquad\mathcal Y_\parallel(r_\perp)\equiv\int y_\parallel^2\dd^2b\,.
\eeq
To compute $\mathcal Y_\perp$ and $\mathcal Y_\parallel$, it is convenient to consider the Fourier transform
\beq
\tilde{\vec y}(\vec k)=\int\vec y(\vec b)e^{-i\vec k\cdot\vec b}\dd^2b=2\pi i\,\tilde\rho_*(k)(1-e^{-i\vec k\cdot\vec r_\perp})\,\frac{\vec k}{k^2}\,,
\eeq
where
\beq
\tilde\rho_*(k)=k\int_0^\infty\mathcal I(b)J_1(kb)\dd b=\int\frac{\rho_*(x)}{m_*}e^{-i\vec k\cdot\vec x}\dd^3x
\label{eqn:rho(k)}
\eeq
is the mass-normalized Fourier transform of the perturber's density profile, and $J_1$ is a Bessel function of the first kind. Parseval's theorem then gives
\beq
\int y_\parallel^2\dd^2b=\int2\tilde\rho_*(k)^2(1-\cos(\vec k\cdot\vec r_\perp))\frac{\vec k_\parallel^2}{k^4}\dd^2k\,,
\eeq
and similarly for $y_\perp$. These integrals can be evaluated directly in polar coordinates, with the final result
\begin{align}
\label{eqn:Yperp+Yparallel}
\mathcal Y_\perp+\mathcal Y_\parallel&=4\pi\int_0^\infty \tilde\rho_*(k)^2\big(1-J_0(kr_\perp)\big)\frac{\dd k}{k}\,,\\
\label{eqn:Yperp-Yparallel}
\mathcal Y_\perp-\mathcal Y_\parallel&=4\pi\int_0^\infty \tilde\rho_*(k)^2J_2(kr_\perp)\frac{\dd k}k\,,
\end{align}
where $J_0$ and $J_2$ are Bessel functions of the first kind. Once the perturber's density profile $\tilde\rho_*(k)$ is known, one can evaluate $\mathcal Y_\perp$ and $\mathcal Y_\parallel$ from \eqref{eqn:Yperp+Yparallel}--\eqref{eqn:Yperp-Yparallel}, then substitute the result in \eqref{eqn:Qr}--\eqref{eqn:Q} and perform the integral over $\dd\theta$ to get the kick covariance tensor $\mathcal Q^{\alpha\beta}$.

It is worth noting that, for any regular $\tilde\rho_*(k)$, the integrals \eqref{eqn:Yperp+Yparallel}--\eqref{eqn:Yperp-Yparallel} always converge around $k=0$, implying that no infrared cutoff---such as an upper limit $b\ped{max}$ on the impact parameter---is needed. This is because, at large distances, the Coulomb divergences due to the two objects cancel each other. On the other hand, the integral \eqref{eqn:Yperp+Yparallel} does diverge at $k\to\infty$ for the point-particle case $\tilde\rho_*(k)=1$. The reason is that the straight-trajectory approximation fails when the perturber passes too close to one of the objects. In reality, this ultraviolet divergence is regulated by the $\SI{90}{\degree}$ deflection scale $k_{90}^{-1}=b_{90}$, analogous to standard treatments of Coulomb logarithms available in the literature. We will return to this point in Sec.~\ref{sec:point-mass}.

\subsection{Drift and diffusion coefficients}
\label{sec:impulsive-coefficients}

Having computed the components $\mathcal Q^{\alpha\beta}$, we now simply need to perform the orbital averages in \eqref{eqn:Bmu-impulsive} and \eqref{eqn:Dmunu-impulsive} to obtain $B^\mu$ and $D^{\mu\nu}$.

We will find it useful to write each coefficient as an explicit orbit average involving $\mathcal Q^{\alpha\beta}(r)$ and some orbital variables.
As in the case of tidal perturbations, the diffusion matrix is block diagonal, decoupling semi-major axis and eccentricity from the orbital orientation variables. Let us start with the $(a,e)$ sector. By reading $G^\mu_\alpha$ off Gauss' equations \eqref{eqn:dot-a}--\eqref{eqn:dot-M0}, and substituting it in \eqref{eqn:Bmu-impulsive} and \eqref{eqn:Dmunu-impulsive}, we find
\begin{widetext}
\begin{align}
\label{eqn:Ba-impulsive}
B^a&=\frac{a^2}{Gm}\cdot\,\overline{Q_r+2Q_t+4(\tilde v_r^2Q_r+\tilde v_\phi^2Q_t)}\,,\\
B^e&=\frac{a}{Gm}\cdot\,\overline{\frac{1-e^2}{2e}(Q_r+2Q_t)-\frac{(1-e^2)^2}{2e^3}(\tilde v_r^2Q_r+\tilde v_\phi^2Q_t)-\frac{1+e^2}{2e^3}\tilde r^2Q_t+\frac{1-e^4}{e^3}Q_t}\,,\\
D^{aa}&=\frac{4a^3}{Gm}\cdot\,\overline{\tilde v_r^2Q_r+\tilde v_\phi^2Q_t}\,,\\
D^{ae}&=\frac{2a^2(1-e^2)}{Gm\,e}\cdot\,\overline{\tilde v_r^2Q_r+\tilde v_\phi^2Q_t-Q_t}\,,\\
\label{eqn:Dee-impulsive}
D^{ee}&=\frac{a}{Gm}\cdot\,\overline{\frac{(1-e^2)^2}{e^2}(\tilde v_r^2Q_r+\tilde v_\phi^2Q_t)+\frac{1-e^2}{e^2}\tilde r^2Q_t-2\frac{(1-e^2)^2}{e^2}Q_t}\,,
\end{align}
\end{widetext}
where $\tilde r=1-e\cos E$ and
\beq
\tilde v_r=\frac{e\sin E}{1-e\cos E}\,,\qquad\tilde v_\phi=\frac{\sqrt{1-e^2}}{1-e\cos E}\,.
\eeq
Although the overlines may make Eqs.~\eqref{eqn:Ba-impulsive}--\eqref{eqn:Dee-impulsive} look complicated, in truth they can all be written in terms of a single one-dimensional integral that involves the specific $Q_r$ and $Q_t$.  We will perform this calculation explicitly in Sec.~\ref{sec:point-mass} for the case of point-mass perturbers.

Using instead the energy $\mathcal E$ and angular momentum $J$ as variables, the coefficients take the following simpler forms,
\begin{align}
B^{\mathcal E}&=\frac12\,\overline{Q_r+2Q_t}\,,\\
B^{J}&=\frac1{2J}\,\overline{r^2Q_t}\,,\\
D^{\mathcal E\mathcal E}&=-2\mathcal E\,\overline{\tilde v_r^2Q_r+\tilde v_\phi^2Q_t}\,,\\
D^{\mathcal EJ}&=J\,\overline{Q_t}\,,\\
D^{JJ}&=\overline{r^2Q_t}\,.
\end{align}

Let us now move to the orbital orientation sector. As discussed in Sec.~\ref{sec:tidal}, the expressions of $B^\mu$ and $D^{\mu\nu}$ are rather involved in the basis of Euler angles $(\Omega,i,\omega)$, due to geometric factors. We therefore leave those available on GitHub \cite{stochastic-binaries:github} and only report here the result in the basis of the binary's axes $(\hat e,\hat q,\hat J)$ plus the mean anomaly at epoch $M_0$, where the diffusion matrix is diagonal except for the usual cross-term $D^{\hat J M_0}$. The explicit expressions can be obtained from those in the basis of Euler angles using the transformation described in \eqref{eqn:Jhatmu}--\eqref{eqn:Dhatmuhatmu}. After orbit averaging, all drift coefficients vanish, $B^{\hat e}=B^{\hat q}=B^{\hat J}=B^{M_0}=0$, while the diffusion coefficients, as explicit orbit averages involving the true anomaly $\phi$, read
\begin{widetext}
\begin{align}
\label{eqn:Dhatehate-impulsive}
D^{\hat e\hat e}&=\frac{\overline{r^2\cos^2\phi\,Q_t}}{Gma(1-e^2)}\,,\\
D^{\hat q\hat q}&=\frac{\overline{r^2\sin^2\phi\,Q_t}}{Gma(1-e^2)}\,,\\
D^{\hat J\hat J}&=\frac{a(1-e^2)}{Gme^2}\cdot\,\overline{\cos^2\phi\,Q_r+\bigg(\frac{2+e\cos\phi}{1+e\cos\phi}\bigg)^2\sin^2\phi\,Q_t}\,,\\
\label{eqn:DM0hatJ-impulsive}
D^{\hat J M_0}&=-\frac{a(1-e^2)^{3/2}}{Gme}\cdot\,\overline{\Big(\frac{\cos\phi}{e}-\frac{2}{1+e\cos\phi}\Big)\cos\phi\,Q_r+\frac1e\Big(\frac{2+e\cos\phi}{1+e\cos\phi}\Big)^2\sin^2\phi\,Q_t}\,,\\
\label{eqn:DM0M0-impulsive}
D^{M_0M_0}&=\frac{a(1-e^2)^2}{Gm}\cdot\,\overline{\bigg(\frac{\cos\phi}e-\frac{2}{1+e\cos\phi}\bigg)^2Q_r+\frac1{e^2}\bigg(\frac{2+e\cos\phi}{1+e\cos\phi}\bigg)^2\sin^2\phi\, Q_t}\,.
\end{align}
\end{widetext}
Equations \eqref{eqn:Dhatehate-impulsive}--\eqref{eqn:DM0M0-impulsive}, together with \eqref{eqn:Ba-impulsive}--\eqref{eqn:Dee-impulsive},
completely specify the stochastic evolution of all binary parameters. Different from the case of tidal perturbations, the diffusion time $T\sped{d}$ defined in \eqref{eqn:Td-impulsive} does not explicitly factorize in the expressions of the coefficients. All coefficients do involve a timescale of the same order as $T\sped{d}$, but its dependence on the eccentricity and mass ratio of the binary, as well as on the perturber's form factor, needs to be calculated numerically unless a simple form for the components $\mathcal Q^{\alpha\beta}$ is available. That is the case for either very large or very small perturbers, which we analyze next.

\subsection{Very large perturbers: recovering the tidal limit}
\label{sec:large-perturbers}

In the previous sections, we have laid down the complete framework for the binary stochastic perturbations induced by impulsive encounters with perturbers with a given spherical density profile $\rho_*(x)$.  The drift and diffusion coefficients can be calculated in closed analytic form in some specific limits, such as that of objects much larger than the size of the binary. This case is particularly interesting for us because the gravitational field produced by very large objects is approximately tidal on the binary's scale, meaning that we should recover the results of Sec.~\ref{sec:tidal}. More precisely, because impulsive three-body encounters have by definition $\tau\ped{c}\ll T$, we expect to find the coefficients derived in Sec.~\ref{sec:white-noise} for white-noise tidal perturbations. This is a highly nontrivial consistency check of our previous results.

Consider for concreteness perturbers with a gaussian density profile,
\beq
\rho_*(x)=\frac{m_*e^{-x^2/(2R^2)}}{(2\pi)^{3/2}R^3}\,,\qquad\tilde\rho_*(k)=e^{-k^2R^2/2}\,,
\eeq
and take the limit $R\gg a$. Let us apply the machinery of Sec.~\ref{sec:kick-covariance-tensor}. Since $r_\perp<r<2a\ll R$, the Bessel function kernels in \eqref{eqn:Yperp+Yparallel}--\eqref{eqn:Yperp-Yparallel} can be expanded to leading order, $1-J_0(x)\approx x^2/4$ and $J_2(x)\approx x^2/8$, obtaining
\begin{align}
\mathcal Y_\perp+\mathcal Y_\parallel&=4\pi\int_0^\infty \tilde\rho_*(k)^2\frac{(kr_\perp)^2}4\frac{\dd k}{k}=\frac{\pi r_\perp^2}{2R^2}\,,\\
\mathcal Y_\perp-\mathcal Y_\parallel&=4\pi\int_0^\infty \tilde\rho_*(k)^2\frac{(kr_\perp)^2}8\frac{\dd k}k=\frac{\pi r_\perp^2}{4R^2}\,.
\end{align}
Solving for $\mathcal Y_\perp$ and $\mathcal Y_\parallel$ and substituting in \eqref{eqn:Qr}--\eqref{eqn:Q}, we obtain
\beq
\begin{split}
Q_r(r)&=\frac{2\sqrt2G^2m_*^2n}{\sqrt\pi\sigma}\int_0^\pi\frac{3\pi(r\sin\theta)^2}{8R^2}\sin^3\theta\dd\theta\\
&=\frac{4\sqrt{2\pi}G^2m_*^2nr^2}{5\sigma R^2}\,,
\end{split}
\label{eqn:Qr-large}
\eeq
and similarly
\beq
Q_t(r)=\frac{4\sqrt{2\pi}G^2m_*^2nr^2}{15\sigma R^2}=\frac{Q_r(r)}3\,.
\label{eqn:Qt-large}
\eeq
We thus have $\mathcal Q^{\alpha\beta}\propto(r^2\delta^{\alpha\beta}+2r^\alpha r^\beta)$, which is the same tensorial structure found in the tidal case, as we will see later in \eqref{eqn:tensorial-structure-tidal}.

To compute the Fokker--Planck coefficients given in \eqref{eqn:Ba-impulsive}--\eqref{eqn:Dee-impulsive} and \eqref{eqn:Dhatehate-impulsive}--\eqref{eqn:DM0M0-impulsive}, we simply need to plug in \eqref{eqn:Qr-large}--\eqref{eqn:Qt-large} and perform all orbit averages. Using $\overline{r^2}=a^2(2+3e^2)/2$, $\overline{r^4}=a^4(8+40e^2+15e^4)/8$, and $\overline{3r^2\tilde v_r^2+r^2\tilde v_\phi^2}=a^2(2+e^2)/2$, we recover \emph{exactly all ten coefficients} given in \eqref{eqn:Ba-white-noise}--\eqref{eqn:Dee-white-noise} and \eqref{eqn:Dhatehate-white-noise}--\eqref{eqn:DM0M0-white-noise}, with a diffusion time
\beq
T\sped{d}=\frac{Gm}{15a^3}\bigg(\frac{Q_t(r)}{r^2}\bigg)^{-1}=\frac{\sigma R^2m}{4\sqrt{2\pi}Gm_*^2na^3}\,.
\label{eqn:Td-large}
\eeq
We emphasize again that recovering the same functional dependence on the eccentricity of all Fokker--Planck coefficients is a nontrivial consistency check. To complete the verification, we further need to confirm whether, starting with the tidal approximation, the normalization of $T\sped{d}$ is the same as in \eqref{eqn:Td-large}. This is indeed the case, but we defer the calculation to Sec.~\ref{sec:density-fluctuations}.

\subsection{Very small and pointlike perturbers}
\label{sec:point-mass}

We can also take the opposite limit of small or pointlike perturbers. As we will show, in this case the Fokker--Planck coefficients can again be expressed in closed analytical form. However, as anticipated in Sec.~\ref{sec:kick-covariance-tensor}, some subtleties related to ultraviolet divergences arise, and they need to be treated carefully.

Using $\tilde\rho_*(k)=1$, the integral \eqref{eqn:Yperp-Yparallel} follows from a property of Bessel functions,
\beq
\mathcal Y_\perp-\mathcal Y_\parallel=4\pi\int_0^\infty J_2(kr_\perp)\frac{\dd k}k=2\pi\,.
\eeq
On the other hand, the integral \eqref{eqn:Yperp+Yparallel} is ultraviolet divergent due to the point-particle nature of the perturber. This divergence comes from a classic Coulomb integral $\sim\int\dd b/b$, and we can treat it in two different ways.

First, similar to the approach of Sec.~\ref{sec:large-perturbers}, we can consider a Gaussian $\tilde\rho_*(k)=e^{-k^2R^2/2}$, and then let $R/a\to0$. The integral \eqref{eqn:Yperp+Yparallel} can be solved exactly,
\beq
\mathcal Y_\perp+\mathcal Y_\parallel=2\pi\bigg[\log\bigg(\frac{r_\perp^2}{4R^2}\bigg)+\gamma\sped{e}+E_1\bigg(\frac{r_\perp^2}{4R^2}\bigg)\bigg]\,,
\label{eqn:YpYp-gaussian}
\eeq
where $\gamma\sped{e}\approx0.577$ is the Euler–Mascheroni constant and the exponential integral $E_1(x)=\int_x^\infty e^{-t}/t\dd t$ can be neglected for $R\ll a$. However, since \eqref{eqn:YpYp-gaussian} depends on the size $R$ of the perturber, the reader may still wonder what happens physically when $R/a\to0$. In that limit, the assumption of straight uniform motion breaks down for perturbers passing close enough to one of the two binary components. As is well-known from other treatments of the Coulomb divergence available in the literature (e.g., see \cite{2008gady.book.....B}), taking into account this deflection introduces a cutoff at the impact parameter corresponding to a $\SI{90}{\degree}$ deflection angle, $b_{90,i}\equiv G(m_i+m_*)/V^2$, with $i\in\{1,2\}$, where $m_i$ is the mass of the $i$-th binary component. To properly implement this cutoff, we need to go back to \eqref{eqn:Delta-v} and regulate each term with its own $b_{90,i}$,
\beq
\frac{\mathcal I(b_i)}{b_i^2}\vec b_i\to\frac{\mathcal I(b_i)}{b_i^2}\vec b_i\Theta(b_i-b_{90,i})\,.
\eeq
The regulated Fourier transform of the kick is then
\beq
\tilde{\vec y}=2\pi i\,\tilde\rho_*(k)\big(J_0(kb_{90,1})-J_0(kb_{90,2})e^{-i\vec k\cdot\vec r_\perp}\big)\frac{\vec k}{k^2}
\eeq
and the integral \eqref{eqn:Yperp+Yparallel} evaluates to
\beq
\begin{split}
\mathcal Y_\perp+\mathcal Y_\parallel&=2\pi\int_0^\infty\Big[J_0(kb_{90,1})^2+J_0(kb_{90,2})^2\\
&\qquad-2J_0(kb_{90,1})J_0(kb_{90,2})J_0(kr_\perp)\Big]\frac{\dd k}k\\
&=2\pi\log\bigg(\frac{r_\perp^2}{b_{90,1}b_{90,2}}\bigg)\,.
\end{split}
\label{eqn:YpYp-pp}
\eeq
Henceforth, we carry both \eqref{eqn:YpYp-gaussian} and \eqref{eqn:YpYp-pp} forward, so that the reader can readily pick the appropriate one depending on which among $R$ and $b_{90}$ is smaller.

Substituting these results in \eqref{eqn:Qr}--\eqref{eqn:Q} and performing the integrals in $\dd\theta$, we get the explicit components of the kick covariance tensor,
\begin{align}
Q_r(r)&=\frac{8\sqrt{2\pi} G^2m_*^2n}{3\sigma}\bigg(\mathcal L(r)-\frac23\bigg)\,,\\
Q_t(r)&=\frac{8\sqrt{2\pi} G^2m_*^2n}{3\sigma}\bigg(\mathcal L(r)-\frac83\bigg)\,,
\end{align}
where for small Gaussian perturbers
\beq
\mathcal L(r)=\log\bigg(\frac{r^2}{R^2}\bigg)+\gamma\sped{e}\,,
\eeq
while for pointlike ones\footnote{There is one subtlety here. In formulae \eqref{eqn:Qr}--\eqref{eqn:Q}, we factored out $\Braket{V^{-1}}$, since $\vec y$ does not depend on $\vec V$. However, the cutoff $b_{90,i}= G(m_i+m_*)/V^2$ does depend on $V$, so the actual velocity integral involves a term $\Braket{V^{-1}\log V^4}$ $=(\log(4\sigma^4)-2\gamma\sped{e})\Braket{V^{-1}}$.}
\beq
\mathcal L(r)=\log\bigg(\frac{16\sigma^4r^2}{G^2(m_1+m_*)(m_2+m_*)}\bigg)-2\gamma\sped{e}\,.
\eeq

Remarkably, all orbital averages needed to compute the drift and diffusion coefficients, \eqref{eqn:Ba-impulsive}--\eqref{eqn:Dee-impulsive} and \eqref{eqn:Dhatehate-impulsive}--\eqref{eqn:DM0M0-impulsive}, reduce to integrals of the form $\int_0^{2\pi}\cos^n\!E\log(1-e\cos E)\dd E$, which can be solved analytically in closed form. The full expressions of the nonzero drift and diffusion coefficients for all orbital elements are
\begin{widetext}
\begin{align}
\label{eqn:Ba-point-mass}
B^{a}&=\frac{a}{15T\sped{d}^*}\bigg(7\log\Lambda-\frac{32}{3}-6\sqrt{1-e^2}\bigg)\,,\\
B^{e}&=\frac1{15T\sped{d}^*}\bigg(\frac{5(1-3e^2)}{4e}\log\Lambda+\frac{33e^4-29e^2+2}{12e^3}+\frac{34e^4-3e^2-1}{6e^3}\sqrt{1-e^2}\bigg)\,,\\
D^{aa}&=\frac{a^2}{15T\sped{d}^*}\bigg(4\log\Lambda-\frac{32}3\bigg)\,, \\
D^{ae}&=\frac{a}{15T\sped{d}^*}\bigg(-\frac{4(1-e^2)\big(1-\sqrt{1-e^2}\big)}{e}\bigg)\,,\\
\label{eqn:Dee-point-mass}
D^{ee}&=\frac1{15T\sped{d}^*}\bigg(\frac52(1-e^2)\log\Lambda-\frac{(1-e^2)(e^2+2)}{6e^2}-\frac{(1-e^2)(16e^2-1)}{3e^2}\sqrt{1-e^2}\bigg)\,,\\
\label{eqn:Dhatehate-point-mass}
D^{\hat e\hat e}&=\frac1{15T\sped{d}^*}\bigg(\frac{4e^2+1}{2(1-e^2)}\log\Lambda-\frac{14e^4-13e^2+2}{6e^2(1-e^2)}-\frac{6e^4+10e^2-1}{3e^2(1-e^2)}\sqrt{1-e^2}\bigg)\,,\\
D^{\hat q\hat q}&=\frac1{15T\sped{d}^*}\bigg(\frac12\log\Lambda-\frac{5e^2-2}{6e^2}-\frac{2e^2+1}{3e^2}\sqrt{1-e^2}\bigg)\,,\\
D^{\hat J\hat J}&=\frac1{15T\sped{d}^*}\bigg(\frac{5-4e^2}{2e^2}\log\Lambda+\frac{2e^4-7e^2+2}{6e^4}+\frac{12e^4-14e^2-1}{3e^4}\sqrt{1-e^2}\bigg)\,,\\
\label{eqn:DM0hatJ-point-mass}
D^{\hat J M_0}&=\frac1{15T\sped{d}^*}\bigg(-\frac{(2e^2+5)\sqrt{1-e^2}}{2e^2}\log\Lambda-\frac{20e^4-19e^2-1}{3e^4}-\frac{20e^4+5e^2+2}{6e^4}\sqrt{1-e^2}\bigg)\,,\\
\label{eqn:DM0M0-point-mass}
D^{M_0M_0}&=\frac1{15T\sped{d}^*}\bigg(\frac{4e^4+11e^2+5}{2e^2}\log\Lambda+\frac{-2e^6+93e^4+15e^2+2}{6e^4}-\frac{4e^6+30e^4+25e^2+1}{3e^4}\sqrt{1-e^2}\bigg)\,,
\end{align}
\end{widetext}
where we defined the effective Coulomb logarithm $\log\Lambda$, which for small Gaussian perturbers reads
\beq
\log\Lambda=\log\bigg(\frac{a^2(1+\sqrt{1-e^2})^2}{4R^2}\bigg)+\gamma\sped{e}\,,
\eeq
and for pointlike perturbers reads
\beq
\log\Lambda=\log\bigg(\frac{4\sigma^4a^2\big(1+\sqrt{1-e^2}\big)^2}{G^2(m_1+m_*)(m_2+m_*)}\bigg)-2\gamma\sped{e}\,.
\label{eqn:logLambda}
\eeq
The time $T\sped{d}^*$ is obtained from the diffusion time by conveniently stripping it of its $\log\Lambda$ and eccentricity dependence, but retaining the conventional factor of 15,
\beq
\begin{split}
T\sped{d}^*&\equiv\frac{m\sigma}{40\sqrt{2\pi}\,Gm_*^2na}\\
&=\frac{T\sped{d}}{\sqrt{1-e^2}}\bigg[\log\Lambda-\frac83+4\log\frac{2\sqrt{1-e^2}}{1+\sqrt{1-e^2}}\bigg]\,.
\end{split}
\label{eqn:Td*}
\eeq

Except for $D^{ae}$, all coefficients \eqref{eqn:Ba-point-mass}--\eqref{eqn:DM0M0-point-mass} contain a ``leading logarithm'' term, plus a finite remainder. The former dominates in the impulsive limit $\sigma^2\gg Gm/a$. However, logarithms grow slowly so the other terms may still be important even rather deep in the impulsive regime. A crucial aspect of our calculation is that our Coulomb logarithm $\log\Lambda$ contains \emph{no arbitrary choice}.\footnote{The reader may doubt this statement, since the introduction of the cutoff $b_{90}$ in our perturbative calculation might seem like an arbitrary choice. However, it is possible to calculate the effect of strong deflections \emph{nonperturbatively}, rather than cutting them off. It turns out that, when one does this, the answer is exactly the same as that found by introducing the cutoff $b_{90}$. This is a well-known result in the context of single-single star scattering~\cite{2008gady.book.....B}. Remarkably, it also holds for scattering of an object off a binary, because each close encounter reduces to an exact two-body problem: the deflection induced by the companion, while not negligible along individual trajectories, enters the impact-parameter integral only as an area-preserving relabeling, so its net effect vanishes in the impulsive limit and formula \eqref{eqn:logLambda} becomes exact, with corrections suppressed by $b_{90}/a$. With the help of Anthropic's Claude, we have obtained a proof of this fact, which this footnote is too narrow to contain. The proof is however accessible on GitHub together with the scripts validating all the results of this paper \cite{stochastic-binaries:github}.} All coefficients \eqref{eqn:Ba-point-mass}--\eqref{eqn:DM0M0-point-mass} are exact in the impulsive limit---we will return to this point when comparing with previous works in Sec.~\ref{sec:lit-impulsive}.

\subsubsection{Semi-major axis and eccentricity}

Let us look in more detail at the $(a,e)$ sector of the Fokker--Planck equation. First of all, we report for completeness the coefficients in terms of energy and angular momentum variables, where they take a slightly more compact form,
\begin{align}
B^{\mathcal E}&=\frac{-2\mathcal E}{15T\sped{d}^*}\bigg(\frac32\log\Lambda-3\sqrt{1-e^2}\bigg)\,,\\
B^{J}&=\frac{D^{JJ}}{2J}\,,\\
D^{\mathcal E\mathcal E}&=\frac{4\mathcal E^2}{15T\sped{d}^*}\bigg(\log\Lambda-\frac83\bigg)\,,\\
D^{\mathcal EJ}&=\frac{-2\mathcal EJ}{15T\sped{d}^*}\bigg(\log\Lambda-\frac23-2\sqrt{1-e^2}\bigg)\,,\\
\begin{split}
D^{JJ}&=\frac{-2\mathcal Ea^2}{15T\sped{d}^*}\bigg[\bigg(1+\frac32e^2\bigg)\log\Lambda+1-\frac32e^2\\
&\qquad\qquad\qquad\quad-\frac{4e^2+11}3\sqrt{1-e^2}\bigg]\,.
\end{split}
\end{align}
As in the case of white-noise tidal perturbations (Sec.~\ref{sec:white-noise}), we have $B^{\mathcal E}>0$, implying that the binaries are on average heated by impulsive encounters.\footnote{Technically, this is only true if $\log\Lambda>2$, but this is essentially always satisfied in practice for small perturbers in the impulsive regime.}

\begin{figure*}
\centering
\includegraphics[width=\textwidth]{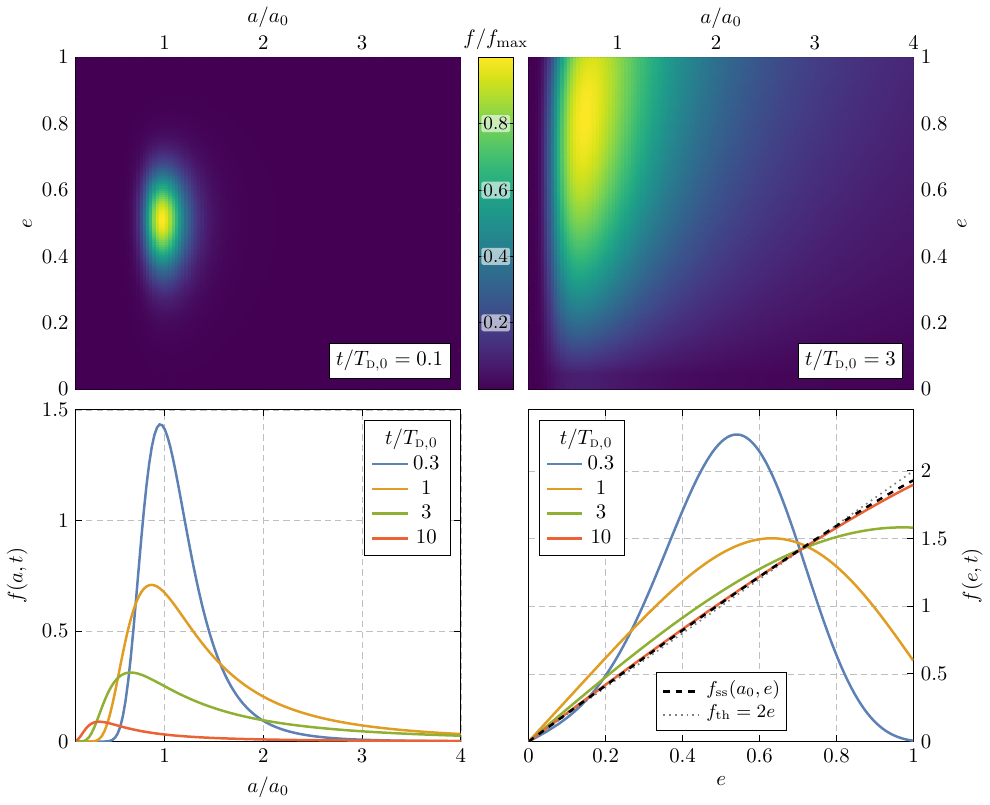}
\caption{Numerical solution of the $(a,e)$ sector of the Fokker--Planck equation \eqref{eqn:fp-ae} for impulsive encounters with point-mass perturbers, using the drift and diffusion coefficients \eqref{eqn:Ba-point-mass}--\eqref{eqn:Dee-point-mass} and using $\log\Lambda_0=15$. Time is in units of the diffusion time $T\sped{d,0}\equiv T\sped{d}(a=a_0,e=0.5)$, evaluated from \eqref{eqn:Td*} with $\log\Lambda_0=15$. Initial conditions, boundary conditions, and structure of the figure are the same as in Fig.~\ref{fig:f-white-noise}. The eccentricity marginal $f(e,t)$ relaxes to the slightly sub-thermal distribution $f\ped{ss}(a,e)$ given in \eqref{eqn:f-ss-point-mass}, which we evaluate at $a=a_0$.}
\label{fig:f-point-mass}
\end{figure*}

The expressions for the currents $J^a$ and $J^e$ (which we do not report explicitly here) retain the same structure discussed for white-noise tidal perturbations in Sec.~\ref{sec:white-noise}: $e=0$, $e=1$, and $a=0$ are inaccessible boundaries, while $a\to\infty$ requires a choice which we take to be absorbing. Because $B^a>0$, the binaries soften on average. On the other hand, the eccentricity marginal relaxes to a quasi-stationary distribution. We can calculate it analytically introducing the constant part of the effective Coulomb logarithm \eqref{eqn:logLambda}, $\log\Lambda_0\equiv\log\Lambda(a=a_0,e=0)$, and working perturbatively in $1/\log\Lambda_0$. We then proceed similarly to the constant-current solution shown in the case of tidal perturbations---see Eq.~\eqref{eqn:f-steady-state} and the discussion around it. We consider the ansatz $f\ped{ss}(a,e)=g(e)/(a^2\log\Lambda(a,e))$, which at order $\mathcal O(1/\log\Lambda_0)$ makes $\partial_aJ^a$ vanish and the cross term $D^{ae}f$ $a$-independent. The steady-state condition then becomes $J^e=0$, or
\beq
B^eg=\frac12\partial_e(D^{ee}g)\,.
\eeq
This ordinary differential equation for $g$ can be solved at first order in $1/\log\Lambda_0$, giving
\beq
\begin{split}
f\ped{ss}\propto\frac{2e}{a^2}\bigg(1&+\frac{-2\log(a/a_0)+4\sqrt{1-e^2}}{\log\Lambda_0}\\
&+\frac{-4\log\big(1+\sqrt{1-e^2}\big)+2\log2}{\log\Lambda_0}\bigg)\,.
\end{split}
\label{eqn:f-ss-point-mass}
\eeq
To zeroth order in $1/\log\Lambda_0$, the steady-state eccentricity distribution is thermal. However, a correction of order $\mathcal O(1/\log\Lambda_0)$ makes the actual steady-state distribution slightly sub-thermal at any $a$.

We show in Fig.~\ref{fig:f-point-mass} a numerical solution to the $(a,e)$ sector of the Fokker--Planck equation, using $\log\Lambda_0=15$. The numerical result in Fig.~\ref{fig:f-point-mass} indeed converges to \eqref{eqn:f-ss-point-mass}.

\subsubsection{Orbital orientation and mean anomaly at epoch}

The $(\hat e,\hat q,\hat J,M_0)$ sector features properties similar to the ones discussed in the tidal case, Secs.~\ref{sec:orientation-adiabatic} and~\ref{sec:orientation-white-noise}. For $e\to0$, both $D^{\hat e\hat e}$ and $D^{\hat q\hat q}$ tend to the same limit, $((1/2)\log\Lambda-4/3)/(15T\sped{d}^*)$. The other coefficients, $D^{\hat J\hat J}$, $D^{\hat J M_0}$, and $D^{M_0M_0}$, all diverge, while the mean longitude at epoch $\lambda_0=M_0+\omega$ does not diffuse. On the other hand, for $e\to1$, we have $D^{\hat q\hat q}=D^{\hat J\hat J}=(\log\Lambda-1)/(30T\sped{d}^*)$, while $D^{\hat e\hat e}$ diverges as expected.

\subsection{Perturbers of generic size}
\label{sec:generic-size}

For perturbers that are neither very large nor very small compared to the binary's semi-major axis, the calculation of Fokker--Planck coefficients must in general be performed numerically. It is interesting to compare the numerical result to the analytic ones we derived in specific limits in Secs.~\ref{sec:large-perturbers} and~\ref{sec:point-mass}.

\begin{figure}
\centering
\includegraphics[width=0.48\textwidth]{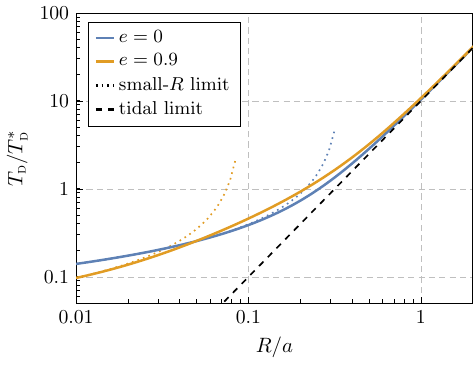}
\caption{Diffusion time $T\sped{d}$, as defined in \eqref{eqn:Td-impulsive} and normalized by $T\sped{d}^*=(m\sigma)/(40\sqrt{2\pi}\,Gm_*^2na)$, for a binary undergoing impulsive encounters with perturbers with Gaussian density profile $\rho_*(x)\propto e^{-x^2/(2R^2)}$, as a function of $R/a$. The tidal limit is given in \eqref{eqn:Td-large} and recovers the Fokker--Planck coefficients in \eqref{eqn:Ba-white-noise}--\eqref{eqn:Dee-white-noise} and \eqref{eqn:Dhatehate-white-noise}--\eqref{eqn:DM0M0-white-noise}. The small-$R$ limit is given in \eqref{eqn:Td*} and recovers the Fokker--Planck coefficients in \eqref{eqn:Ba-point-mass}--\eqref{eqn:DM0M0-point-mass}.}
\label{fig:Td(R)}
\end{figure}

Rather than comparing every single coefficient, for simplicity we focus only on the diffusion time $T\sped{d}$, as defined in \eqref{eqn:Td-impulsive}. We show in Fig.~\ref{fig:Td(R)} the diffusion time as a function of perturber size, for Gaussian perturbers $\tilde\rho_*(k)=e^{-k^2R^2/2}$, normalized by the $R$- and $e$-independent $T\sped{d}^*$ as defined in \eqref{eqn:Td*}. As expected, for large $R/a$ the diffusion time reduces to the tidal limit $T\sped{d}\to10T\sped{d}^*(R/a)^2$ given in \eqref{eqn:Td-large}. The approximation is remarkably good, with errors of less than 9\% for any value $R/a>1$ and any eccentricity $e<1$. The small-$R$ limit also recovers the small-perturber limit from \eqref{eqn:Td*}, which approaches zero logarithmically according to
\beq
T\sped{d}\to\frac{T\sped{d}^*\sqrt{1-e^2}}{\log(a^2/(4R^2))}\,.
\eeq
The higher the eccentricity, the lower the value of $R/a$ where the small-$R$ formula \eqref{eqn:Td*} becomes accurate. For instance, an accuracy better than $10\%$ is achieved for $R/a<0.16$ ($R/a\lesssim 0.032$) when $e=0$ ($e=0.9$).

\section{Application to physical cases}
\label{sec:applications}

In the previous sections, we classified binary perturbations as tidal or impulsive, based on the form of the stochastic force $\vec F\sped{s}$. We remained agnostic on the physical origin of that stochastic force. In this section, we apply our results to explicit and astrophysically relevant examples. We start in Sec.~\ref{sec:applicability} by evaluating in physical units the range of applicability of our results. We then consider a generic stochastic density field in Sec.~\ref{sec:density-fluctuations}, and specialize it to the case of ultralight dark matter in Sec.~\ref{sec:uldm}. In Sec.~\ref{sec:dm-substructures} we study the effects of generic (spherical) dark matter subhaloes, and end in Sec.~\ref{sec:ism} with perturbations from the gaseous interstellar medium.

\subsection{Range of applicability}
\label{sec:applicability}

\subsubsection{Slow evolution}
\label{sec:slow-evolution}

The core assumption of the whole formalism presented in Sec.~\ref{sec:formalism}, as already expressed in \eqref{eqn:Ttauc<<Td}, is that the diffusion timescale $T\sped{d}$ is much longer than both the binary's period $T=2\pi\sqrt{a^3/(Gm)}$ and the coherence time $\tau\ped{c}$. The first of these inequalities, $T\ll T\sped{d}$, means that the perturbation is \emph{weak}. This justifies the use of orbital averages,\footnote{It is nevertheless possible to track the evolution of orbital elements beyond the orbit average approximation. This may lead to enhanced sensitivity to the properties of the perturbing field~\cite{Foster:2025nzf,Foster:2025csl}.} since the fractional variation of the orbital parameters over the course of an orbit is small.

Now that we have calculated $T\sped{d}$ as a function of the parameters of the binary and its environment, we can quantitatively verify when the assumption holds. Let us first consider the case of perturbers with size $R>a$,\footnote{We write the tidal limit simply as $R>a$, rather than $R\gg a$, due to the exceptional accuracy of the expression for $R\approx a$, as shown in Fig.~\ref{fig:Td(R)} and discussed in Sec.~\ref{sec:generic-size}.} coinciding with the tidal limit, whose $T\sped{d}$ is given in \eqref{eqn:Td-tidal-R}. Requiring that $T\ll T\sped{d}$, we obtain\footnote{We write these inequalities keeping $a^{5/2}$ (rather than $a$) on the LHS, to emphasize that a correction of order $\mathcal O(T/T\sped{d})$ can be estimated by taking the ratio of LHS and RHS of~\eqref{eqn:T<<Td-large} and~\eqref{eqn:T<<Td-small}.}
\beq
\begin{split}
a^{5/2}\ll{}&(\SI{3.6}{pc})^{5/2}\bigg(\frac{R}a\bigg)^2\bigg(\frac{\sigma}{\SI{100}{km/s}}\bigg)\bigg(\frac{m}{M_\odot}\bigg)^{3/2}\\
&\times\bigg(\frac{M_\odot/\si{pc}^3}{\bar\rho}\bigg)\bigg(\frac{M_\odot}{m_*}\bigg)
\end{split}
\label{eqn:T<<Td-large}
\eeq
where $\bar\rho=m_*n$ is the average mass density of the perturbers. Considering that, e.g., the widest stellar binaries in the Milky Way have separation $a\lesssim\mathcal O(\SI{1}{pc})$, essentially any set of perturbations with $R/a > 1$ will satisfy this inequality. A mass density of $\bar\rho=1\,M_\odot/\si{pc}^3$ is higher than that in the solar neighborhood, so that even perturbers with mass $m_*>10^3M_\odot$ satisfy the requirement.

For small perturbers, with $R\ll a$, we get a slightly tighter condition,
\beq
\begin{split}
a^{5/2}\ll{}&(\SI{0.39}{pc})^{5/2}\bigg(\frac{\sigma}{\SI{100}{km/s}}\bigg)\bigg(\frac{m}{M_\odot}\bigg)^{3/2}\\
&\times\bigg(\frac{M_\odot/\si{pc}^3}{\bar\rho}\bigg)\bigg(\frac{M_\odot}{m_*}\bigg)\bigg(\frac{25}{\log\Lambda}\bigg)
\end{split}
\label{eqn:T<<Td-small}
\eeq
where we neglected the eccentricity-dependent factors in \eqref{eqn:Td*}. This inequality may be violated by very wide binaries immersed in a dense bath of massive but small perturbers.

\subsubsection{Negligible shot noise}
\label{sec:shot-noise}

Besides $T\ll T\sped{d}$, there is another sense in which perturbations must be weak. To model impulsive encounters as a smooth diffusive process, each individual encounter must induce a small change in the binary's parameters; mathematically this is because our truncation of the Kramers--Moyal expansion \eqref{eqn:kramers-moyal} assumes the variance of orbital increments scales linearly with time. 

We can perform a simple consistency check that there is negligible shot noise, by requiring that the average time between two consecutive penetrative encounters is much shorter than the diffusion timescale for semi-major axis evolution,
\beq
\frac1{na^2\sigma}\ll T\sped{d}\,.
\eeq
In other words, by the time the orbital parameters change significantly, the binary has undergone a large number of three-body encounters. For $R>a$ we use the diffusion time estimate \eqref{eqn:Td-large}, and this condition gives
\beq
a\gg\SI{4.3e-6}{pc}\,\bigg(\frac{a}R\bigg)^2\bigg(\frac{\SI{100}{km/s}}\sigma\bigg)^2\frac{m_*^2}{M_\odot m}\,,
\label{eqn:poisson-large}
\eeq
while for $R\ll a$ we use the diffusion time \eqref{eqn:Td*} and find
\beq
a\gg\SI{1.1e-3}{pc}\,\bigg(\frac{\SI{100}{km/s}}\sigma\bigg)^2\bigg(\frac{\log\Lambda}{25}\bigg)\frac{m_*^2}{M_\odot m}\,.
\eeq
When these inequalities do fail, the binaries' evolution will be a shot noise-dominated ``jump process'', rather than a smooth diffusion; Refs.~\cite{Penarrubia:2019wei,2024MNRAS.532.2425H,Bhalla:2025xce} referred to this region of semi-major axis-space as ``the fringe''. In the fringe, encounters are sometimes referred to as ``catastrophic'' because a single encounter is sufficient to unbind the binary \cite{2008gady.book.....B}.

Another physical requirement for our framework to apply is that binaries undergo a large number of penetrative encounters within their typical lifetime $1/(na^2\sigma)\ll T\ped{life}$, see \cite{2024MNRAS.532.2425H}. Binaries that violate this condition \emph{only} experience distant impulsive encounters. In a homogeneous sea of perturbers, the corresponding relative velocity kick from a distant encounter scales as $\Delta v \propto b^{-2}$, meaning its variance scales as $\int b^{-4} b\dd b \propto \int b^{-3}\dd b$. This integral is entirely dominated by its lower limit, so the cumulative effect is not a smooth diffusion, but a Lévy flight dictated by the single nearest passing body \cite{Collins:2008kf,Collins:2010pv}, which may need to be modeled with fractional diffusion equations \cite{Seoane:2025kcr}. See Sec.~\ref{sec:lit-adiabatic} for further comments.

\subsubsection{High eccentricity}

While the condition $T\ll T\sped{d}$ generally ensures that the orbital parameters evolve slowly, the perturbative picture eventually always breaks down at sufficiently high eccentricities. The reason is that the relative variation of the angular momentum over one orbit,
\beq
\frac{\Delta J}{J}\sim\frac{e\,\Delta e}{1-e^2}\,,
\eeq
diverges for $e\to1$. It is then natural to ask for what values of $e$ our formalism remains valid.

For adiabatic tidal perturbations, we can estimate
\beq
\Delta e\sim T\dot e\sim\sqrt{\frac{a(1-e^2)}{Gm}}T_{ij}a\sim\sqrt{1-e^2}\frac{T}{\sqrt{\tau\ped{c}T\sped{d}}}\,,
\eeq
so the requirement $\Delta J/J\ll1$ gives
\beq
\sqrt{1-e^2}\gg\frac{T}{\sqrt{\tau\ped{c}T\sped{d}}}\,.
\label{eqn:validity-adiabatic}
\eeq
The right-hand side is always parametrically small, because $T\ll\tau\ped{c}\ll T\sped{d}$, and the resulting limiting eccentricities are exceedingly high.

Conversely, for white-noise tidal perturbations and impulsive three-body encounters, the evolution of the eccentricity is diffusive even within one orbital period, because the perturbation loses coherence well before the binary completes one orbit. We can then estimate
\beq
\begin{split}
&\Delta e\sim\sqrt{D^{ee}T}\sim\sqrt{1-e^2}\sqrt{\frac{T}{T\sped{d}}}\\
&\;\;\implies\sqrt{1-e^2}\gg\sqrt{\frac{T}{T\sped{d}}}\,.
\end{split}
\eeq
While this condition is appreciably more restrictive than \eqref{eqn:validity-adiabatic}, for concrete values of $T/T\sped{d}$ the validity limit still extends to extremely high eccentricities (e.g., up to $e\approx0.995$ for $T/T\sped{d}=10^{-2}$).

\subsection{Tidal perturbations from density fluctuations}
\label{sec:density-fluctuations}

Before jumping straight into a concrete astrophysical example, we consider a general setup where the binary is immersed in a stochastic density field $\rho(\vec x,t)$ and is perturbed by the tidal forces generated by that environment. Suppose that the stochastic properties (such as the power spectrum $P_\rho$) of $\rho(\vec x,t)$ are known. To apply the formalism of Sec.~\ref{sec:tidal}, we need to compute the diffusion time $T\sped{d}$, as defined in Eq.~\eqref{eqn:Td-tidal}, in terms of $P_\rho$. In particular, we have to evaluate the correlator
\beq
\int_{-\infty}^\infty\braket{T_{ij}(t)T_{kl}(t+\tau)}\dd\tau\,.
\label{eqn:fdm-integral-TijTkl}
\eeq
The calculation is most easily done in momentum space. As in Sec.~\ref{sec:impulsive}, we indicate Fourier transforms with a tilde; however, we denote here the wavenumber with $\vec p$ rather than $\vec k$, to avoid confusion with the cartesian indices of the tidal tensor. From Poisson's equation $\nabla^2\Phi=4\pi G\rho$ we have
\beq
\tilde\Phi(\vec p,t)=-\frac{4\pi G}{p^2}\tilde\rho(\vec p,t)\,,
\eeq
and, from the definition of the tidal tensor,
\beq
\tilde T_{ij}(\vec p,t)=-p_ip_j\tilde\Phi(\vec p,t)=4\pi G\,\frac{p_ip_j}{p^2}\tilde\rho(\vec p,t)\,.
\eeq
We can thus write the tidal field correlator entering \eqref{eqn:fdm-integral-TijTkl} as
\beq
\begin{split}
&\braket{T_{ij}(\vec x,t)T_{kl}(\vec x,t+\tau)}\\
&=\int\frac{\dd^3p}{(2\pi)^3}\int\frac{\dd^3p'}{(2\pi)^3}e^{i(\vec p-\vec p')\cdot\vec x}\braket{\tilde T_{ij}(\vec p,t)\tilde T_{kl}^*(\vec p',t+\tau)}\\
&=(4\pi G)^2\int\frac{\dd^3p}{(2\pi)^3}\frac{p_ip_jp_kp_l}{p^4}P_\rho(p,\tau)\\
&=(4\pi G)^2\frac{\mathds{1}_{ijkl}}{15}\int_0^\infty\frac{p^2\dd p}{2\pi^2}P_\rho(p,\tau)\,,
\end{split}
\label{eqn:calculating-TijTkl}
\eeq
where we defined the unequal-time power spectrum of density fluctuations $P_\rho$ from
\beq
\braket{\tilde\rho(\vec p,t)\tilde\rho^*(\vec p',t+\tau)}=(2\pi)^3\delta^{(3)}(\vec p-\vec p')P_\rho(p,\tau)\,.
\label{eqn:P-rho}
\eeq

A few comments are in order. First, note that in \eqref{eqn:calculating-TijTkl} we evaluated $T_{ij}$ and $T_{kl}$ at the same spatial point $\vec x$. This does not need to be the case, as one could take spatial decoherence into account by calculating $\braket{T_{ij}(\vec x,t)T_{kl}(\vec x+\vec r,t+\tau)}$ for a generic $\vec r$. However, the formalism of Sec.~\ref{sec:tidal} assumes that the acceleration can be written as $a_i=-T_{ij}r_j$, where $T_{ij}$ is independent of the position. Only those binaries for which spatial decoherence can be neglected fall within that regime: in other words, we must have $a\ll\ell\ped{c}$, where $\ell\ped{c}$ is the coherence length of density perturbations. The correlator \eqref{eqn:calculating-TijTkl} is thus all we need. Second, we assumed that the density field is
\begin{itemize}
\item statistically homogeneous, i.e., $\braket{\rho(\vec x,t)\rho(\vec y,t+\tau)}$ is only a function of $\vec x-\vec y$, hence the delta function $\delta^{(3)}(\vec p-\vec p')$ in \eqref{eqn:P-rho};
\item statistically isotropic, so $P_\rho$ is only a function of $p$ rather than $\vec p$.
\end{itemize}
Third, we used isotropy to factorize the radial and angular parts of the integral over $\dd^3p$. In particular, the angular integral must give a rank-4 isotropic tensor symmetric under all permutations of $i, j, k, l$, and the only such tensor is $\mathds{1}_{ijkl}$. The normalization of $1/15$ can be easily determined after having fixed the tensorial structure, and justifies the inclusion of this numerical factor in the definition \eqref{eqn:Td-tidal}.

By comparing with Eq.~\eqref{eqn:Td-tidal}, we can immediately write down the formula for the diffusion time $T\sped{d}$ in terms of the unequal-time power spectrum $P_\rho$,
\beq
T\sped{d}=\frac{Gm}{(4\pi G)^2a^3}\bigg(\int_{-\infty}^\infty\dd\tau\int_0^\infty\frac{p^2\dd p}{2\pi^2}P_\rho(p,\tau)\bigg)^{-1}\,.
\label{eqn:Td-Prho}
\eeq
This procedure is completely general and can be applied to any physical case where the density fluctuations induce a tidal perturbation (either adiabatic or white-noise) and the assumptions discussed in Sec.~\ref{sec:applicability} are satisfied.

As an example, let us explicitly evaluate the result for a bath of $N$ perturbers with number density $n$, each with mass $m_*$ and spherical density profile $\rho_*(x)$, moving with velocities $\vec V_i$ relative to the binary's center of mass drawn from a Maxwellian distribution. This will also allow us to make contact with the tidal limit considered in Sec.~\ref{sec:large-perturbers} and verify that the diffusion time matches that obtained from the impulsive encounter route. The density field of the bath can be written as
\beq
\rho(\vec x,t)=\sum_{i=1}^N\rho_*(\abs{\vec x-\vec x_i(t)})\,,
\eeq
where $\vec x_i(t)=\vec x_{0,i}-\vec V_it$ and $\vec x_{0,i}$ is the initial position of the perturber. In Fourier space, the density correlator reads
\beq
\begin{split}
&\braket{\tilde\rho(\vec p,t)\tilde\rho^*(\vec p',t+\tau)}\\
&=\Braket{\sum_{i,j}^Ne^{-i\vec p\cdot\vec x_i(t)+i\vec p'\cdot\vec x_j(t+\tau)}\tilde\rho_*(p)\tilde\rho_*(p')}\,.
\end{split}
\eeq
For statistically independent initial positions $\vec x_{0,i}$, the exponential contains a random phase, which is killed by the ensemble average, unless $i=j$. Replacing the remaining sum over diagonal terms with an integral over the distribution of initial positions and velocities,
\beq
\sum_{i=1}^N\quad\longrightarrow\quad n\int\dd^3x_0\int\dd^3V\,\frac{e^{-V^2/(2\sigma^2)}}{(2\pi)^{3/2}\sigma^3}\,,
\eeq
we can put the correlator in the form \eqref{eqn:P-rho}, with power spectrum
\beq
P_\rho(p,\tau)=m_*^2n\,\tilde\rho_*(p)^2e^{-p^2\sigma^2\tau^2/2}\,.
\eeq
We can now plug this into \eqref{eqn:Td-Prho} and perform the integrals. For Gaussian perturbers with profile $\tilde\rho_*(p)=e^{-p^2R^2/2}$, we obtain
\beq
T\sped{d}=\frac{\sigma R^2m}{4\sqrt{2\pi}Gm_*^2na^3}\,,
\label{eqn:Td-tidal-R}
\eeq
which coincides with \eqref{eqn:Td-large} exactly. This completes the proof, initiated in Sec.~\ref{sec:large-perturbers}, that the white-noise regime of tidal stochastic perturbations is recovered as a limit of impulsive three-body encounters with spatially large perturbers.

\subsection{Ultralight Dark Matter}
\label{sec:uldm}

A scenario where our formalism applies directly is that of an ensemble of binaries immersed in a halo of ultralight dark matter (ULDM), also known as Fuzzy or
Wave Dark Matter \cite{Hu:2000ke,Hui:2016ltb,Hui:2021tkt}. ULDM consists of a real scalar field $\Psi$ of mass $\mu\sim10^{-22}\text{--}10^{-20}\,\si{eV}$, obeying the Klein-Gordon equation
\beq
\bigg(\Box-\frac{\mu^2c^2}{\hbar^2}\bigg)\Psi=0\,.
\eeq
Different from models of particle dark matter, ULDM haloes generically exhibit order-unity density fluctuations, which can be interpreted as wave interference patterns \cite{Schive:2014dra,Hui:2020hbq,Hui:2021tkt}. We denote the dark matter density by $\rho_\Psi=\bar\rho+\rho$, where $\bar\rho$ is its mean value and $\rho$ is the fluctuating stochastic part, with $\rho/\bar\rho\sim\mathcal O(1)$ and only $\rho$ contributing to the stochastic tidal field. The typical spatial extent of the density fluctuations is the de Broglie wavelength,
\beq
\lambda\ped{dB}\equiv\frac\hbar{\mu\sigma}=\SI{0.96}{pc}\,\bigg(\frac{\SI{e-20}{eV}}{\mu}\bigg)\bigg(\frac{\SI{200}{km/s}}{\sigma}\bigg)\,,
\label{eqn:lambda-dB}
\eeq
and they have a typical coherence time
\beq
\tau\ped{c}\equiv\frac\hbar{\mu\sigma^2}=\SI{4.7e3}{yr}\,\bigg(\frac{\SI{e-20}{eV}}{\mu}\bigg)\bigg(\frac{\SI{200}{km/s}}{\sigma}\bigg)^2\,,
\eeq
where $\sigma$ is the dark matter one-dimensional velocity dispersion. Although we kept $\hbar$ explicit until now, we will set $\hbar=1$ in the remainder of this section.

Such ULDM density fluctuations generate a stochastic and time-varying tidal field, which perturbs the binaries within the halo. We laid down the theory of stochastic isotropic tidal perturbations in Sec.~\ref{sec:tidal}, and connected it to the power spectrum of density fluctuations in Sec.~\ref{sec:density-fluctuations}. We now wish to apply everything to the case of ULDM. In order to use those results, we simply need to
\begin{enumerate}
\item check that the perturbation is well approximated as a tidal field;
\item check whether we are in the adiabatic ($T\ll\tau\ped{c}$) or white-noise ($\tau\ped{c}\ll T$) regime;
\item compute the density power spectrum $P_\rho(p,\tau)$, and the diffusion time $T\sped{d}$ from \eqref{eqn:Td-Prho}.
\end{enumerate}
Once we have performed these steps, we can read off the Fokker--Planck drift and diffusion coefficients given in Eqs.~\eqref{eqn:Ba-adiabatic}--\eqref{eqn:Dee-adiabatic} and~\eqref{eqn:Dhatehate-adiabatic}--\eqref{eqn:DM0M0-adiabatic} for the adiabatic regime, and in Eqs.~\eqref{eqn:Ba-white-noise}--\eqref{eqn:Dee-white-noise} and~\eqref{eqn:Dhatehate-white-noise}--\eqref{eqn:DM0M0-white-noise} for the white-noise regime.

Let us start with the first check. Because the stochastic part $\rho$ of the dark matter density fluctuates over a characteristic distance $\lambda\ped{dB}$, the same is true for its gravitational field $\Phi$ and for the tidal field $T_{ij}=\partial_i\partial_j\Phi$. In order to apply the formalism of Sec.~\ref{sec:tidal}, where the relative acceleration between the binary components is assumed to be $a_i=-T_{ij}r_j$, we therefore need
\beq
a\ll\lambda\ped{dB}\,.
\label{eqn:tidal-limit}
\eeq
This ensures that the perturbation along the entire binary orbit can be derived from a given tidal field. For the reference values of $\mu$ and $\sigma$ used in \eqref{eqn:lambda-dB}, the condition $a\ll\lambda\ped{dB}$ applies to essentially all binaries in a Milky Way-like galaxy, because semi-major axes larger than the typical interstellar distance $\mathcal O(\SI{1}{pc})$ are prone to quick unbinding due to three-body interactions.

The second condition listed above amounts to a simple numerical check. In terms of the binary's semi-major axis, we are in the adiabatic regime ($T\ll\tau\ped{c}$) whenever
\beq
a\ll\SI{1.4}{mpc}\,\bigg(\frac{m}{M_\odot}\bigg)^{1/3}\bigg(\frac{\SI{e-20}{eV}}{\mu}\bigg)^{2/3}\bigg(\frac{\SI{200}{km/s}}{\sigma}\bigg)^{4/3}\,,
\eeq
while in the opposite limit we are in the white-noise regime ($T\gg\tau\ped{c}$).

Finally, we need to evaluate $P_\rho(p,\tau)$ for the case of ULDM. We report the explicit calculation here, although similar derivations are also available in the literature, see e.g.~\cite{Bar-Or:2018pxz,Foster:2017hbq,Centers:2019dyn,Boddy:2025oxn,Kim:2023pkx, Kim:2024xcr}. We start by decomposing the ULDM field in a way that takes advantage of the nonrelativistic velocities typical of dark matter haloes ($\sigma\ll c$), namely
\beq
\Psi(\vec x,t)=\frac1{\sqrt{2\mu}}\big(\psi(\vec x,t)e^{-i\mu c^2t}+\psi(\vec x,t)^*e^{i\mu c^2t}\big)\,.
\label{eqn:nr-ansatz}
\eeq
Here, the complex ``wavefunction'' $\psi$ varies on the slow timescale $\dot\psi/\psi\sim1/\tau\ped{c}=\mu\sigma^2$, while the exponentials oscillate with the much faster Compton frequency $\mu c^2$. The ULDM density can be read off as the 00 component of the corresponding energy-momentum tensor. Using the free Schrödinger equation $\dot\psi=i\nabla^2\psi/(2\mu)$, and expanding to next-to-leading order in $\sigma^2/c^2$, we obtain
\beq
\begin{split}
\rho_\Psi={}&\frac1{2c^2}\bigg(\frac1{c^2}\dot\Psi^2+(\vec\nabla\Psi)^2+\mu^2c^2\Psi^2\bigg)\\
={}&\rho\ped{slow}+\rho\ped{fast}\,,
\end{split}
\eeq
where
\begin{align}
\rho\ped{slow}&=\mu\abs{\psi}^2+\frac{\abs{\vec\nabla\psi}^2}\mu-\frac{\nabla^2\abs{\psi}^2}{4\mu}+\mathcal O\bigg(\frac{\sigma^4}{c^4}\bigg)\,,\\
\rho\ped{fast}&=\frac{e^{-2i\mu c^2t}\vec\nabla\cdot(\psi\vec\nabla\psi)+\text{c.c.}}{4\mu}+\mathcal O\bigg(\frac{\sigma^4}{c^4}\bigg)\,.
\end{align}
Intuitively, the two terms $\rho\ped{slow}$ and $\rho\ped{fast}$ evolve with typical timescales $\tau\ped{c}$ and $1/\mu c^2$ respectively. Given that $\rho\ped{fast}\sim\mathcal O(\sigma^2/c^2)$, the overall leading term (zeroth order in $\sigma^2/c^2$) appearing in $\rho_\Psi$ is $\mu\abs{\psi}^2$.

We now wish to connect these expressions to the dark matter properties, such as the average density $\bar\rho$ and velocity dispersion $\sigma$. Let us assume again that $\psi$ obeys the free Schrödinger equation, so that $\tilde\psi(\vec p,t)\propto e^{-i\omega_pt}$, where $\omega_p=p^2/(2\mu)$. Furthermore, let us treat $\psi$ as a homogeneous, isotropic Gaussian random field, with $\braket{\psi}=\braket{\psi\psi}=0$ and wavefunction correlator
\beq
\braket{\tilde\psi(\vec p,t)\tilde\psi^*(\vec p',t+\tau)}=(2\pi)^3\delta^{(3)}(\vec p-\vec p')P_\psi(p)e^{i\omega_p\tau}\,.
\label{eqn:psi-correlator}
\eeq
It is then easy to express the mean density as the equal-time physical-space correlator
\beq
\bar\rho=\braket{\rho_\Psi}\approx\mu\braket{\abs{\psi(\vec x,t)}^2}=\mu\int\frac{\dd^3p}{(2\pi)^3}P_\psi(p)\,.
\eeq
We can identify this formula with the classical expression of the density as a phase space integral of the dark matter distribution function $f(\vec x,\vec v)$,
\beq
\rho_\Psi(\vec x)=\int\dd^3v\,f(\vec x,\vec v)\,.
\eeq
Assuming a homogeneous medium of density $\bar\rho$, with $f$ being a Maxwellian distribution, and using $\vec p=\mu\vec v$, we get
\beq
P_\psi(p)=\frac{(2\pi)^{3/2}\bar\rho}{\mu^4\sigma^3}e^{-p^2/(2\mu^2\sigma^2)}\,.
\eeq

From here, it is straightforward to compute the density correlators we need, by repeatedly expressing $\rho_\Psi$ in terms of $\psi$ and using \eqref{eqn:psi-correlator}. First, we notice that any cross-term like $\braket{\rho\ped{slow}\rho\ped{fast}}$ contains an unequal number of $\psi$ and $\psi^*$, hence it vanishes. We can then evaluate the slow and fast pieces separately, $P_\rho=P_\rho\ap{slow}+P_\rho\ap{fast}$. By only retaining the leading term $\rho\ped{slow}\approx\mu\abs{\psi}^2$, we find
\beq
\tilde\rho\ped{slow}(\vec p,t)=\mu\int\frac{\dd^3q}{(2\pi)^3}\tilde\psi(\vec q,t)\tilde\psi^*(\vec q-\vec p,t)\,,
\eeq
and from there
\beq
\begin{split}
&\braket{\tilde\rho\ped{slow}(\vec p,t)\tilde\rho\ped{slow}^*(\vec p',t+\tau)}\\
&=(2\pi)^6\delta^{(3)}(\vec p)\delta^{(3)}(\vec p')\bar\rho^2+(2\pi)^3\delta^{(3)}(\vec p-\vec p')P_\rho\ap{slow}(p,\tau)\,.
\end{split}
\eeq
The first term just signals the presence of a nonzero average density $\bar\rho$. But we are only interested in the stochastic part of $\rho_\Psi$, so that term cancels out when, e.g., considering the correlator of $\rho\ped{slow}-\bar\rho$. The second term contains instead the power spectrum of the slow density fluctuations,
\beq
\begin{split}
P_\rho\ap{slow}(p,\tau)&=\mu^2\int\frac{\dd^3q}{(2\pi)^3}P_\psi(q)P_\psi(\abs{\vec q-\vec p})e^{i(\omega_q-\omega_{\abs{\vec q-\vec p}})\tau}\\
&=\frac{\pi^{3/2}\bar\rho^2}{\mu^3\sigma^3}\exp\bigg[-\frac{p^2}{4\mu^2\sigma^2}\bigg(1+\frac{\tau^2}{\tau\ped{c}^2}\bigg)\bigg]\,,
\end{split}
\eeq
which agrees with the result derived in \cite{Bar-Or:2018pxz}. The procedure for the fast oscillatory terms is similar, but slightly more technically involved. The result is
\beq
\begin{split}
&P_\rho\ap{fast}(p,\tau)\\
&=\Re\bigg[\frac{p^4e^{2i\mu c^2\tau}}{16\mu^2c^4}\!\int\!\frac{\dd^3q}{(2\pi)^3}P_\psi(q)P_\psi(\abs{\vec q-\vec p})e^{i(\omega_q+\omega_{\abs{\vec q-\vec p}})\tau}\bigg]\\
&=\frac{\pi^{3/2}\bar\rho^2p^4}{16\mu^7\sigma^3c^4}\Re\bigg[\frac{e^{2i\mu c^2\tau}}{(1-i\tau/\tau\ped{c})^{3/2}}\,e^{-\frac{p^2(1-i\tau/\tau\ped{c})}{4\mu^2\sigma^2}}\bigg]\,.
\end{split}
\label{eqn:Prhofast-uldm}
\eeq

We are now finally ready to calculate the diffusion time $T\sped{d}$ due to the ULDM tidal perturbations. Note that the power spectrum $P_\rho$ only enters \eqref{eqn:Td-Prho} through a time integral over the entire real line. The fast piece given in \eqref{eqn:Prhofast-uldm} has singularities only on the negative imaginary axis (at $\tau=-i\tau\ped{c}$), while the factor $e^{2i\mu c^2\tau}$ decays for $\Im\tau>0$. By closing the integration contour in the upper complex plane, we conclude that the fast density oscillations of ULDM do not contribute to the stochastic tidal perturbations of a binary.\footnote{Here we have only considered the Newtonian tidal field, $T_{ij}=\partial_i\partial_j\Phi$. More generally, the tidal tensor perturbing the binary is the Fermi-frame Riemann tensor $R_{0i0j}=T_{ij}+\ddot\Psi\delta_{ij}$, where $-2\Psi\delta_{ij}$ is the perturbation to the spatial part of the metric in conformal Newtonian gauge. For the slow density fluctuations, this term is suppressed by $\sigma^2/c^2$, while for the fast fluctuations it is unsuppressed. The fast perturbation is however narrow-band, at frequency $2\mu$, which can only lead to secular effects near resonance~\cite{Blas:2019hxz,Foster:2025nzf,Foster:2025csl}. For wide stellar binaries, the fundamental resonance is at $\mu=\pi/T\approx6.6\times\SI{e-29}{eV}\,(T/\si{ Myr})^{-1}$, far below standard cosmological lower bounds on the ULDM mass.} On the other hand, the slow piece does contribute, giving the final result
\beq
\begin{split}
T\sped{d}={}&\frac{m\mu\sigma^2}{32\pi^2G\bar\rho^2a^3\hbar}\\
={}&\SI{3.8}{Gyr}\,\bigg(\frac{m}{M_\odot}\bigg)\bigg(\frac{\mu}{\SI{e-21}{eV}}\bigg)\bigg(\frac{\sigma}{\SI{100}{km/s}}\bigg)^2\\
&\times\bigg(\frac{M_\odot/\si{pc}^3}{\bar\rho}\bigg)^2\bigg(\frac{\SI{0.1}{pc}}{a}\bigg)^3\,,
\end{split}
\label{eqn:Td-uldm}
\eeq
where we reinserted a factor of $\hbar$ for clarity. Note that, apart from numerical prefactors, this formula for $T\sped{d}$ coincides with that for a bath of large perturbers, given in Eq.~\eqref{eqn:Td-tidal-R}, with mass $m_*=\bar\rho\lambda\ped{dB}^3$, size $R=\lambda\ped{dB}$ and number density $n=\bar\rho/m_*$.

This diffusion time induced by ULDM tidal perturbations is typically long. In \eqref{eqn:Td-uldm} we have chosen ``optimistic'' reference values of the parameters (especially, a high dark matter density $\bar\rho$), but, in most of the plausible parameter space, $T\sped{d}$ is much longer than a Hubble time. On the upside, \eqref{eqn:Td-uldm} clearly delineates which astrophysical systems are most sensitive to the effect: they are low-mass wide binaries in ultra-faint dwarf galaxies, which typically have the lowest known values of $\sigma^2/\bar\rho^2$. Such systems have recently been detected and characterized for the first time, and used to place constraints on primordial black holes \cite{Shariat:2025dxs}. We can thus anticipate that a population of such binaries can also become a new probe of ULDM. 

\subsection{Dark Matter Substructures}
\label{sec:dm-substructures}

Another interesting application of our formalism is the impact of dark matter substructures on a population of Galactic binaries. Such substructures may range widely in size and mass, e.g., from subhaloes to massive compact halo objects (MACHOs) to primordial black holes.

This idea was proposed and developed in \cite{Ramirez:2022mys}, which used Gaia wide binaries \cite{El_Badry_2021,Gaia:2021gsq,Gaia:2016zol} to constrain size, masses and density of such substructures. To model the evolution of the ensemble of binaries, Ref.~\cite{Ramirez:2022mys} employed Monte Carlo simulations of impulsive three-body encounters. Within the regime of validity of our formalism, we can reproduce those findings analytically, providing a clean parametric scaling of the results, as well as trading their Monte Carlo sampling of random impulsive kicks for the solution of a Fokker--Planck equation.\footnote{To avoid confusion, we emphasize that the Monte Carlo method employed in Ref. \cite{Ramirez:2022mys} is used purely to randomly sample individual velocity kicks \eqref{eqn:Delta-v}; it is \emph{not} a Monte Carlo approach to solve a Fokker--Planck equation, as was employed in, e.g., Ref. \cite{2024MNRAS.532.2425H}.}

For instance, the observation from \cite{Ramirez:2022mys} that wider binaries are more easily diffused toward larger semi-major axes is immediately understood as an inverse power-law dependence of $T\sped{d}$ on $a$. More specifically, in the tidal limit ($R>a$), which is most appropriate in the case of subhaloes, from \eqref{eqn:Td-large} we have
\beq
\begin{split}
T\sped{d}={}&\SI{0.87}{Gyr}\,\bigg(\frac{\sigma}{\SI{100}{km/s}}\bigg)\bigg(\frac{\bar\rho\sped{dm}}{\bar\rho}\bigg)\\
&\times\bigg(\frac{R}{\SI{0.1}{pc}}\bigg)^2\bigg(\frac{\SI{0.1}{pc}}{a}\bigg)^3\bigg(\frac{m}{M_\odot}\bigg)\bigg(\frac{10^3M_\odot}{m_*}\bigg)
\end{split}
\label{eqn:Td-large-substructures}
\eeq
where $\bar\rho=m_*n$ is the mass density of the perturbers and $\bar\rho\sped{dm}=0.026M_\odot/\si{pc}^3$ is the dark matter density in the solar neighborhood. For small perturbers ($R\ll a$), such as MACHOs or primordial black holes, we have instead
\beq
\begin{split}
T\sped{d}={}&\SI{3.5}{Gyr}\,\bigg(\frac{\sigma}{\SI{100}{km/s}}\bigg)\bigg(\frac{\bar\rho\sped{dm}}{\bar\rho}\bigg)\\
&\times\bigg(\frac{\SI{0.1}{pc}}{a}\bigg)\bigg(\frac{m}{M_\odot}\bigg)\bigg(\frac{M_\odot}{m_*}\bigg)\bigg(\frac{25}{\log\Lambda}\bigg)
\end{split}
\label{eqn:Td-small-substructures}
\eeq
where we neglected the eccentricity-dependent factors in \eqref{eqn:Td*}. We see immediately that, other parameters being equal, small perturbers lead to a faster diffusion, and can therefore be more powerfully constrained through statistics of orbital parameters of wide binaries. This indeed qualitatively matches the constraints as a function of $R$ from \cite{Ramirez:2022mys}.

\begin{figure*}
    \centering
    \includegraphics[width=\textwidth]{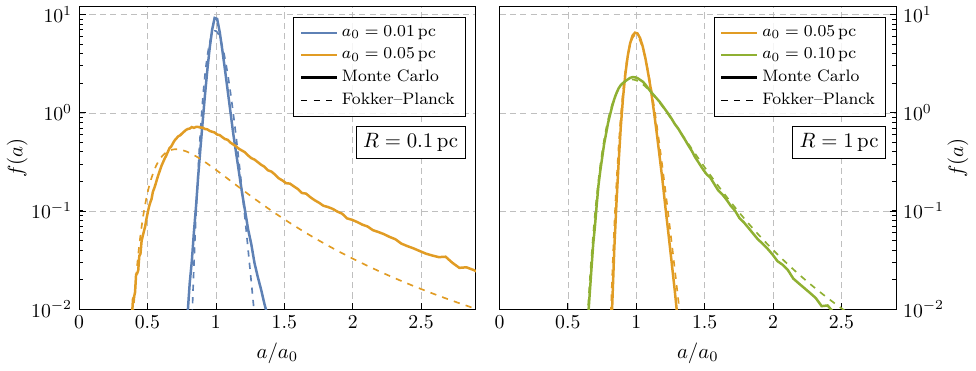}
    \caption{Marginalized distribution of the semi-major axis of an ensemble of binaries undergoing impulsive three-body encounters, after $\SI{10}{Gyr}$ of evolution. We compare the brute-force Monte Carlo employed in \cite{Ramirez:2022mys} to our analytical Fokker--Planck approach. To allow for a better comparison with \cite{Ramirez:2022mys}, we match their choices of velocity dispersion $\sigma=\SI{200}{km/s}$, mean density $\bar\rho=\num{0.0104}\,M_\odot/\si{pc^3}$, and spherical constant-density perturbers with mass $m_*=10^3M_\odot$. The perturber size is set to $R=\SI{0.1}{pc}$ in the left panel, and to $R=\SI{1}{pc}$ in the right panel. All binaries have the same initial eccentricity $e_0=0.5$ and semi-major axis $a_0$, and we show results for different values of $a_0$. Monte Carlo and Fokker--Planck results diverge for $a_0=\SI{0.05}{pc}$ because the Poisson noise cannot be ignored, as per \eqref{eqn:poisson-large}.}
    \label{fig:fp-vs-mc}
\end{figure*}

We can make the comparison with \cite{Ramirez:2022mys} more precise by reproducing their Monte Carlo simulation. In Fig.~\ref{fig:fp-vs-mc} we show the marginalized probability distribution function $f(a)$ obtained through a Monte Carlo approach and through our Fokker--Planck equation. We consider spherical constant-density perturbers and scan a similar parameter range as in \cite{Ramirez:2022mys}, with $m_*=10^3M_\odot$, perturber radii from $R=\SI{0.1}{pc}$ to $R=\SI{1}{pc}$, and initial semi-major axes ranging from $a_0=\SI{0.01}{pc}$ to $a_0=\SI{0.1}{pc}$. The two approaches agree remarkably well.

The only instance where we observe a meaningful difference is the case $a_0=\SI{0.05}{pc}$ and $R=\SI{0.1}{pc}$, shown in the left panel of Fig.~\ref{fig:fp-vs-mc}. This is due to Poisson noise being significant: as quantified in \eqref{eqn:poisson-large}, we expect the Fokker--Planck description to be accurate only for $a\ll \SI{0.02}{pc}$, which in this case turns out to be comparable to the initial semi-major axis of the binaries. The low number of three-body encounters experienced by the binary during its evolution makes the Kramers--Moyal truncation inaccurate.

Within its regime of validity, we consider the Fokker--Planck approach to be a significant improvement, due to its lower computational cost and, especially, the analytical insights that come with it. Comparing the Gaia wide binaries to the Fokker--Planck solutions, it is possible to reproduce the bounds on dark matter substructures placed in \cite{Ramirez:2022mys}, and easily extend them to different perturber profiles, velocity distributions, galactic environments (such as dwarf galaxies \cite{Shariat:2025dxs}), etc. We plan to accomplish this task in a future work.

\subsection{Interstellar medium}
\label{sec:ism}

Another source of stochastic perturbations for Galactic binaries is the interstellar medium (ISM).  It exhibits density fluctuations that vary over many different timescales, and over spatial scales from several kpc down to a few pc \cite{Girichidis_2016,Kim_2017,Green_2019,2023ApJ...946....3K,2026ApJ...999..136M}. To model their effect on binaries, we rely on the statistical characterization extracted by Ref.~\cite{2026ApJ...999..136M} from the state-of-the-art TIGRESS-NCR simulations of the ISM \cite{2023ApJ...946....3K}. Each of these simulations focuses on a periodic shearing box whose center rotates around the Galaxy at the local circular frequency. The physical parameters chosen for the simulations are supposed to mimic two distinct Galactic environments: one,  named R8, is calibrated on the solar neighborhood today, and the other, named LGR4, represents a more strongly turbulent, higher-surface density environment, as may have existed in the inner Milky Way earlier in its lifetime. 

Reference~\cite{2026ApJ...999..136M} showed that, in both R8 and LGR4, the power spectra of vertically-integrated surface density fluctuations exhibit a spatial power-law scaling, from several hundred-pc scales down to the grid scale of the simulation (a few pc). Here we will naively extrapolate the same power laws all the way down to the smallest scales in our problem, although as we will see these smallest scales do not significantly contribute to the perturbation. Reference~\cite{2026ApJ...999..136M} also measured temporal spectra (in the frame of the box) at each spatial wavenumber $p$. These are approximately flat (i.e., white noise) at frequencies $\vert \omega\vert < \omega_0(p)$, and cut off sharply for $\vert \omega \vert > \omega_0(p)$, where $\omega_0^2(p)=\tau_0^{-2}+v\ped{eff}^2p^2$, with $\tau_0$ the characteristic lifetime of the largest gas structures and $v\ped{eff}$ the characteristic velocity of the turbulent gas. These findings are packaged into a full spatio-temporal power spectrum for surface density fluctuations in Eq.~(32) of Ref.~\cite{2026ApJ...999..136M}. When we Fourier transform their result to the time domain, we obtain
\beq
P\sap{2d}_\Sigma(p,\tau)=\bar\Sigma^2L^2P_{\delta,0}(pL)^{-n}(1+\omega_0\abs{\tau})e^{-\omega_0\abs{\tau}}\,,
\eeq
where $\bar\Sigma$ is the average surface density, $L$ is the size of the simulation box, $P_{\delta,0}$ is a dimensionless amplitude, and $n$ is the power-law index.

The only length scale in this power spectrum is $\ell\ped{c} = v_\mathrm{eff}\tau_0$, with the power spectrum being exponentially suppressed for $p\gg\ell\ped{c}^{-1}$, at any $\tau>0$. Using characteristic values from \cite{2026ApJ...999..136M} we find that $\ell_\mathrm{c}\sim\SI{60}{pc}$ for both R8 and LGR4, much larger than any binary we are considering. ISM perturbations are thus in the tidal regime. The corresponding coherence time at $p\sim\ell\ped{c}^{-1}$ is $\tau_\mathrm{c}\sim \tau_0 \sim\SI{5}{Myr}$, which is comparable to or larger than the typical binary period (recall $T=\SI{3}{Myr}$ for a solar-mass binary with a semi-major axis of $\SI{0.1}{pc}$). For $a\lesssim\SI{0.1}{pc}$, we thus expect ISM-binary interactions to lie in the adiabatic tidal regime, which we studied in Sec.~\ref{sec:adiabatic} and whose Fokker--Planck coefficients are given in \eqref{eqn:Ba-adiabatic}--\eqref{eqn:Dee-adiabatic} and \eqref{eqn:Dhatehate-adiabatic}--\eqref{eqn:DM0M0-adiabatic}. However, the timescales $\tau_\mathrm{c}$ and $T$ are not well-separated for the widest binaries, so non-adiabatic effects may well be important.

Of course, the ISM fluctuations measured in Ref.~\cite{2026ApJ...999..136M} are not homogeneous and isotropic, due to both the Galactic shear and the flattened geometry of the disk. However, since $\ell_\mathrm{c}$ is  much smaller than the gas scale height (a few hundred $\si{pc}$), and shear can be largely ignored on those scales~\cite{2026ApJ...999..136M}, we expect that the perturbations will be nearly isotropic as measured locally in the binary's center-of-mass frame. 
Given this, the only way the binary can experience inhomogeneity is if its center of mass oscillates in and out of the gas layer, which would add a slow oscillatory time dependence to the diffusion coefficients.   Within the gas layer, we can model the ISM as a slab of thickness $D$, where $D=\SI{354}{pc}$ for the R8 model and $D=\SI{363}{pc}$ for the LGR4 model is chosen to give 
the correct midplane density $\bar\Sigma/D$ in each case. Binaries that live within the gas layer then experience a 3D ISM power spectrum which is approximately isotropic and reads
\beq
P_\rho(p,\tau)=\frac{P\sap{2d}_\Sigma(p,\tau)}{D}\,.
\label{eqn:ism-ps}
\eeq
This is the result we will employ in the rest of this section. For binaries that \textit{do} oscillate in and out of the gas layer, our results should still be informative, but the power spectrum above ought to be multiplied by some $\mathcal{O}(1)$ reduction factor (see also Sec.~\ref{sec:homogeneity-isotropy-stationarity}).

Plugging \eqref{eqn:ism-ps} in \eqref{eqn:Td-Prho}, which is valid for generic weak tidal perturbations (both adiabatic and white-noise), we get 
\beq
T\sped{d}=\frac{2\sqrt\pi}{\Gamma(\frac{3-n}2)\Gamma(\frac{n-2}2)}\,\frac{mDv\ped{eff}^{3-n}\tau_0^{2-n}}{32Ga^3\bar\Sigma^2L^{2-n}P_{\delta,0}}\,.
\eeq
Note that the normalization of the power spectrum in Ref.~\cite{2026ApJ...999..136M} is chosen such that $L^{2-n}P_{\delta,0}$ is a constant, so the diffusion time $T\sped{d}$ does not actually depend on the size of the simulation box $L$. This confirms that it is fluctuations on the scale $\sim \ell_\mathrm{c}$ and not those on the larger scales (where most of the power is concentrated), that have the greatest effect on tidally-perturbed binary dynamics.

For the solar neighborhood model R8, for which $\bar\Sigma=\num{12}\,M_\odot/\si{pc^2}$, $L=\SI{1024}{pc}$, $P_{\delta,0}=2.4$, $n=2.3$, $\tau_0=\SI{5}{Myr}$, and $v\ped{eff}=\SI{12}{km/s}$, we obtain
\beq
T\sped{d}=\SI{45}{Gyr}\,\bigg(\frac{m}{M_\odot}\bigg)\bigg(\frac{\SI{0.1}{pc}}{a}\bigg)^3\,.
\label{eqn:Td-ism}
\eeq
This estimate is remarkably close to that derived under the traditional assumption that the ISM consists of discrete clouds and that they interact impulsively with binaries (see, e.g., \cite{1983MNRAS.204..603B,1985AJ.....90.1548H}, and Eq.~(17) of \cite{2024MNRAS.532.2425H}). In retrospect this is not surprising, because we know that the white-noise tidal limit and the impulsive, large-perturber limit are equivalent (Sec.~\ref{sec:large-perturbers}), and that equation \eqref{eqn:Td-Prho} works for both white-noise and adiabatic interactions. However, there is a key physical difference, which is that adiabatic perturbations do not change binary semi-major axes, as per \eqref{eqn:Ba-adiabatic}--\eqref{eqn:Dee-adiabatic}. Thus, while Eq.~\eqref{eqn:Td-ism} gives the characteristic diffusion time for the eccentricity and the orbital orientation, if the interactions are adiabatic then the semi-major axis would not evolve, opposite to the traditional calculation. Since binaries with $a\sim\SI{0.1}{pc}$ sit somewhere between the adiabatic and white-noise limits, it is not obvious whether the ISM will truly act to unbind them or just change their eccentricities and orientations. This could be tested numerically by subjecting binaries to the TIGRESS potential fluctuations~\cite{modak2026interstellarmediumdrivenorbitaltransport}. 

In any case, Eq.~\eqref{eqn:Td-ism} confirms the traditional belief that, for the majority of wide binaries in the solar neighborhood, the ISM is likely not an important source of dynamical perturbations.  On the other hand, for the more strongly turbulent model LGR4, where $\bar\Sigma=\num{50}\,M_\odot/\si{pc^2}$, $L=\SI{512}{pc}$, $P_{\delta,0}=1.7$, $n=2.2$, $\tau_0=\SI{5}{Myr}$, and $v\ped{eff}=\SI{10}{km/s}$, we have
\beq
T\sped{d}=\SI{1.6}{Gyr}\,\bigg(\frac{m}{M_\odot}\bigg)\bigg(\frac{\SI{0.1}{pc}}{a}\bigg)^3\,,
\eeq
suggesting that binaries may have been significantly affected by the ISM at earlier epochs when the disk's gas density was higher. We emphasize again that, for $a\lesssim\SI{0.1}{pc}$, these interactions were predominantly adiabatic, meaning the diffusion time for semi-major axis would be much longer.

\section{Discussion}
\label{sec:discussion}

We now discuss some limitations and possible extensions of our work, as well as its relation to work by others.
First in Sec.~\ref{sec:homogeneity-isotropy-stationarity} we discuss the extension of our results to inhomogeneous, anisotropic and nonstationary perturbation fields. Then in Sec.~\ref{sec:deterministic} we estimate the role of deterministic perturbations compared to the stochastic ones we have focused on here. In Sec.~\ref{sec:disruption} we comment on the applicability of the formalism to binary disruption. Finally, in Sec.~\ref{sec:literature} we give a detailed description of how our work fits into---and to some extent, unifies---the existing literature on binary dynamics.

\subsection{Homogeneity, isotropy, and stationarity}
\label{sec:homogeneity-isotropy-stationarity}

We have so far assumed that the properties of the environment, such as its density and coherence time, remain constant throughout the binary ensemble evolution. This hypothesis can be violated due to a combination of the binary's motion and the inhomogeneity of the environment. For example, if the perturbing environment is given by the Galactic dark matter halo, then its density depends on the distance from Galactic center. If a binary's orbit within the Galaxy features significant radial oscillations, then the average density and velocity dispersion of the environment seen by the binary changes over time. As a consequence, the diffusion time $T\sped{d}$ also varies.

This complication would be relatively straightforward to implement if one was interested in the evolution of the orbital elements of a specific binary whose Galactic orbit is known: simply update $T\sped{d}$ over time, according to the binary's motion. But such a correction cannot be easily applied when dealing with a large ensemble of binaries, each with its own different motion in the Galaxy. This makes it more difficult to reliably use our framework to study real datasets of binaries.

There are two ways out. The simplest fix is to apply filters to the ensemble of binaries, only retaining those whose orbit does not explore wildly different regions of the Galaxy. The more sophisticated correction is to perform a simulation where each binary is initialized on a certain orbit, which dictates the Fokker--Planck evolution of the orbital parameters of that binary. Averaging over the simulated binaries returns the evolution of the whole ensemble. This is closely analogous to the method of orbit-averaged Fokker--Planck codes that are supposed to predict the evolution of star clusters \cite{1978ApJ...226.1087C}.

The binary's motion also introduces another subtle correction to our treatment: the environment no longer appears isotropic. For instance, the perturber encounter rate given in \eqref{eqn:dGamma} becomes \emph{directional}, requiring the velocity distribution to be boosted according to the binary's motion. Isotropy can also be broken when the perturbing field itself has a preferred direction. This effect has been included in, e.g., numerical calculations of binary evolution in globular clusters when the binary is subjected to encounters from passing stars \cite{Rasskazov:2023tal}. On the other hand, the role of this directionality in determining properties of perturbed stellar binaries in the Galaxy has, to our knowledge, never been tested explicitly.

Finally, in the case of perturbations from a stationary, stochastic tidal field, we note that the \emph{effective} coherence time as seen by the binary is reduced by the binary's own motion. Suppose for example that the tidal field correlator takes the form
\beq
\braket{T_{ij}(\vec x,t)T_{kl}(\vec x+\vec r,t+\tau)}\propto\exp\bigg(-\frac{r^2}{2\ell\ped{c}^2}-\frac{\tau^2}{2\tau\ped{c}^2}\bigg)\,,
\eeq
where $\ell\ped{c}$ and $\tau\ped{c}=\ell\ped{c}/\sigma$ are the coherence length and coherence time of the field. If the binary moves with velocity $\vec v$, we can substitute $\vec r=\vec v\tau$ and obtain the effective coherence time
\beq
\tau\ped{c}\ap{eff}\equiv\frac{\tau\ped{c}}{\sqrt{1+v^2/\sigma^2}}\,.
\label{eqn:tauc-eff}
\eeq
If the binary and the field share the same one-dimensional velocity dispersion $v\sim\sigma$, then the typical correction given by \eqref{eqn:tauc-eff} is a reduction of $\tau\ped{c}$ by a factor of about $\sqrt2$. One astrophysical example where such corrections might be important is for binaries on Galactic halo-like orbits \cite{Hwang2022b,GaiaBH3}, as they pass through the Galactic disk.

\subsection{Competition with deterministic perturbations}
\label{sec:deterministic}

In Sec.~\ref{sec:slow-evolution} we quantified the regime where the assumption $T\ll T\sped{d}$ holds. There is another inequality underlying the treatment of Sec.~\ref{sec:formalism}, that is, $\tau\ped{c}\ll T\sped{d}$. Perturbations with coherence time longer than the diffusion time appear as constant as far as the binary's evolution is concerned. They thus fall in the same category as deterministic perturbations, which we have intentionally left out of our treatment so far. An example of such a perturbation is the Galactic tide. Another is the perturbation due to a single distant flyby encounter on a known orbit, relevant for instance for understanding the eccentricity excitation in millisecond pulsar binaries \cite{Rasio:1995kf,Heggie:1995yk,Hamers_2019,Hamers:2019pbp}.

In principle, deterministic perturbations can be simply included in our Fokker--Planck equation \eqref{eqn:fp} as an additive contribution to the drift coefficient. However, we caution that linear superposition in the drift vector may fail to capture highly non-linear synergies between deterministic and stochastic forcing. For example, Galactic tides can drive von Zeipel--Lidov--Kozai eccentricity oscillations, which violently change a wide binary's vulnerability to stochastic impulsive kicks \cite{Kaib:2013naa}. Furthermore, different binaries follow different center-of-mass orbits, so in general they feel different deterministic perturbations. As already discussed in Sec.~\ref{sec:homogeneity-isotropy-stationarity}, the inclusion of such perturbations then requires one to track these orbits (or orbit-average them) alongside the Fokker--Planck evolution of their internal orbital elements.

Alternatively, we can restrict ourselves to cases where stochastic perturbations dominate over deterministic ones. First of all, we note that any perturbation with very long coherence time will leave the semi-major axis unaffected, meaning that diffusion in $a$ is always controlled by stochastic perturbations. As for the other orbital variables, we can make a comparison by considering, for concreteness, the tidal field of the Galactic disk $T_{ij}\sim4\pi G\rho_0$, where $\rho_0\sim0.1\,M_\odot/\si{pc^3}$ in the solar neighborhood, and the corresponding characteristic time $T\ped{gal}\sim 2\pi/\sqrt{4\pi G\rho_0}$. The perturbation makes the orbital elements evolve over a timescale
\beq
\begin{split}
T\ped{coh}&\sim\frac{T\ped{gal}^2}{T}\\
&=\SI{2.4}{Gyr}\,\bigg(\frac{0.1\,M_\odot/\si{pc^3}}{\rho_0}\bigg)\bigg(\frac{m}{M_\odot}\bigg)^{1/2}\bigg(\frac{\SI{0.1}{pc}}{a}\bigg)^{3/2}\,.
\end{split}
\label{eqn:Tcoh}
\eeq
This timescale should be compared with the diffusion time $T\sped{d}$, which we evaluated in Sec.~\ref{sec:applications} in concrete physical scenarios. Whenever $T\ped{coh}<T\sped{d}$, the evolution of orbital elements other than the semi-major axis is primarily deterministic, not stochastic.

It is worth noting that $T\ped{coh}$ scales as $a^{-3/2}$. This generally means that, in the case of large perturbers ($R>a$), stochastic perturbations dominate for wider binaries, as $T\sped{d}\propto a^{-3}$. The opposite holds for small perturbers ($R\ll a$), where $T\sped{d}\propto a^{-1}$.

\subsection{Binary disruption}
\label{sec:disruption}

Throughout this paper, we have focused on the evolution of the orbital elements of binaries that remain bound. In many applications, however, the primary interest is whether the binaries survive at all. Binary disruption (or ``ionization'') \cite{1985ApJ...290...15B, Jiang:2009ax,Yoo:2003fr, Quinn:2009zg, Penarrubia:2016ltr, 1975AJ.....80..809H, 2025ApJ...981..124Y} is in fact already incorporated in our formalism, through the choice of boundary conditions at $a\to\infty$, equivalently $\mathcal E\to0$, discussed in Secs.~\ref{sec:white-noise} and~\ref{sec:point-mass}. Because the probability current $J^a$ does not vanish at large $a$, we impose an absorbing boundary there, so that binaries crossing it are permanently removed from the ensemble: the outgoing probability flux is the disruption rate. The fraction of surviving binaries at time $t$ is then obtained by solving the Fokker--Planck equation \eqref{eqn:fp} and marginalizing $f(\vec w,t)$ over all orbital variables. The deficit $1-\int f\,\dd\vec w$ is the fraction of disrupted binaries. This is directly visible in Figs.~\ref{fig:f-white-noise} and~\ref{fig:f-point-mass}, where the semi-major axis marginal progressively loses area.

No new local transport coefficient is required to describe diffusive
disruption, as the same Fokker--Planck coefficients that govern the
evolution of the orbital elements also determine the flux
through the absorbing boundary. The corresponding survival time is therefore
parametrically controlled by $T\sped{d}$, although the precise
average disruption time depends on the initial distribution, the position of the
absorbing boundary, and the $a$- and $e$-dependence of the coefficients. It is also worth noting that, as binaries evolve towards disruption, they inevitably enter a regime where the shot-noise is no longer negligible, i.e., the inequalities in Sec.~\ref{sec:shot-noise} are violated. This introduces a small correction to the disruption time as determined from our Fokker--Planck equation~\cite{Penarrubia:2019wei}.

An important exception is the case of adiabatic tidal perturbations, for which
$B^a=D^{aa}=D^{ae}=0$; see Sec.~\ref{sec:adiabatic}. In this regime the semi-major axis, and hence the binary binding energy, does not evolve. The
absorbing boundary at $\mathcal E=0$, or equivalently $a\to\infty$, is therefore
never reached within the idealized Keplerian problem and binaries are not ionized, even though the remaining orbital elements keep diffusing. We also stress that diffusive disruption should be distinguished from catastrophic ionization by a single strong encounter. The latter lies outside the Kramers--Moyal truncation used here unless the encounter distribution is restricted to weak kicks, as discussed in Sec.~\ref{sec:shot-noise}. 

Finally, we caution that the absorbing boundary treats disruption as irreversible. In physical environments, unbound stars can later ``recombine'' into new binaries \cite{Pe_arrubia_2020}, a process that is not captured by our formalism.

\subsection{Comparison with previous works}
\label{sec:literature}

The study of binary evolution in the presence of stochastic perturbations has a rich history in astrophysics. In this section, we situate our unified framework within the existing literature. In doing so, we demonstrate how our results reduce to known limits, correct existing discrepancies, and strictly generalize previously available models. However, the binary evolution literature as a whole is far too large to be reviewed thoroughly here, so we begin with a brief historical sketch and then focus on more recent papers that are directly related to our results.

Historically this subject traces its roots at least as far as Jeans \cite{1919MNRAS..79..408J} and Ambartsumian \cite{1937AZh....14..207A}, who demonstrated the importance of thermal eccentricity distribution $f(e) = 2e$.  Chandrasekhar \cite{1944ApJ....99...54C} subsequently studied the stability
and disruption of binaries under random perturbations from passing stars. 

In the 1970s, Heggie~\cite{Heggie:1975tg} and
Hills~\cite{1975AJ.....80..809H} established many of the basic principles of
binary--single scattering, including the distinction between hard and soft
binaries and the associated energy-exchange rules. This program was developed
further in the binary--single-star scattering series by Hut and collaborators,
who computed and fitted scattering cross sections for hard binaries and used
them to study the steady-state distribution of binary binding energies in
stellar backgrounds~\cite{1983ApJ...268..319H,1983ApJ...268..342H,Hut:1992cx,Heggie:1992cz,Goodman:1992cy,McMillan:1996bt,Heggie:1996bs}. These works are different from the weak-perturbation diffusion problem
studied here: they treat strong, often resonant encounters and binary creation
or destruction, whereas our Fokker--Planck treatment assumes small orbital
changes accumulated over many perturbations.

A related but distinct diffusion tradition focused on the cumulative evolution
and survival of wide or soft binaries. Retterer~\cite{1980AJ.....85..249R}
used a Fokker--Planck approach to model the continuous evolution of the binary
binding-energy distribution. Bahcall, Hut \& Tremaine \cite{1985ApJ...290...15B} applied similar ideas to understand the survival of wide binaries, carefully delineating where the smooth accumulation of weak encounters gives way to single catastrophic kicks, thereby constraining the maximum mass of unseen perturbers in the Galactic disk. Weinberg et al.~\cite{1987ApJ...312..367W} applied Fokker--Planck techniques to study wide binaries in the solar neighborhood perturbed by both passing stars and giant molecular clouds (GMCs). Treating the GMCs as extended spherical perturbers, they derived diffusion coefficients for the binary energy. Though they did not comment on it explicitly, their numerical results exhibit the $f \propto a^{-2}$ steady-state scaling that we found analytically in Sec.~\ref{sec:point-mass} for the marginalized semi-major axis distribution. 

For a textbook summary of these classic results, we refer the reader to~\cite{2008gady.book.....B}.

\subsubsection{Impulsive encounters}
\label{sec:lit-impulsive}

More recently, Hamilton \& Modak \cite{2024MNRAS.532.2425H} derived Fokker--Planck coefficients for binaries undergoing impulsive encounters, which can be directly compared with ours from Sec.~\ref{sec:impulsive}. In fact, our treatment is strictly more general than the one presented in \cite{2024MNRAS.532.2425H}, for four different reasons.
First, while we employ perturbers with generic spherical density profiles $\rho_*(x)$, Ref.~\cite{2024MNRAS.532.2425H} considered point-mass perturbers only, and must therefore be compared with the results in Sec.~\ref{sec:point-mass}.
Second, Ref.~\cite{2024MNRAS.532.2425H} simplified their calculation by assuming that each encounter leaves one star unaffected. As a consequence, their diffusion coefficients strictly diverge due to perturbations with large impact parameters (via the Coulomb logarithm), whereas ours do not (see the discussion after \eqref{eqn:Yperp+Yparallel}--\eqref{eqn:Yperp-Yparallel}). Third, Ref.~\cite{2024MNRAS.532.2425H} predicted that the steady-state marginal eccentricity distribution is the thermal distribution $f(e)=2e$. However, because we do not assume one star is unaffected in every encounter, we are able to resolve the finite eccentricity-dependent corrections to the drift and diffusion coefficients, and thereby derive a steady-state that is slightly \emph{sub}-thermal, see Eq.~\eqref{eqn:f-ss-point-mass}. This may help explain the fact that some simulations of binary evolution have produced  sub-thermal eccentricity distributions where thermal ones were expected \cite{Geller_2019,Pe_arrubia_2020}; however, it also deepens the mystery surrounding the observed \emph{super}-thermal wide binaries in the Galaxy \cite{Hwang_2022,2024MNRAS.532.2425H}.
Fourth, we studied all six orbital variables including the sector $(\hat e,\hat q,\hat J,M_0)$, while \cite{2024MNRAS.532.2425H} only considered $(a,e)$.
Despite these differences, we do recover the results of \cite{2024MNRAS.532.2425H} in the appropriate limit: specifically, as $\log\Lambda\to\infty$ the five coefficients \eqref{eqn:Ba-point-mass}--\eqref{eqn:Dee-point-mass} reduce to theirs exactly.

\subsubsection{Distant adiabatic encounters}
\label{sec:lit-adiabatic}

In the regime where a discrete point-mass perturber's velocity is slow compared to the binary's orbital speed, interactions are adiabatic rather than impulsive. Heggie \& Rasio \cite{Heggie:1995yk} and later Hamers \& Samsing \cite{Hamers_2019,Hamers:2019pbp} studied the deterministic variations in orbital elements induced by individual, distant, slow-moving perturbers. 

However, attempts to model the cumulative effect of these distant adiabatic encounters as a continuous stochastic process must be treated with extreme caution. The reason is that the strength of the perturbation scales very steeply with the encounter periapsis distance $r_\mathrm{p}$. The diffusion coefficients derived in this way then diverge as $r_\mathrm{p}\to 0$. Physically, this divergence is telling us that the binary's evolution is not driven by a smooth cumulative background, but by the single closest passing body. 
If one integrates over the full physical range of periapsis distances, the occasional deep encounter can produce a macroscopic jump in, e.g., binary angular momentum, $\Delta J \sim \mathcal{O}(J)$. The presence of such heavy-tailed jumps violates the core assumption of microscopic steps underlying the Kramers--Moyal expansion, rendering the standard Fokker--Planck equation physically invalid for this regime.

References~\cite{Hamers_2019,Hamers:2019pbp} derived a steady-state distribution for these encounters using a Fokker--Planck equation---and numerically verified it with Monte Carlo simulations. However, this agreement is likely an artifact of their adopted inner spatial cutoff, $r_{p,\min} = 2a$. By enforcing a hard inner boundary, they artificially truncate the jump distribution and cap the maximum step size to be strictly microscopic, $\Delta J \ll J$. Because the variance integral is entirely dominated by its lower limit, their simulated binary is effectively driven exclusively by a sequence of weak encounters skimming the $r_{p,\min}$ boundary. Accumulating many small, artificially capped kicks forces the system to satisfy the central limit theorem, restoring the mathematical validity of the Fokker--Planck equation.

This highlights a fundamental physical distinction regarding our adiabatic tidal regime (Sec.~\ref{sec:adiabatic}). Our continuous Fokker--Planck framework is mathematically rigorous precisely because it models perturbations from continuous stochastic fields (such as ULDM or the ISM density power spectra) without the need for artificial truncation. It usually cannot be applied to adiabatic scattering by discrete point masses, unless there is good reason to believe their periapsis distances are truncated at some distance larger than $a$.

\subsubsection{White-noise tidal perturbations}
\label{sec:lit-white-noise}

Peñarrubia also derived the Fokker--Planck coefficients for $\mathcal E$ and $\vec J$ in~\cite{Penarrubia:2019wei}. His ``impulsive regime'' corresponds to our white-noise tidal case, discussed in Sec.~\ref{sec:white-noise}, and it is therefore useful to compare the two calculations directly.

Our results differ from Peñarrubia's in several ways. The most fundamental discrepancy is that \cite{Penarrubia:2019wei} builds a model of the tidal field based on \cite{Penarrubia:2017nzw}, where the tensor structure of the tidal force is intentionally ignored.\footnote{This is presumably to simplify the algebra, as \cite{Penarrubia:2017nzw} writes \emph{``Unfortunately, the resulting field is not spherically isotropic, which greatly complicates the mathematical analysis [\ldots] Yet, at leading order one can isotropize the tidal field''}. However, the isotropic and non-isotropic parts of the tidal velocity kick correlator are of the same order, so one cannot ignore one compared to the other.} Specifically, the tidal acceleration $a_i=-T_{ij}r_j$ is replaced in \cite{Penarrubia:2017nzw} by $a_i=\Lambda_ir$, where $\Lambda_i$ is a fluctuating random vector with $\braket{\Lambda_i\Lambda_j}\propto\delta_{ij}$. This expression of the acceleration depends only on the magnitude, and not on the direction of the separation $r_i$. As a consequence, the ``velocity kicks'' $\Delta v_i=a_i\Delta t$ induced by such a force satisfy
\beq
\frac{\braket{\Delta v_i\Delta v_j}}{\Delta t}\bigg|_{\text{Peñ.}}\propto r^2\delta_{ij}\,.
\eeq
Using the full expression of the tidal acceleration, we get instead
\beq
\frac{\braket{\Delta v_i\Delta v_j}}{\Delta t}\bigg|_{\text{ours}}\propto\mathds{1}_{ikjl}r_kr_l=r^2\delta_{ij}+2r_ir_j\,.
\label{eqn:tensorial-structure-tidal}
\eeq

Interestingly, not all Fokker--Planck coefficients are affected by this difference. While changes in energy $\Delta\mathcal E=\vec v\cdot\Delta\vec v$ are genuinely sensitive to the full tensor structure, those in angular momentum $\Delta\vec J=\vec r\times\Delta\vec v$ project out components parallel to $\vec r$. It is then not a surprise that (after matching the normalization) our results for the energy coefficients differ from \cite{Penarrubia:2019wei},
\beq
\frac{B^{\mathcal E}\ped{ours}}{B^{\mathcal E}_{\text{Peñ.}}}=\frac53\,,\qquad\frac{D^{\mathcal E\mathcal E}\ped{ours}}{D^{\mathcal E\mathcal E}_{\text{Peñ.}}}=\frac{2+e^2}{2-e^2}\,,
\eeq
while $D^{JJ}\ped{ours}=D^{JJ}_{\text{Peñ.}}$ matches perfectly, which is a nontrivial check of \eqref{eqn:DJJ-white-noise}.

Based on this discussion, however, we should also expect our $B^J$ and $D^{\mathcal EJ}$ to match Peñarrubia's, while \cite{Penarrubia:2019wei} gives $B^J=0$ and $D^{\mathcal EJ}=0$. As it turns out, these are separate discrepancies, independent from the previous one. Specifically, Ref.~\cite{Penarrubia:2019wei} invokes the isotropy assumption $\overline{\braket{\Delta\vec J}}=0$ to argue that $B^J$ and $D^{\mathcal EJ}$ must vanish. While the full angular momentum vector has indeed no drift, the same is not true for its \emph{magnitude} $J$, which is a convex function of $\vec J$. In fact, as we show from first principles in Appendix~\ref{sec:first-principles}, geometric constraints dictate that one must have $B^J=D^{JJ}/(2J)$, analogous to the radial drift of a two-dimensional random walk.

The final way our treatment differs from \cite{Penarrubia:2019wei} is that we study the evolution of all six orbital elements, while \cite{Penarrubia:2019wei} focuses solely on the $(\mathcal E,J)$ sector.

\subsubsection{Stochastic gravitational-wave backgrounds}
\label{sec:lit-gw-background}

Blas \& Jenkins derived Fokker--Planck coefficients for a binary perturbed by a stochastic gravitational-wave (GW) background~\cite{Blas:2021mpc,Blas:2021mqw}. This can be mapped onto our results for the tidal regime, with two crucial modifications.

When the GW wavelength is much larger than the binary, the relative acceleration is $a_i=-\ddot h_{ij}r_j/2$, where $h_{ij}$ is the transverse-traceless metric perturbation, with correlator
\beq
\int_{-\infty}^\infty\braket{\ddot h_{ij}(t)\ddot h_{kl}(t+\tau)}\dd\tau\propto\mathds{1}_{ijkl}-\frac53\delta_{ij}\delta_{kl}\,.
\label{eqn:traceless}
\eeq
The subtraction of the trace originates from the absence of a local density perturbation carried by the GW. This results in the following structure of the velocity kick tensor,
\beq
\frac{\braket{\Delta v_i\Delta v_j}}{\Delta t}\bigg|\sped{gw}\propto r^2\delta_{ij}+\frac13r_ir_j\,,
\eeq
which differs from the one given in \eqref{eqn:tensorial-structure-tidal}. As argued in Sec.~\ref{sec:lit-white-noise}, this only affects \emph{some} of the Fokker--Planck coefficients. Repeating the procedure of Sec.~\ref{sec:tidal-force} with $\mathds{1}_{ijkl}$ replaced by the traceless part \eqref{eqn:traceless}, for adiabatic tidal perturbations we find the following updated values of the coefficients 
\begin{align}
\label{eqn:DhatJhatJ-traceless}
D^{\hat J\hat J}&=\frac{7(1-e^2)}{15T\sped{d}}\,,\\
D^{\hat J M_0}&=-\frac{\sqrt{1-e^2}(8+7e^2)}{15T\sped{d}}\,,\\
D^{M_0M_0}&=\frac{31+48e^2+21e^4}{45T\sped{d}}\,,
\end{align}
while for white-noise tidal perturbations we have
\begin{align}
B^a&=\frac{11a(2+e^2)}{45T\sped{d}}\,,\qquad B^{\mathcal E}=-\frac{\mathcal E(2+3e^2)}{9T\sped{d}}\,,\\
B^e&=\frac{64-213e^2-229e^4}{720eT\sped{d}}\,,\\
D^{aa}&=\frac{4a^2(3-e^2)}{45T\sped{d}}\,,\qquad D^{\mathcal E\mathcal E}=\frac{4\mathcal E^2(3-e^2)}{45T\sped{d}}\,,\\
D^{ae}&=-\frac{11ae(1-e^2)}{45T\sped{d}}\,,\\
D^{ee}&=\frac{64+61e^2-125e^4}{360T\sped{d}}\,,\\
D^{\hat J\hat J}&=\frac{64+15e^2-58e^4}{360e^2T\sped{d}}\,,\\
D^{\hat J M_0}&=-\frac{\sqrt{1-e^2}(64+175e^2+62e^4)}{360e^2T\sped{d}}\,,\\
\label{eqn:DM0M0-traceless}
D^{M_0M_0}&=\frac{64+399e^2+487e^4+58e^6}{360e^2T\sped{d}}\,.
\end{align}
The coefficients not listed here remain unchanged compared to those in Sec.~\ref{sec:tidal}.

The other difference is related to the gauge invariance of General Relativity. Rather than taking the adiabatic and white-noise limits, Refs.~\cite{Blas:2021mpc,Blas:2021mqw} write the result for a generic stochastic GW background intensity spectrum $\Omega\sped{gw}(f)$, which reads as a sum over the harmonics,
\begin{align}
B^\mu&=\sum_{n=1}^\infty B^\mu_n\,\frac{n}{T}\,\Omega\sped{gw}(n/T)\,,\\
D^{\mu\nu}&=\sum_{n=1}^\infty D^{\mu\nu}_n\,\frac{n}{T}\,\Omega\sped{gw}(n/T)\,.
\end{align}
Crucially, the $n=0$ term is absent from the sum. That is because a zero-frequency GW produces a vanishing tidal field, as $\ddot h_{ij}\propto f^2$---in other words, it is pure gauge.

As a consequence, the adiabatic (i.e., low-frequency) limit of tidal perturbations studied in Sec.~\ref{sec:adiabatic} has no GW counterpart: the tidal field is vanishingly small. On the other hand, all other harmonics do contribute. For a white-noise GW tidal field, $\Omega\sped{gw}\propto f^{-1}$, we should in fact expect to recover our white-noise tidal coefficients, \emph{minus} the low-frequency $n=0$ contribution. After matching the normalizations and converting to the same set of variables, \emph{all} Fokker--Planck coefficients (including the ones for the orbital orientation) from \cite{Blas:2021mpc} should be related to ours by
\begin{align}
\label{eqn:B-check-gw}
B^\mu\sped{gw}(\Omega\sped{gw}\propto f^{-1})&=B^\mu_{\text{white-noise}}-B^\mu\ped{adiabatic}\,,\\
\label{eqn:D-check-gw}
D^{\mu\nu}\sped{gw}(\Omega\sped{gw}\propto f^{-1})&=D^{\mu\nu}_{\text{white-noise}}-D^{\mu\nu}\ped{adiabatic}\,,
\end{align}
where the coefficients on the right-hand side are \eqref{eqn:DhatJhatJ-traceless}--\eqref{eqn:DM0M0-traceless}, rather than the ones listed in Sec.~\ref{sec:tidal}. By evaluating numerically the expressions that \cite{Blas:2021mpc} gives for $D^{\mu\nu}$ in terms of the Hansen coefficients, we have verified that \eqref{eqn:D-check-gw} indeed holds exactly for any eccentricity. On the other hand, \eqref{eqn:B-check-gw} does not. We suspect this is an error in Eq.~(4.1) of \cite{Blas:2021mpc}, because, when combined with their Eq.~(4.2), it gives $B^{\mathcal E}<0$, implying that the stochastic GW background removes energy from the binary on average.  This seems to be erroneous because $B^{\mathcal E}=\tr[\Delta v_i\Delta v_j]/(2\Delta t)>0$.\footnote{We have also confirmed with a direct Monte Carlo simulation, similar to the one in Appendix~\ref{sec:numerical-check}, that the difference $B^\mu_{\text{white-noise}}-B^\mu\ped{adiabatic}$ does reproduce the simulated drift within numerical errors.}

In our language, the diffusion time induced by a stochastic GW background with $\Omega\sped{gw}\propto f^{-1}$ (normalized to the cosmological critical density) is
\beq
T\sped{d}=\frac{Gm}{9\pi^2a^3H_0^2\,f\,\Omega\sped{gw}(f)}\,,
\eeq
where $H_0$ is the Hubble constant.

\subsubsection{Galactic nuclei}
\label{sec:lit-galactic-nuclei}

The Fokker--Planck approach also has long been the standard tool for studying loss-cone
dynamics and two-body relaxation around supermassive black holes
\cite{Magorrian:1999vm,2008gady.book.....B,2013degn.book.....M,Kaur:2024ofj}. Any Keplerian
orbit around a supermassive black hole can be thought of as a ``binary'', so it is
natural that our results resemble these works. There are, however, two distinct
classes of effects in such environments that fall outside the scope of the
equations derived here, which we now delineate.

The first concerns the mutual interaction of orbits. In our treatment, each
binary interacts with its environment but \emph{not} with other binaries. In a
galactic nucleus, by contrast, the different orbits can torque one another
coherently over long timescales---in other words, the ``binaries'' can interact with
each other. This gives rise to \emph{resonant relaxation}
\cite{Rauch:1996fb,Hopman:2006qr,Kocsis_2015,Bar_Or_2018},
which possesses its own nontrivial statistical mechanics
\cite{Roupas:2019fll,Flores:2024njd}. In principle, this mutual coupling
could still be recast in a Fokker--Planck form, but the drift and diffusion
coefficients would then depend self-consistently on the distribution function
$f$ itself \citep{Bar_Or_2018}, rendering the equation nonlinear.

The second concerns the breakdown of the diffusive picture itself. Our framework
relies on the accumulation of many weak, rapidly decorrelating perturbations
($\tau\ped{c}\ll T\sped{d}$, together with the shot-noise condition of
Sec.~\ref{sec:shot-noise}), which guarantees ordinary Gaussian diffusion. In
sufficiently dense systems, however, the evolution can instead be dominated by
rare but violent events. This occurs, for instance, in strong, comparable-mass
three-body encounters \cite{Quinlan:1996vp,Sesana:2006xw}, in massive-perturber-driven evolution when rare large-angle encounters must be
separated from the diffusive component~\cite{Perets:2006bz,Perets:2007nc}, or when a binary
experiences only distant fly-bys so that its evolution is controlled by the
single closest passage (Sec.~\ref{sec:shot-noise}). In such cases the orbital
increments are heavy-tailed~\cite{Bhalla:2025xce}, the Gaussian assumption underlying the
Kramers--Moyal expansion fails, and the evolution is better described as a
jump process or anomalous (L\'evy-type) transport rather than as smooth diffusion~\cite{Collins:2008kf,Collins:2010pv,2013ApJ...764...52B}.

\section{Conclusions}
\label{sec:conclusions}

The statistical evolution of Keplerian binaries under stochastic perturbations has previously been studied in a number of special limits. In this work we have shown that many of these problems can be treated within a \emph{unified} Fokker--Planck framework. For weak perturbations, the binaries diffuse in phase space. Once a set of variables is chosen, such as the Keplerian elements $\vec w=(a,e,\Omega,i,\omega,M_0)$, the drift and diffusion coefficients follow directly from the stochastic force $\vec F\sped{s}=\dot{\vec w}$ and its time correlation function. The calculation is algorithmic and mechanical.

We have applied this framework to three physically distinct regimes: adiabatic tidal perturbations, white-noise tidal perturbations, and impulsive encounters with a third body of arbitrary density profile. In each case, we have provided the complete set of drift and diffusion coefficients, together with the characteristic diffusion timescale $T\sped{d}$. The paper can thus be used not only as a reference for the theoretical framework, but also as a dictionary of ready-to-evaluate Fokker--Planck coefficients: once the statistical properties of the perturbing environment are specified, the stochastic evolution of the binary follows directly.

Thanks to the systematic nature of the calculation, we have extended previous results in several directions. For instance, we have obtained the coefficients governing the orientation of the binary orbit, retained the complete tensor structure of tidal correlators, treated impulsive encounters with non-pointlike perturbers, and kept the full eccentricity dependence. As an example, we have derived finite-eccentricity corrections to the thermal eccentricity distribution previously obtained for impulsive encounters with pointlike perturbers.

The framework developed here therefore provides both a thorough mathematical description of stochastic Keplerian dynamics and a practical tool for applications to data. For instance, the orbital-element distribution of wide binaries observed by Gaia~\cite{El_Badry_2021,Gaia:2021gsq,Gaia:2016zol} can be used to constrain dark matter substructure, the distribution of compact objects, or the nature of stochastic fluctuations in the interstellar medium. More generally, the coefficients derived here can be incorporated as
subgrid prescriptions for cumulative weak perturbations into population-synthesis models~\cite{Hurley:2002rf, Breivik:2019lmt} or cluster-evolution calculations~\cite{Rodriguez:2021qhl, Giersz:2011em}, and can be used in forward-model or likelihood analyses of
binary catalogs~\cite{Pittordis:2022qrz, Ramirez:2022mys, Hwang_2022}. Natural extensions include combining the stochastic terms with deterministic secular effects from smooth Galactic tides, von Zeipel--Lidov--Kozai torques, or relativistic precession, and treating spatially varying environments in which the diffusion time changes along the binary's orbit through the Galaxy. In this sense, the formalism turns stochastic binary evolution into a calculable, data-facing problem: given an environment, it predicts the flow of binaries through orbital-element space; given a binary population, it provides a route to infer the properties of the environment.

\section*{Acknowledgements}

We thank Scott Tremaine and Shaunak Modak for helpful discussions. We also thank Alexander C. Jenkins, Hyungjin Kim, Jorge Peñarrubia, Hagai Perets, and Cheyanne Shariat for useful comments on the draft. A.C.\ is supported by an ERC STG grant (``AstroDarkLS'', grant No. 101117510). G.M.T.\ gratefully acknowledges support from the Rubicon Fellowship, awarded by the Netherlands Organisation for Scientific Research (NWO), Grant ID \href{https://doi.org/10.61686/WYKDB06497}{https://doi.org/10.61686/WYKDB06497}. C.H.\ was partly supported by the John N.\ Bahcall Fellowship Fund at the Institute for Advanced Study.

\appendix

\section{Kramers--Moyal expansion}
\label{sec:kramers-moyal}

In this appendix we review the Kramers--Moyal derivation of the Fokker--Planck equation, assuming knowledge of the averages $\braket{\Delta w^\mu}$ and $\braket{\Delta w^\mu\Delta w^\nu}$.

Define the probability density $f(\vec w,t)\equiv\braket{\delta(\vec w-\vec w(t))}$. After a time step $\Delta t$, the density at $\vec w$ receives contributions from all initial points $\vec w_0$ that evolve to $\vec w$ in time $\Delta t$,
\beq
f(\vec w,t+\Delta t)=\Braket{\delta\big(\vec w-\vec w(t)-\Delta\vec w\big)}\,,
\eeq
which can be expanded in a Taylor series in $\Delta\vec w$ as
\beq
\begin{split}
\delta\big(\vec w-\vec w(t)-\Delta\vec w\big)=\sum_{n=0}^\infty&\frac{(-1)^n}{n!}\Delta w^{\mu_1}\cdots\Delta w^{\mu_n}\\
&\times\partial_{\mu_1}\cdots\partial_{\mu_n}\delta(\vec w-\vec w(t))\,,
\end{split}
\label{eqn:taylor}
\eeq
where we used the notation $\partial_\mu\equiv\partial/\partial w^\mu$ for short. We now take the ensemble average. To evaluate the right-hand side, we write
\beq
\begin{split}
&\int\dd\vec w_0\,f(\vec w_0,t)\braket{\Delta w^{\mu_1}\cdots\Delta w^{\mu_n}\partial_{\mu_1}\cdots\partial_{\mu_n}\delta(\vec w-\vec w(t))}\\
&=\braket{\Delta w^{\mu_1}\cdots\Delta w^{\mu_n}\partial_{\mu_1}\cdots\partial_{\mu_n}\,\delta(\vec w-\vec w(t))}\,,\\
\end{split}
\eeq
where the ensemble average within the integral is conditioned on $\vec w(t)=\vec w_0$. Pulling $\partial_{\mu_1}\cdots\partial_{\mu_n}\delta(\vec w-\vec w(t))$ out of the average and integrating by parts, we can write \eqref{eqn:taylor} as
\beq
\begin{split}
&\frac{f(\vec w,t+\Delta t)-f(\vec w,t)}{\Delta t}\\
&=\sum_{n=1}^\infty\frac{(-1)^n}{n!}\,\partial_{\alpha_1}\cdots\partial_{\alpha_n}\bigg[\frac{\braket{\Delta w^{\alpha_1}\cdots\Delta w^{\alpha_n}}}{\Delta t}\,f(\vec w,t)\bigg]\,.
\end{split}
\label{eqn:kramers-moyal}
\eeq
This is the Kramers--Moyal expansion. Truncating at $n=2$ yields the Fokker--Planck equation
\beq
\frac{\partial f}{\partial t}=-\frac{\partial}{\partial w^\mu}\big(B^\mu f\big)+\frac12\frac{\partial^2}{\partial w^\mu\partial w^\nu}\big(D^{\mu\nu}f\big)\,,
\label{eqn:fp-general}
\eeq
with the drift and diffusion coefficients
\beq
B^\mu\equiv\frac{\braket{\Delta w^\mu}}{\Delta t}\,,\qquad D^{\mu\nu}\equiv\frac{\braket{\Delta w^\mu\Delta w^\nu}}{\Delta t}\,.
\label{eqn:drift-diffusion-km}
\eeq
The truncation is justified when $\Delta t$ is chosen much longer than the correlation time $\tau\ped{c}$ but much shorter than the diffusion timescale $T\sped{d}$, so that the increments $\Delta w^\alpha$ are small and their distribution is approximately Gaussian by the central limit theorem.

\section{Drift and diffusion coefficients for the Euler angles and $M_0$}
\label{sec:coeffs-euler-angles}

In Secs.~\ref{sec:orientation-adiabatic} and \ref{sec:orientation-white-noise}, we presented the orientation diffusion coefficients in the body-frame basis $(\hat e,\hat q,\hat J)$, where they take a compact, diagonal form. In this appendix, we give the corresponding expressions in the Euler angle parametrization $(\Omega,i,\omega)$, together with the cross-terms involving the mean anomaly at epoch $M_0$. These formulae are needed for numerical integration of the Fokker--Planck equation, since the Euler angles, unlike the body-frame axes, provide a global coordinate system on SO(3) (away from the coordinate singularities at $i=0$ and $\pi$).

The two sets of coefficients are related by the change-of-basis matrix $M^{\hat\mu}{}_\mu$ given in \eqref{eqn:change-of-basis-matrix}. One structural feature that persists in both the adiabatic (Sec.~\ref{sec:adiabatic}) and white-noise (Sec.~\ref{sec:white-noise}) limits is the relation $D^{\Omega\omega}=-D^{\Omega\Omega}\cos i$, which holds because the body-frame diffusion matrix is diagonal (i.e.\ $D^{\hat e\hat q}=D^{\hat e\hat J}=D^{\hat q\hat J}=0$). The dependence of all coefficients on $\omega$ and $i$ is purely geometric: it encodes the orientation of the body-frame axes relative to the fixed reference frame, and disappears upon switching to the $(\hat e,\hat q,\hat J)$ basis.

The cross-term $D^{\omega M_0}$ is the only nonzero element coupling $M_0$ to the Euler angles, since $D^{\Omega M_0}=D^{i M_0}=0$ in both limits. In the body-frame basis this corresponds to $D^{\hat J M_0}=D^{\omega M_0}$, as follows from the third row of the change-of-basis matrix \eqref{eqn:change-of-basis-matrix}.

\subsection{Adiabatic tidal perturbations}

The drift coefficients for the Euler angles are
\begin{align}
B^\Omega&=-\frac{e^2(2+3e^2)\sin(2\omega)}{24(1-e^2)\tan i\sin i\,T\sped{d}}\,,\\
B^i&=\frac{2+6e^2+17e^4-5e^2(2+3e^2)\cos(2\omega)}{240(1-e^2)\tan i\,T\sped{d}}\,,\\
B^\omega&=\frac{e^2(2+3e^2)(3+\cos(2i))\sin(2\omega)}{96(1-e^2)\sin^2i\,T\sped{d}}\,,
\end{align}
and the diffusion coefficients are
\begin{align}
D^{\Omega\Omega}&=\frac{2+6e^2+17e^4-5e^2(2+3e^2)\cos(2\omega)}{120(1-e^2)\sin^2i\,T\sped{d}}\,,\\
D^{\Omega i}&=\frac{e^2(2+3e^2)\sin(2\omega)}{24(1-e^2)\sin i\,T\sped{d}}\,,\\
D^{\Omega\omega}&=-D^{\Omega\Omega}\cos i\,,\\
D^{ii}&=\frac{2+6e^2+17e^4+5e^2(2+3e^2)\cos(2\omega)}{120(1-e^2)T\sped{d}}\,,\\
D^{i\omega}&=-\frac{e^2(2+3e^2)\sin(2\omega)}{24(1-e^2)\tan i\,T\sped{d}}\,,\\
D^{\omega\omega}&=\frac{43(1-e^2)}{60T\sped{d}}+D^{\Omega\Omega}\cos^2i\,.
\end{align}
The $M_0$--orientation cross-term is
\beq
D^{\omega M_0}=-\frac{\sqrt{1-e^2}(67+43e^2)}{60T\sped{d}}\,.
\eeq

\subsection{White-noise tidal perturbations}

The drift coefficients for the Euler angles are
\begin{align}
B^\Omega&=-\frac{7e^2(2+e^2)\sin(2\omega)}{80(1-e^2)\tan i\sin i\,T\sped{d}}\,,\\
B^i&=\frac{8+5e^2(8+3e^2)-21e^2(2+e^2)\cos(2\omega)}{480(1-e^2)\tan i\,T\sped{d}}\,,\\
B^\omega&=\frac{7e^2(2+e^2)(3+\cos(2i))\sin(2\omega)}{320(1-e^2)\sin^2i\,T\sped{d}}\,.
\end{align}
The diffusion coefficients share the same structural features as the adiabatic case, with different numerical prefactors:
\begin{align}
D^{\Omega\Omega}&=\frac{8+5e^2(8+3e^2)-21e^2(2+e^2)\cos(2\omega)}{240(1-e^2)\sin^2i\,T\sped{d}}\,,\\
D^{\Omega i}&=\frac{7e^2(2+e^2)\sin(2\omega)}{80(1-e^2)\sin i\,T\sped{d}}\,,\\
D^{\Omega\omega}&=-D^{\Omega\Omega}\cos i\,,\\
D^{ii}&=\frac{8+5e^2(8+3e^2)+21e^2(2+e^2)\cos(2\omega)}{240(1-e^2)T\sped{d}}\,,\\
D^{i\omega}&=-\frac{7e^2(2+e^2)\sin(2\omega)}{80(1-e^2)\tan i\,T\sped{d}}\,,\\
D^{\omega\omega}&=\frac{28+25e^2-46e^4}{120e^2T\sped{d}}+D^{\Omega\Omega}\cos^2i\,.
\end{align}
The $M_0$--orientation cross-term is
\beq
D^{\omega M_0}=-\frac{\sqrt{1-e^2}(28+145e^2+44e^4)}{120e^2T\sped{d}}\,.
\eeq

\section{First-principles derivation of drift and diffusion coefficients}
\label{sec:first-principles}

In the main text, we follow an algorithmic procedure where we write out the stochastic force $\vec F\sped{s}$ as a function of $\vec w$ through Gauss' equations, and then use formulae \eqref{eqn:Deltaw1}, \eqref{eqn:Deltaw2}, \eqref{eqn:Deltaww1}, and \eqref{eqn:Deltaww2} to determine the drift and diffusion coefficients. While very general and straightforward, this may not always be the route that provides the best physical intuition for the problem at hand. This is especially true for the $(a,e)$ sector, where working with energy and angular momentum $(\mathcal E,J)$ can provide both more direct (albeit less generalizable) derivations and more compact expressions of the Fokker--Planck coefficients. In this appendix, we provide such a derivation of the drift and diffusion coefficients for $(\mathcal E,J)$ for the case of non-adiabatic tidal perturbations, given in \eqref{eqn:BE-white-noise}--\eqref{eqn:DJJ-white-noise}.

Consider the energy first. The instantaneous power input from the tidal field is $\dot{\mathcal E}=-\dot r_iT_{ij}r_j$. Since $\braket{T_{ij}}=0$, the leading contribution to $\braket{\dot{\mathcal E}}$ comes from the velocity perturbation $\delta\dot r_i(\Delta t)=-\int_0^{\Delta t} T_{ik}(t)r_k(t)\dd t$. We thus obtain
\beq
\begin{split}
\braket{\dot{\mathcal E}}&=r_j(0)\int_0^{\Delta t}\braket{T_{ik}(t)T_{ij}(0)}r_k(t)\dd t\\
&=\frac12r_jr_k\frac{Gm}{15a^3T\sped{d}}\mathds{1}_{ikij}\\
&=\frac{r^2}6\frac{Gm}{a^3T\sped{d}}\,,
\end{split}
\eeq
where we used $r_k(t)\approx r_k(0)$ when $t<\Delta t\ll T$. We can then identify $B^{\mathcal E}=\braket{\Delta\mathcal E}/\Delta t=\braket{\dot{\mathcal E}}$, and perform an orbit average to recover the result \eqref{eqn:BE-white-noise} exactly.

The diffusion coefficient follows just as simply: the energy variation is
\beq
\Delta\mathcal E=\int_0^{\Delta t}\dot r_i(t)\,T_{ij}(t)\,r_j(t)\dd t\,,
\eeq
so that
\beq
\begin{split}
D^{\mathcal E\mathcal E}&=\frac{\braket{(\Delta\mathcal E)^2}}{\Delta t}=\dot r_i r_j\dot r_k r_l
\frac{Gm}{15a^3T\sped{d}}\mathds{1}_{ijkl}\,.
\end{split}
\eeq
Using
\beq
\dot r_i r_j\dot r_k r_l\,\mathds{1}_{ijkl}=(\vec r\cdot\dot{\vec r})^2+r^2\dot r^{2}+(\vec r\cdot\dot{\vec r})^2
=
J^2+3r^2\dot r_r^{2}\,,
\eeq
where $\dot r_r\equiv\dot{\vec r}\cdot\hat r$ is the radial velocity, we obtain
\beq
D^{\mathcal E\mathcal E}=\big(J^2+3r^2\dot r_r^{2}\big)\frac{Gm}{15a^3T\sped{d}}\,,
\eeq
which reduces to \eqref{eqn:DEE-white-noise} after using the orbit-averaged relation $\overline{r^2\dot r_r^{2}}=Gmae^2/2$.

\begin{figure*}
    \centering
    \includegraphics[width=\textwidth]{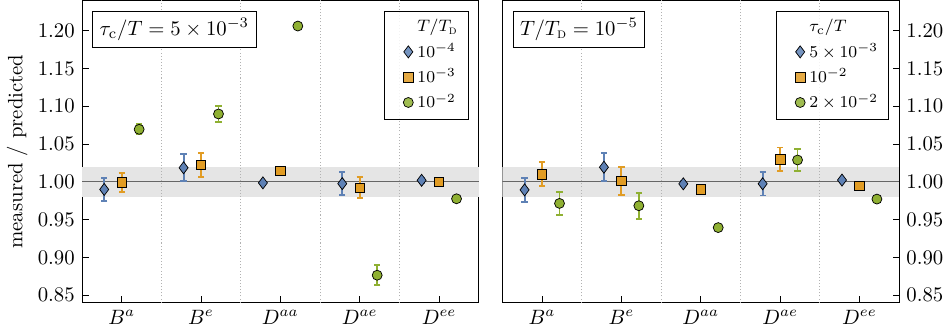}
    \caption{Ratio of the Fokker--Planck coefficients extracted numerically to the analytical predictions \eqref{eqn:Ba-white-noise}--\eqref{eqn:Dee-white-noise} in the case of white-noise tidal perturbations. Left: varying $T/T\sped{d}$ at fixed $\tau\ped{c}/T=\num{5e-3}$. Right: varying $\tau\ped{c}/T$ at fixed $T/T\sped{d}=\num{e-5}$. The error bars are $1\sigma$ Monte Carlo statistical uncertainties, and are smaller than the marker sizes in some cases. The gray band marks the $\pm2\%$ region. As expected, the ratios approach $1$ when $T/T\sped{d}\to0$ and $\tau\ped{c}/T\to0$.}
    \label{fig:white-noise-numerics}
\end{figure*}

We can obtain the cross-diffusion coefficient similarly. Since $\dot J=r^2T_{ij}\hat r_i\hat\phi_j$, one finds
\beq
\begin{split}
D^{\mathcal EJ}
&=\frac{\braket{\Delta\mathcal E\,\Delta J}}{\Delta t}\\
&=\dot r_i r_j\,r^2\hat r_k\hat\phi_l
\frac{Gm}{15a^3T\sped{d}}\mathds{1}_{ijkl}\\
&=Jr^2\frac{Gm}{15a^3T\sped{d}}\,,
\end{split}
\eeq
whose orbit average gives \eqref{eqn:DEJ-white-noise}. For the angular momentum itself, $\vec J=\vec r\times\dot{\vec r}$, we can first compute its diffusion coefficient as
\beq
\begin{split}
D^{JJ}&=\frac{\braket{(\Delta J)^2}}{\Delta t}\\
&=\frac1{\Delta t}\Braket{\bigg(r^2\int_0^{\Delta t}T_{ij}(t)\hat r_i\hat\phi_j\dd t\bigg)^2}\\
&=r^4\hat r_i\hat\phi_j\hat r_k\hat\phi_l\mathds{1}_{ijkl}\frac{Gm}{15a^3T\sped{d}}\\
&=r^4\frac{Gm}{15a^3T\sped{d}}\,,
\end{split}
\eeq
which gives \eqref{eqn:DJJ-white-noise}. To obtain the drift coefficient, we first compute
\beq
\begin{split}
B^{J^2}&=\Braket{\frac{\dd J^2}{\dd t}}\\
={}&2\braket{r^2\dot r_i\ddot r_i-r_i\dot r_ir_j\ddot r_j}\\
={}&2r^2\int_0^{\Delta t}\braket{T_{ik}(t)r_k(t)T_{ij}(0)r_j(0)}\dd t\\
&-2r_ir_j\int_0^{\Delta t}\braket{T_{ik}(t)r_k(t)T_{jl}(0)r_l(0)}\dd t\\
={}&(r^2r_kr_j\mathds{1}_{ikij}-r_ir_jr_kr_l\mathds{1}_{ikjl})\frac{Gm}{15a^3T\sped{d}}\\
={}&2r^4\frac{Gm}{15a^3T\sped{d}}\,.
\end{split}
\eeq
The drift $B^J$ can then be obtained from \eqref{eqn:B-transform} as 
\beq
B^J=\frac{B^{J^2}-D^{JJ}}{2J}=\frac{r^4}{2J}\frac{Gm}{15a^3T\sped{d}}\,,
\eeq
whose orbit average matches \eqref{eqn:BJ-white-noise}. Because of the relation $B^J=D^{JJ}/(2J)$, the evolution of the angular momentum $J$ is equivalent to the distance from the origin in a two-dimensional Brownian motion. Even though the average torque vanishes, $\braket{\dot{\vec J}}=0$, the norm of the angular momentum still drifts to larger values on average.

\section{Numerical verification of the white-noise tidal coefficients}
\label{sec:numerical-check}

The second-order truncation of the Kramers--Moyal expansion and the resulting Fokker--Planck equation hold in the limit of weak perturbation, $T/T\sped{d}\to0$. Furthermore, in the main text we derived the drift and diffusion coefficients either in the $\tau\ped{c}/T\to0$ or $T/\tau\ped{c}\to0$ limit. To verify the accuracy of our results in these limits, in this appendix we integrate the binary's motion directly and extract the coefficients from the ensemble statistics. For concreteness, we focus on the white-noise tidal regime, described in Sec.~\ref{sec:white-noise}.

\subsection{Tidal field construction}

First, we numerically construct a stochastic tidal tensor $T_{ij}(t)$ whose integrated two-point function reproduces the correlator \eqref{eqn:Td-tidal}, which we copy here for ease of reference
\beq
\int_{-\infty}^\infty\braket{T_{ij}(t)T_{kl}(t+\tau)}\dd\tau=\frac{Gm}{a^3T\sped{d}}\,\frac{\mathds{1}_{ijkl}}{15}\,.
\eeq
Interpreted as a linear map on the six-dimensional space of symmetric $3\times3$ matrices, $S_{ij}\mapsto\mathds{1}_{ijkl}S_{kl}=\delta_{ij}\tr S+2S_{ij}$, the tensor $\mathds{1}_{ijkl}$ has eigenvalue $5$ on the identity and $2$ on the five traceless modes. We thus write
\beq
T_{ij}(t)=\zeta(t)\,\delta_{ij}+\sum_{A=1}^5\xi_A(t)\,e_{ij}^A\,.
\eeq
where $\{e_{ij}^A\}$, $A=1,\ldots,5$ is an orthonormal basis of symmetric traceless matrices, while $\zeta(t)$ and $\xi_A(t)$ are six independent Ornstein--Uhlenbeck (OU) processes with autocorrelation $e^{-\abs{\tau}/\tau\ped{c}}$ and variances
\beq
\sigma_\zeta^2=\frac{Gm}{18a^3T\sped{d}\tau\ped{c}}\,,\qquad\sigma_\xi^2=\frac{Gm}{15a^3T\sped{d}\tau\ped{c}}\,.
\eeq
This choice reproduces \eqref{eqn:Td-tidal} exactly, as can be verified by direct substitution. Given a time step $\delta t$, we numerically advance the OU processes according to $\xi_{n+1}=e^{-\delta t/\tau\ped{c}}\,\xi_n+\sigma\sqrt{1-e^{-2\delta t/\tau\ped{c}}}\,Z$, where $Z$ is a normally distributed random variable.

\subsection{Orbit integration}

We then consider an ensemble of binaries, all having the same initial semi-major axis and eccentricity $e_0=0.3$, and solve the equation of motion of the relative separation, $\ddot r_i=-Gm\,r_i/r^3-T_{ij}(t)r_j$. The mean anomalies at epoch $M_0$ are drawn uniformly in $[0,2\pi)$, so that the ensemble average also naturally provides the orbit average. Each binary evolves under an independent realization of $T_{ij}(t)$. After a time $\Delta t$ corresponding to 3 orbital periods, we extract from $(\vec r,\dot{\vec r})$ the values of $(a,e)$ and measure
\begin{align}
B^\mu&=\frac{\braket{\Delta w^\mu}}{\Delta t}\,,\\
\label{eqn:Dmunu-numerical}
D^{\mu\nu}&=\frac{\braket{\Delta w^\mu\Delta w^\nu}-\braket{\Delta w^\mu}\braket{\Delta w^\nu}}{\Delta t}\,,
\end{align}
for $w^\mu\in\{a,e\}$, which can be directly compared with the analytical results \eqref{eqn:Ba-white-noise}--\eqref{eqn:Dee-white-noise}.

Note that in \eqref{eqn:Dmunu-numerical} we compute $D^{\mu\nu}$ from the covariance of the increments, i.e., subtracting the term $\braket{\Delta w^\mu}{\braket{\Delta w^\nu}}$ in the numerator. This term is not present in \eqref{eqn:drift-diffusion-def} because it is higher order in $\Delta t$. For the purpose of numerical extraction, however, its inclusion is useful to mitigate finite-$(\Delta t/T\sped{d})$ effects. The extraction of the drift coefficients $B^\mu$ also requires care. These are second order in $T_{ij}$, but ride on top of a much larger first-order effect due to the particular noise realization. We suppress this noise with \emph{antithetic variates}~\cite{1956PCPS...52..449H}: we evolve each binary twice, under the tidal fields generated by $(\zeta,\xi_A)$ and $(-\zeta,-\xi_A)$. The aforementioned first-order effect cancels when taking the average of $\Delta w^\mu$ over the two antithetic integrations. The diffusion coefficients use instead the raw $\Delta w^\mu$.

\subsection{Results}

We show in Fig.~\ref{fig:white-noise-numerics} the ratio of the coefficients extracted numerically to the analytical expressions. We calculate the error bars using the standard error of the mean for the drift coefficient, and bootstrapping the uncertainties for the diffusion coefficients~\cite{Efron:1979bxm}. In the left panel, we hold $\tau\ped{c}/T=\num{5e-3}$ fixed, and approach the weak-perturbation limit by decreasing $T/T\sped{d}$. In the right panel, we hold $T/T\sped{d}=\num{e-5}$ fixed, and approach the white-noise limit by decreasing $\tau\ped{c}/T$. In both limits, the measured coefficients match the analytical predictions within percent-level error bars. Away from the weak-perturbation or white-noise limits, deviations from the analytical predictions become apparent when $T/T\sped{d}\gtrsim\num{e-2}$ or $\tau\ped{c}/T\gtrsim \num{2e-2}$.

\bibliography{main}

\end{document}